\documentclass[12pt]{iopart}
\usepackage{iopams}  
\usepackage{graphicx}
\usepackage{epsfig}

\usepackage[dvipsnames,table]{xcolor}  
\usepackage{wrapfig}
\usepackage[font={small,it}]{caption}
\usepackage{tcolorbox}

\usepackage[inline]{enumitem}
\usepackage{tabularx}
\usepackage[numbers]{natbib}
\usepackage{hyperref}
\usepackage{todonotes} 
\usepackage{rotating}
\usepackage{outlines}
\graphicspath{{./figures/}}

\hypersetup{
  colorlinks   = true,     
  urlcolor     = NavyBlue, 
  linkcolor    = NavyBlue, 
  citecolor   =  NavyBlue  
}

\usepackage{nuclides}
\usepackage{units}

\begin{document}

\pagenumbering{roman}

\title[Horizons: Nuclear Astrophysics in the 2020s and Beyond]{Horizons: Nuclear Astrophysics in the 2020s and Beyond}

\author{
H   Schatz$^{  1,  2,  3}$,
A D Becerril Reyes$^{  2,  3}$,
A   Best$^{  4,  5}$,
E F Brown$^{  1,  2,  6,  3}$,
K   Chatziioannou$^{  7,  8}$,
K A Chipps$^{  9, 10}$,
C M Deibel$^{ 11}$,
R   Ezzeddine$^{ 12,  3}$,
D K Galloway$^{ 13, 14, 15}$,
C J Hansen$^{ 16, 17, 18}$
% ...
}

\address{$^{  1}$ Department of Physics and Astronomy, Michigan State University, East Lansing, MI 48824, USA}
\address{$^{  2}$ Facility for Rare Isotope Beams, Michigan State University, East Lansing, MI 48824, USA}
\address{$^{  3}$ Joint Institute for Nuclear Astrophysics -- Center for the Evolution of the Elements (JINA-CEE), USA}
\address{$^{  4}$ Department of Physics, University of Naples "Federico II", 80126 Napoli, Italy}
\address{$^{  5}$ INFN - Sezione di Napoli, 80126 Napoli, Italy}
\address{$^{  6}$ Department of Computational Mathematics, Science and Engineering, Michigan State University, East Lansing, MI 48824, USA}
\address{$^{  7}$ Department of Physics, California Institute of Technology, Pasadena, California 91125, USA}
\address{$^{  8}$ LIGO Laboratory, California Institute of Technology, Pasadena, California 91125, USA}
\address{$^{  9}$ Physics Division, Oak Ridge National Laboratory, Oak Ridge, TN 37831, USA}
\address{$^{ 10}$ Department of Physics and Astronomy, University of Tennessee Knoxville, Knoxville TN 37996, USA}
\address{$^{ 11}$ Department of Physics and Astronomy, Louisiana State University, Baton Rouge, LA 70803, USA}
\address{$^{ 12}$ Department of Astronomy, University of Florida, Gainesville, FL 32601, USA}
\address{$^{ 13}$ School of Physics and Astronomy, Monash University, VIC 3800, Australia}
\address{$^{ 14}$ OzGRav-Monash, School of Physics \& Astronomy, Monash University, Victoria 3800, Australia}
\address{$^{ 15}$ Institute for Globally Distributed Open Research and Education (IGDORE)}
\address{$^{ 16}$ Goethe University Frankfurt, Institute for Applied Physics, Max-von-Laue-Str. 12, 60438 Frankfurt am Main, Germany}
\address{$^{ 17}$ Technical University of Darmstadt, Institute for Nuclear Physics, Germany}
\address{$^{ 18}$ Max Planck Institute for Astronomy, Heidelberg, Germany}
% ...
\address{$^{108}$ NuGrid Collaboration, \url{http://nugridstars.org}}
% ...


\date{\today}

\begin{abstract}
Nuclear Astrophysics is a field at the intersection of nuclear physics and astrophysics,
which seeks to understand the nuclear engines of astronomical objects and the origin of
the chemical elements. This white paper summarizes progress and status of the field, the new open questions 
that have emerged, and the tremendous scientific opportunities that have opened up with major advances in capabilities across an ever growing 
number of disciplines and subfields that need to be integrated. We take a holistic view of the field discussing the unique challenges and opportunities in nuclear astrophysics in regards to science, 
diversity, education,  
and the interdisciplinarity and breadth of the field. Clearly nuclear astrophysics is a dynamic field with a bright future that is entering a new era of discovery opportunities. 
\end{abstract}

\pacs{00.00, 20.00, 42.10}
\submitto{\jpg}
%
%
\maketitle

\clearpage

{\tableofcontents{}}
\title[Horizons: Nuclear Astrophysics in the 2020s and Beyond]{}
\pagenumbering{arabic}
\setcounter{page}{1}

\section{Executive Summary}

Nuclear Astrophysics is a field at the intersection of nuclear physics and astrophysics, which seeks to understand the nuclear engines of astronomical objects and the origin of the chemical elements. These topics are linked to the origin of the building blocks of life and the world we live in and have been at the forefront of science and philosophy since ancient times. They also have been consistently defined as high priority challenges by the scientific community and National Academies studies \citep{NAP10079,NP2010}. 

The last decade has seen major progress in this field and as a result a more complex, and likely more complete, picture of the nuclear processes in the cosmos is emerging. Long standing paradigms are shifting about the physics of low-energy reactions in stars, the mixing of different layers inside stars, the composition of the Sun, the variety of types of processes responsible for the origin of the heavy elements, and the remnants of stellar explosions. These paradigm shifts result in a multitude of new questions and coincide with an extraordinary confluence of transformational advances in capabilities in each of the key fields that make up nuclear astrophysics - nuclear science, astronomy, and computational modeling - that is unique in the history of the field. 

In nuclear science, new rare isotope accelerator facilities such as the Facility for Rare Isotope Beams (FRIB) in the US are beginning to provide experimental access to the short-lived rare isotopes that shape the heavy element composition of the cosmos. A new generation of deep underground accelerators are enabling measurements of the extremely slow nuclear reactions that power the Sun and other stars. 
In astronomy, gravitational wave detection is transforming the field. An important new capability is the ability to carry out multi-messenger observations, large scale coordinated observations using multiple types of radiation, particles, and gravitational waves. These new capabilities have already led to major new insights into neutron star mergers as heavy element nucleosynthesis sites, as well as dense matter physics and black hole properties of key importance for nuclear astrophysics. Other transformational advances in astronomy of great importance for this field are the advent of asteroseismology, greatly improved techniques to analyze stardust, the spectroscopy of late type stars that directly reveal elements created in their centers, and the large-scale spectroscopy of ancient stars that is now providing a detailed “fossil record” of how the Galaxy was enriched with new elements over its history. 
In computational modeling, a key insight from the last decade is that new nuclear pathways for element synthesis can open up, driven by 
complex fluid motions in dynamic and often asymmetric environments such as interior regions of stars, supernova explosions and neutron star mergers.  With increasing computational capabilities, the required multi-dimensional computer simulations are now crossing a threshold for achieving the level of detail needed to understand the nuclear processes. 

The full scientific potential of these capabilities will be unlocked in future work through global interdisciplinary networks (e.g., JINA and IReNA in the US, ChETEC and ChETEC-INFRA in Europe, Ukakuren in Japan) that have proven to be powerful tools for integrating research efforts of the experimental, observational, and theory communities. Together, these advances across fields define a new era in nuclear astrophysics with unprecedented scientific opportunities.

This white paper summarizes the discussions at the JINA Horizons community town meeting held virtually on November 30 – December 4, 2020. The meeting and this white paper are part of a series organized by the Joint Institute for Nuclear Astrophysics (JINA) over the last two decades, including the 1999 town meeting at the University of Notre Dame, and the 2012 town meeting in Detroit. JINA Horizons brought together 575 scientists from across the world and from all relevant fields to discuss the newly emerging scientific opportunities in the era of multi-messenger astronomy and rare isotope nuclear science. It complements community-driven activities within nuclear science and astronomy in that it is an interdisciplinary effort that combines perspectives from both fields to develop a common vision. On the astronomy side, the ASTRO2020 Decadal Survey has been published recently, with field-specific priorities well aligned with this work. 

Identifying pathways towards improving the diversity and work environment of the nuclear astrophysics community were also an important part of the meeting. The community concluded that this goal should be treated on equal footing with the scientific goals for nuclear astrophysics. 
Early career scientists are the future of the field. Therefore, the meeting was preceded by an early career researcher workshop offering professional development opportunities. Early career researchers also constituted a significant fraction of the speakers, and played important roles as working group conveners. The community identified goals for improving training and career perspectives of young scientists. 

This white paper identifies the exciting scientific opportunities in the field and what is needed to take full advantage of them in the coming decade: 

\begin{itemize}
    \item {\bf Experimental nuclear science:} In experimental nuclear science, a priority is the optimal operation of the FRIB rare isotope accelerator facility, including the FRIB400 energy upgrade to 400 MeV/u, and the completion of key equipment for nuclear astrophysics such as the SECAR recoil separator, the FRIB Decay Station (FDS), the Gamma-Ray Energy Tracking Array (GRETA), isotope harvesting, the High Rigidity Spectrometer (HRS), and the Isochronous Spectrometer with Large Acceptances (ISLA). Operation of the ATLAS facility at Argonne National Laboratory with complementary stable and radioactive beam capabilities will also be important, as are the complementary capabilities of future international rare isotope facilities such as FAIR in Germany. Deep underground accelerators provide unique capabilities that are essential for studies of stellar burning. Essential new capabilities for nuclear astrophysics are also provided by a broad range of stable beam, neutron beam, $\gamma$-beam, and laser facilities, including the use of recoil separators, active targets, and high density and temperature plasmas. 
    \item {\bf Nuclear Theory:} Advances in nuclear theory related to nuclear reactions are crucial, especially at the lowest energies, as well as predictions of the properties of the most neutron rich and the heaviest nuclei, including mechanisms for their synthesis in the laboratory, and reliable prediction of various modes of nuclear fission. In addition, theory needs to provide models of dense matter properties across all relevant densities. To achieve these goals, the field should take advantage of synergies with multi-institutional nuclear theory centers and coalitions such as the FRIB Theory Alliance. 
    
    \item {\bf Astronomy:} In astronomy, increases in the sensitivity of gravitational wave and neutrino detectors, expanded samples and improved analysis of spectroscopic data on ancient stars, asteroseismology, time domain observations to identify transient phenomena, and a new nuclear $\gamma$-ray mission are ongoing developments of special importance for nuclear astrophysics. Future capabilities that would be important for the community include an advanced X-ray telescope and a capability for space-based UV spectroscopy such as JWST and LUVOIR (see ASTRO2020 \cite{ASTRO2020}). 
    \item {\bf Cosmochemistry:} In cosmochemistry, advances in analyzing stardust as a complementary messenger from nucleosynthesis sites are salient, as well as more and improved measurements of elemental and isotopic abundances in meteorites and other planetary systems.
    \item {\bf Astrophysics Theory:} 
    Advances in astrophysics theory are needed in all relevant astrophysical scenarios where nuclear
processes play a key role. It is important that the fidelity of the models continues to increase as more
detailed and quantitative observations become available, that the performance of the models is increased
to enable a full exploration of the diversity in stellar and explosive environments, that the expected
advances in quantity and precision of nuclear data can be fully implemented, and that uncertainties can be quantified.
    
     \item {\bf Diversity, Equity, Inclusion, and Accessibility:} Diversity, equity, inclusion, and accessibility are important goals for nuclear astrophysics that should be pursued by the community on equal footing to the science goals (Section~\ref{sec:diversity}). 
    \item {\bf Early Career Researchers:} Improvements are needed in mentoring early career researchers and preparing them for the broad range of exciting careers that training in nuclear astrophysics opens up. 
    \item {\bf Centers:} Center-based networks are essential for creating the collaborative connections and exchange across field boundaries, institutions, and countries that are needed for advancing nuclear astrophysics. They provide an overarching umbrella for experimental, theoretical, and observational developments as well as sustainable software ecosystems, and for shaping the frontiers of the field  as new discoveries are made and new capabilities emerge.  
    \item {\bf Software Instruments:} Software is an integral enabler of experiment, observation, theory, and computation and a primary modality for realizing discoveries and innovations.  Nearly all research relies on free, open-knowledge software written and maintained by a small number of developers. Figuring out how to support a thriving open-knowledge digital infrastructure may seem daunting, but there are plenty of reasons to see the road ahead as an opportunity. 
   \item {\bf Data:} Data evaluation, transformation, and dissemination efforts that enable and facilitate usage of nuclear and astrophysical data across field boundaries are essential for the field. Nuclear data efforts such as JINA REACLIB \cite{Cyburt2016}, Starlib \cite{sallaska_2013_aa}, BRUSLIB \cite{Xu2013}, or pynucastro \cite{pynucastro} that make nuclear data available for astrophysical applications, as well as astronomy efforts such as JINABase \cite{Abohalima2018} that make observational data available to nuclear astrophysicists should be continued and expanded. 

\end{itemize}
\clearpage

\section{Dynamic Nuclear Burning in Stars} \label{sec:stars}



\begin{wrapfigure}{r}{0.5\textwidth}
\vspace{-0.30in}   
  \begin{center}
    \includegraphics[width=0.48\textwidth]{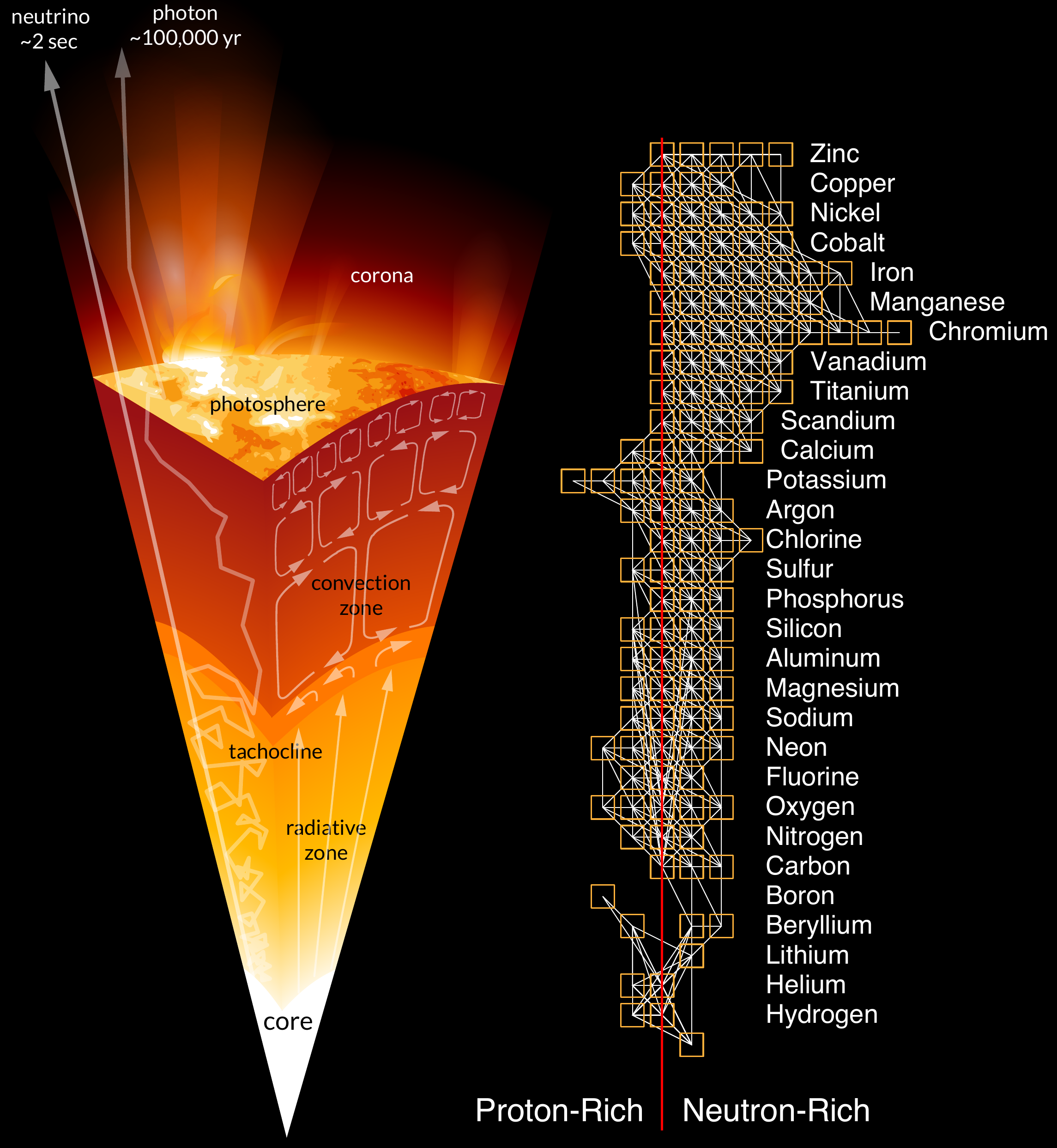}
  \end{center}
  \vspace{-0.20in} 
  \captionsetup{margin=0.2cm}
  \caption{Layers in a Sun-like star and a 
          schematic nuclear reaction network to follow the nuclear energy generation and nucleosynthesis.
          Credit: Adapted from an illustration by Kelvin Ma.}
  \label{fig:wg2:stellar_interior_reactions_black}
\vspace{-0.20in}  
\end{wrapfigure}

\vspace{-0.15in}
A star is a luminous ball of plasma held together by gravity. Stars
are the most commonly observed objects in the universe, and derive
their power from natural nuclear fusion reactors in their cores.  Through
the nuclear reactions that power them, stars have taken primordial
hydrogen and helium forged during the Big Bang and used it to produce the majority of the chemical elements found in nature (see
Figure~\ref{fig:wg2:stellar_interior_reactions_black}). These stars then
dispersed this material -- sometimes by spectacular explosions -- so
it became incorporated in subsequent generations of stars and the
planets that accompany them \citep{Costantini2009,West2013, lrp_2015_aa,bertulani_2016_aa,schatz_2016_aa, arcones_2017_aa, Grimmett2018,Broggini2018,kobayashi_2020_aa,nunes_2020_aa,rapisarda_2021_aa, aziz_2021_aa}. The underlying nuclear reaction processes occurring during the various phases of stellar evolution are guided by specific properties of nuclei that are then reflected in the observed cosmic abundance distributions.

Despite the first models of stars being
created in the early 19$^{\rm th}$ century 
\citep{thomson_1862_aa, jeans_1928_aa, gamow_1938_aa,hufbauer_1981_aa, kragh_2016_aa}
and the nuclear reaction sequences for hydrogen burning being
described by the mid 20$^{\rm th}$ century
\citep{eddington_1920_aa, bethe_1939_aa, burbidge_1957_aa, cameron_1957_aa, wallerstein_1997_aa, bahcall_2003_aa}, 
stars and their explosions remain poorly understood. This begins with the best-studied star, our Sun, where its composition from
surface to core is still an open question \citep{sitterly_1939_aa, ross_1976_aa, anders_1989_aa, asplund_2009_aa, lodders_2009_aa, piersanti_2007_aa, asplund_2021_aa, lodders_2021_aa}. 
One challenge in understanding
stars is the the uncertain rates of the slow nuclear reactions in
their interior \citep[e.g.,][]{sallaska_2013_aa,iliadis_2016_ab}. These reactions determine the neutrino flux from the Sun~\cite{Adel11}, the evolutionary fate of stars, and the masses of black holes~\cite{farmer_2020_aa}. The slowness of the reactions enables stars to live
for extended periods of time ranging from millions of years to longer than the current age of the universe \citep{bouvier_2010_aa}, but also makes experimental measurements
extraordinarily difficult \citep[e.g.,][]{deboer_2017_aa,deboer_2021_aa}. Present reaction rates are therefore mostly based on theoretical model extrapolations, which require a detailed understanding of the quantum mechanical reaction mechanism at the energy threshold, including the effects of screening and nuclear clusters. Other fundamental, yet poorly understood,
aspects of stars and their explosions include convection, mixing, rotation, magnetic fields, mass loss, and the influence of companion stars
\citep{Heger2000,Heger2005,karakas_2014_aa, stancliffe_2015_aa, buldgen_2019_aa, ekstrom_2021_aa}. 
It is becoming clearer that while these effects critically impact element synthesis and nuclear reactions they cannot
always be adequately modeled in spherical symmetry (one spatial
dimension). As computationally challenging multi-dimensional models are developed \citep[e.g.,][]{herwig_2014_aa, woodward_2015_aa, couch_2015_aa, Muller2016, baraffe_2017_aa, Jones2017, pratt_2020_aa, lovekin_2020_aa, fields_2020_aa, fields_2021_aa,Bollig21}, validation of
their predictions using observations of stars becomes crucial \citep{aerts_2010_aa,borexino-collaboration_2018_aa, borexino-collaboration_2020_aa}. 
Such validations require a much improved understanding of the underlying 
nuclear reaction rates, which are determined by the low-energy nuclear cross sections and the plasma
conditions in the stellar interior \citep[e.g,][]{iliadis_2015_aa}. 
Advances in 
experimental nuclear astrophysics facilities \citep{Sher18,Gatu17,Sava00,Aliotta2021}, laboratory plasma experiments \cite{Casey2017}, 
multi-wavelength detector technologies \citep{middleton_2017_aa, kuulkers_2019_aa, meszaros_2019_aa},
gravitational wave detectors \citep{ligo-scientific-collaboration_2015_aa,acernese_2015_aa}, 
neutrino astronomy detectors \citep{araki_2005_aa, andringa_2016_aa, acciarri_2016_aa, brugiere_2017_aa, newstead_2019_aa, simpson_2019_aa}, 
computer processing power, 
ubiquitous cloud capabilities,
and open-source software instruments 
\citep[e.g.,][]{paxton_2011_aa, paxton_2013_aa, paxton_2015_aa, paxton_2018_aa, paxton_2019_aa, azuma_2010_aa, sallaska_2013_aa, 
townsend_2013_aa, townsend_2018_aa, astropy-collaboration_2013_aa,astropy-collaboration_2018_aa, 
vanderplas_2012_aa, foreman-mackey_2013_aa,hogg_2018_aa, vanderplas_2018_aa, baron_2019_aa, timmes_2020_aa, huerta_2021_aa}
are paving the way for significant new approaches to a better understanding of stars in the 21$^{\rm st}$ century.

\begin{tcolorbox}[colback=green!5, colframe=green!40!black, title=Sidebar: Exploring the Heart of Stars]
  \begin{center}
    \includegraphics[width=01\textwidth]{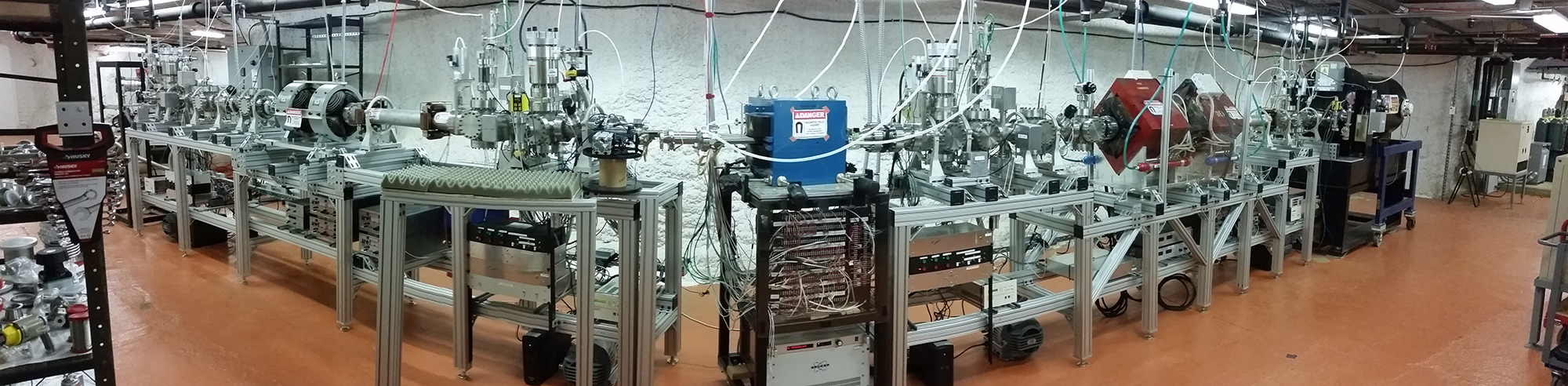}
  \end{center}
  \vspace{-0.20in} 
  \begin{center}
  {\footnotesize {\it CASPAR facility. Credit: University of Notre Dame.}} \label{fig:wg2:frib} \\  
  \end{center}
  \vspace{-0.08in} 
{\footnotesize
New experimental facilities with beams of stable isotopes are dramatically increasing 
the sensitivity for measurements of the slow nuclear reactions that create elements in stars. Deep undergound 
facilities like CASPAR in the US, located 4850 ft below the Earth's surface at the Sanford Underground Research Facility (SURF), LUNA at 
Gran Sasso National Laboratories (LNGS) in Italy, and JUNA at the China Jinping Underground Laboratory,
enable measurements free from backgrounds caused by cosmic radiation. 
Upgrades of above-ground accelerator facilities employ new technologies to achieve 
increased sensitivity. Each of these facilities have unique capabilities 
that offer opportunities for answering fundamental questions about 
the origin of the elements and the evolution of stars. 


}
\label{sidebar:underground}
\end{tcolorbox}

\vspace{-0.15in}
\subsection{Open Questions} \label{sec:stars:questions}


\begin{itemize}

\item What are the rates of the proton, neutron, and $\alpha$-particle capture reactions, photodissociation reactions, and the carbon and oxygen fusion reactions in stars, 
and what are the best means to determine these rates at stellar energies?

\item
How do mixing, rotation, magnetic fields, and mass loss  affect the lives of stars?

\item
What are the nuclear reactions in the first stars formed after the Big Bang and what elements did these stars produce? 

\item 
How do massive stars trigger a supernova explosion?

\item 
How do compact objects form and what controls their mass, radius and spin?

\item
What can photons, neutrinos, gravitational waves, and stardust tell us about stars?

\end{itemize}


\begin{tcolorbox}[colback=blue!5, colframe=blue!40!black, title=Sidebar: Studying Stars in the Laboratory]
  \begin{center}
    \includegraphics[width=0.8\textwidth]{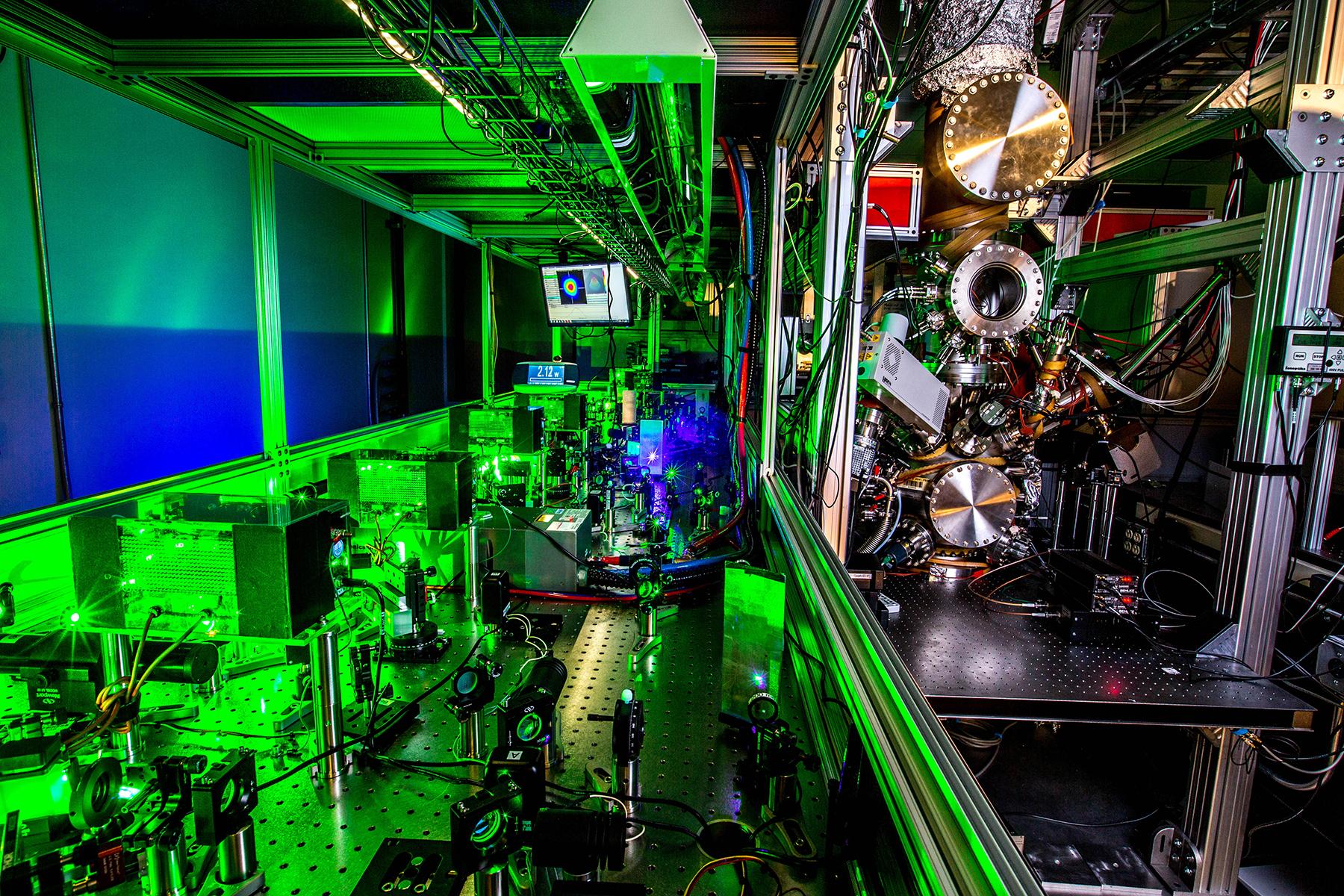}
  \end{center}
  \vspace{-0.20in} 
  \begin{center}
  {\footnotesize {\it The Livermore Ionization of Neutrals (LION) instrument at Lawrence Livermore National
Laboratory (LLNL) allows for trace-isotopic analysis of multiple elements simultaneously in
individual stardust grains. Credit: Lawrence Livermore National Laboratory}} \label{fig:lion} \\  
  \end{center}
  \vspace{-0.08in} 
{\footnotesize
Micrometer-sized stardust grains allow us to study the nuclear fingerprints of stellar outflows in
laboratories on the Earth. While nanoprobes such as the NanoSIMS have been utilized for the last two
decades to analyze the major-element isotopic composition in stardust grains, recent
developments in resonance ionization mass spectrometry (RIMS) enable, for the first time, to
regularly analyze stardust grains for their trace-elements isotopic composition of multiple
elements simultaneously. Such measurements give valuable constraints on nucleosynthesis
pathways and galactic chemical evolution.
}
\label{sidebar:lion}
\end{tcolorbox}

\subsection{How Did We Get Here?} \label{sec:stars:how}

One of the foundations upon which astronomy rests are the fundamental
properties of stars throughout their evolution.  We have arrived at a
threshold of a significant new understanding of stars through the
advent of transformative capabilities at experimental nuclear
astrophysics facilities and a new generation of multi-messenger
telescopes.

On the experimental side, new accelerator laboratories deep
underground (CASPAR in the US \citep{Robertson2016}, LUNA-MV in Italy \citep{Costantini2009,Broggini2018,Sen2019,Prati2020}, and JUNA in China \citep{Zhang2021}) and novel approaches in above-ground facilities have greatly increased the sensitivity of direct reaction-rate measurements. The latter include the use of recoil separators such as ERNA at INFN/Naples \cite{Buompane2018}, St. George at the University of Notre Dame \cite{Meisel2017b}, and DRAGON at TRIUMF \cite{Williams2020,Lennarz2020}; as well as new detector technologies such as the use of high-resolution silicon detector arrays \cite{Matei2020} or active targets that track individual reaction products at TUNL's HI$\gamma$S \cite{Smit2021}, quasi-spectroscopic neutron detectors \cite{FEBBRARO2018189}, or compact coincidence detection configurations at TUNL's LENA \cite{Dermigny2020}. In addition, indirect experimental methods such as the Trojan Horse Method or those used to extract Asymptotic Normalization Coefficients have been developed, enabling the determination of reaction rates that are too low for direct measurements \cite{Tribble14,Kiss2020,Tumino:2021}. In addition, the prediction of reaction rates by \textit{ab initio}  many-body theory (or ``from first principles") has made remarkable progress 
\cite{FRIBTAwhite2018,QuaglioniN20}), and the calculation of reactions beyond the lightest nuclei has now become feasible, e.g., for $\alpha$-capture reactions \cite{DreyfussLESBDD20} and nucleon scattering \cite{LauneyMD_ARNPS21,MecenneLDEQSD21}.

Precise predictions of the solar neutrino spectrum from CNO cycle burning~\cite{borexino-collaboration_2020_aa}, as well as the associated stellar burning lifetimes, have been enabled by measurements employing a variety of techniques. For instance, the key CNO cycle reaction $^{14}{\rm N}(p,\gamma)^{15}$O is now constrained by a combination of direct measurements deep underground and above-ground with high beam intensities, activation measurements, and indirect measurements of the lifetimes of excited states~\cite{Form2004,Li2016,Fren2021,GHK2022}. The s-process neutron sources $^{13}{\rm C}(\alpha,n)^{16}$O and $^{22}${\rm Ne}($\alpha$,n)$^{25}$Mg as well as competing reactions such as $^{22}$Ne($\alpha$,$\gamma$)$^{26}$Mg are being investigated through a combination of direct and indirect measurements that will be connected through comprehensive R-matrix analyses \cite{Heil2008,Longland2012,LaCognata2013,Trippella2017,cristallo_2018_apj, ciani_2021_prl, deboer_2020_prc,Jaeg01,Jaya20,Adsley21}. These measurements have been enabled by new detectors and experimental techniques, for instance total cross section measurements at the Edwards Accelerator Laboratory at Ohio University and deep underground accelerators such as LUNA and JUNA, as well as partial cross section measurements with an ORNL-developed deuterated scintillator array at the University of Notre Dame. Though higher precision is still needed, experimental campaigns to determine the
$^{12}$C($\alpha$,$\gamma$)$^{16}$O reaction have drastically improved estimates of the C/O mass ratio resulting from core helium burning, with significant downstream implications for stellar burning. These measurements relied on recoil separators, precision $\gamma$-spectroscopy, photon beams, and a variety of indirect techniques to arrive at a reaction rate uncertainty on the order of 20\%~\cite[e.g][]{deboer_2017_aa,Schur05,Mate06,Avil15,Smit2021}. At the HI$\gamma$S facility at TUNL, $\gamma$-beams have been successfully employed to develop new techniques for measurements of the $^{12}{\rm C}(\alpha,\gamma)^{16}$O reaction \cite{Smit2021}. Meanwhile, traditional precision $\gamma$-spectroscopy, for instance at the Edwards Accelerator Laboratory, has provided key insights to complement direct measurements~\cite{Mate08,Sayr12}. Advances in neutron beam facilities have led to drastically improved understanding of the neutron-capture rates in the s-process, enabling detailed predictions of the distribution of heavy elements produced. In particular, a plethora of new data on neutron-capture rates in the weak s-process (responsible for element synthesis up to $\sim$Zr) have led to drastic changes in nucleosynthesis predictions. These new data were obtained at Karlsruhe, Argonne National Laboratory (accelerator mass spectroscopy analysis), n-TOF at CERN, SARAF, J-PARC, and LANSCE at Los Alamos National Laboratory (see \cite{Aliotta2021} for an overview). 

On the observational side, the detection of neutrinos emitted by the nuclear reactions in the core of the Sun by Davis \cite{Bahcall1982} and by the collapsing core of supernova 1987A \cite{Mirizzi2016} opened a pathway to directly determine the rate of ongoing nuclear reactions in stellar environments and at the same time confirm and further refine the basic theories describing these stellar objects. In the case of the Sun, precise nuclear reaction data were key in establishing the solar neutrino problem, a deficiency of detected neutrinos compared to expectations, and ultimately led to the discovery of neutrino oscillations and the determination of a neutrino mixing angle \cite{Haxton2013}. The recent detection of neutrinos from the CNO cycle operating in the solar core by BOREXINO \cite{Orebi2021} marks a similar milestone. In this case the combination of neutrino observations with precise nuclear reaction rates, for example for the $^{14}$N(p,$\gamma$)$^{15}$O reaction, open up a pathway forward to constrain the contents of C, N, and O in the solar core, with prospects of resolving the so called solar abundance problem, a long standing discrepancy between the solar composition inferred from spectroscopy of sunlight and from helioseismology \cite{CD2021}. The CNO neutrino flux depends on the rate of proton capture on $^{12}$C, $^{14}$N and $^{16}$O at very low energies. These reactions are presently investigated in deep underground accelerator laboratories such as CASPAR in the United States, JUNA in China, and LUNA in Italy.

In the era of multi-messenger astronomy, asteroseismology with the Kepler, TESS, and future PLATO missions probe the nuclear astrophysics and structural properties of stellar interiors
\citep{bedding_2011_aa,mosser_2011_aa,cantiello_2014_aa,fuller_2019_aa,deheuvels_2021_aa,chidester_2021_aa}, 
gravitational wave detections explore the nuclear physics of neutron star and black hole mergers 
\citep{abbott_2016_ab,abbott_2017_ad,abbott_2017_aj,farmer_2020_aa,mehta_2021_aa}, 
neutrino detectors examine the rich nuclear astrophysics of neutrino emission and 
possibly provide early alerts to the gravitational wave and electromagnetic communities
\citep{yoshida_2016_aa,kato_2017_aa,patton_2017_ab,simpson_2019_aa,li_2020_aa,farag_2020_aa,mukhopadhyay_2020_aa}, and
the spectra of ancient stars discovered in large scale observational surveys 
reveal the chemical traces of the first stars formed after the Big Bang
\citep{gull_2018_aa,ezzeddine_2019_aa,ezzeddine_2020_aa,ji_2020_aa}. 
In addition, recent instrumental advances \citep{Stephan2016} enable trace-element
isotopic abundance measurements of heavy elements in micrometer-sized stardust grains recovered from meteorites \citep{lewis_1990_aa, zinner_2014_aa, nittler_2016_aa, trappitsch_2018_aa, hoppe_2021_aa, lugaro_2020_aa}. Such
measurements in grains from AGB stars allow access to unprecedented precision in
measuring the isotopic composition produced by the s-process. Simultaneous, multi-element
isotopic analyses are now possible on a regular basis \citep[e.g.,][]{Liu2015,Stephan2019}. Furthermore, cosmic rays from distant dynamical astrophysical environments may provide an additional handle on nucleosynthetic sites of heavy elements~\cite{Donn12,Kyut16,Alex16,Komi17}.  

Together, these advances are providing an
unprecedented volume of high-quality measurements that are
significantly strengthening and extending the experimental and
observational data upon which all of astrophysics ultimately
depends. In partnership with this ongoing explosion of activity in
experimental and observational astrophysics, community-driven
open-source software instruments 
\citep[e.g.,][]{paxton_2011_aa, paxton_2013_aa, paxton_2015_aa, paxton_2018_aa, paxton_2019_aa,
azuma_2010_aa, sallaska_2013_aa, townsend_2013_aa, townsend_2018_aa, astropy-collaboration_2013_aa,astropy-collaboration_2018_aa, timmes_2020_aa}
and machine learning techniques \citep[e.g.,][]{vanderplas_2012_aa, foreman-mackey_2013_aa,hogg_2018_aa, vanderplas_2018_aa, baron_2019_aa,huerta_2021_aa} 
are transforming how theory, modeling, and simulations interact with
experiments and observations. Particularly important achievements are: (1) the modeling of large-scale grids of stars as functions of initial mass and metal contents to predict their cumulative element synthesis impact on the Galaxy e.g., \cite{2016ApJS..225...24P, 2016ApJ...825...26K,Ritter18}, (2) the demonstration that hydrogen ingestion in stars can lead to novel nucleosynthesis via an intermediate neutron-capture process \cite{Herwig2011}, and (3) the emergence of the first self-consistent 3D core collapse supernova models that predict stellar explosions that approach observed features \cite{Bollig21}.

\begin{tcolorbox}[colback=red!5, colframe=red!40!black, title=Sidebar: Breakthrough Computations]
  \begin{center}
    \includegraphics[width=0.8\textwidth]{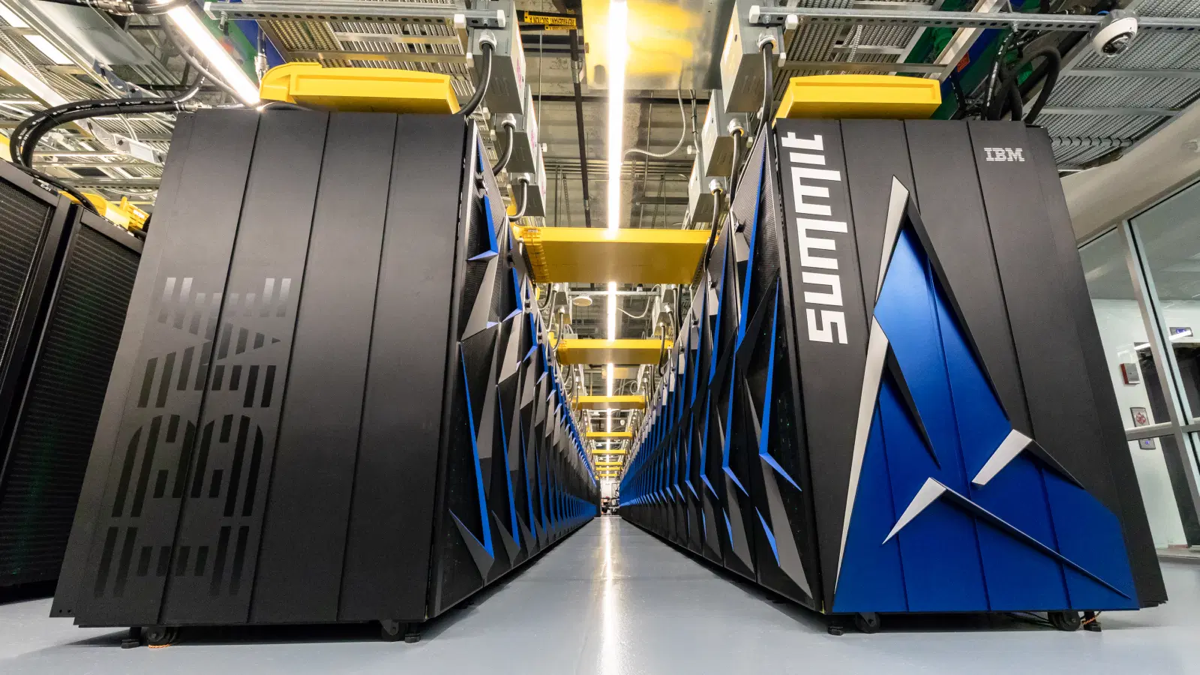}
  \end{center}
  \vspace{-0.20in} 
  \begin{center}
  {\footnotesize {\it Summit supercomputer. Credit: Oak Ridge National Laboratory, U.S. Dept. of Energy}} \label{fig:wg2:summit} \\  
  \end{center}
  \vspace{-0.08in} 
{\footnotesize
Computing is important to nearly all scientific activities in nuclear astrophysics.  
This includes theoretical calculations as well as those calculations that are
relevant to experiments and observations, from simulating data streams from telescopes
to processing the enormous amounts of data being collected by nuclear detectors. 
Computers are the laboratory for astrophysics theory.
Some questions are physically impossible to answer through observation. For example, 
the final outcome of the Sun's evolution would take billions of years to study.
However, programming supercomputers with equations describing the physics of such events 
provides answers on a human timescale. By carefully devising accurate and stable algorithms specifying how
the Sun evolves, researchers can capture the enormous length and time scales of these astronomical processes.
The Leadership-class supercomputers essential for 3D simulations of stars and stellar explosions, such as Summit at Oak Ridge National Laboratory, are 
among the world’s most powerful. 
}
\label{sidebar:computing}
\end{tcolorbox}

\vspace{-0.15in}
\subsection{What Needs To Be Done?} \label{sec:stars:todo}

A key challenge for nuclear astrophysics in understanding stars is the determination of the extremely slow nuclear capture and fusion reactions that govern stellar evolution. Despite decades of progress, the majority of these stellar rates are still based on uncertain extrapolations from measurements at higher energies. To push measurements to lower energies, it will be important to take full advantage of recent technical developments in underground accelerators~\cite{Form2003,Robertson2016,Liu16,STB19,Fels20}, recoil separators~\cite{Hutch03,Coud08,Dile17,Berg2018}, and advanced detection systems ~\citep[e.g.,][]{Avila2017,Randhawa2020,Kosh20,Anastasiou2021}. This includes new separators such as St. George at the University of Notre Dame \cite{Coud08}, SECAR at FRIB \cite{Berg2018}, and EMMA at TRIUMF \cite{Davids2019}, as well as new detection systems \citep[e.g.,][]{Febbraro2020} and accelerator upgrades such as Sta. ANA at the University of Notre Dame, LENA at TUNL, and FRENA at the Saha Institute of Nuclear Physics in India. This will require a concerted effort at a wide range of stable beam, radioactive beam, and $\gamma$-beam facilities. 

In addition, precise neutron-capture rates on stable and long-lived radioactive isotopes will be needed to accurately predict s-process nucleosynthesis. This is also important for understanding mixing processes in red giants, and to disentangle the s-process contributions from the plethora of heavy element nucleosynthesis processes that involve rare isotope reactions (Section \ref{sec:elements:intro}). This will require taking full advantage of existing neutron beam facilities in concert with new facilities such as SARAF-II at the Soreq
Nuclear Research Center in Israel, FRANZ at Frankfurt University in Germany, and at the University of Notre Dame. Additional opportunities will be provided by new isotope harvesting capabilites, for example at FRIB, to produce radioactive targets for measurements of s-process branch points. 

Concurrent advances in the theoretical understanding of all nuclear reactions relevant for stars will be needed to predict nuclear reaction rates from first principles, to consistently and coherently analyze multiple data sets with R-matrix theory, and to extract reliable reaction rate constraints from indirect measurement techniques. Nuclear theory advances are also needed to address the emerging question of  quantum effects that may occur at very low reaction energies \cite{DiazTorres18,Tokieda20}, potentially making current reaction rate estimates more uncertain than previously thought. 

For comparison with observations of element abundance distributions on stellar surfaces, new nuclear physics must be implemented into advanced stellar models that better account for mixing, rotation, magnetic fields, and mass loss. To that end, a robust stellar-model pipeline should be realized that regularly conducts simulations of low- and high-mass stars with a combination of high-resolution 1D and 3D approaches. An investment in such projects will dramatically enhance the science yield of many observational projects, as noted several times in the recently released Astro2020 Decadal Survey \cite{ASTRO2020}.

For comparisons with observations, the field should take advantage of the many new developments at the forefront of astronomy that have now begun to reveal the composition of stars from surface to center. The nucleosynthesis patterns and evolutionary tracks resulting from the simulation pipelines should be 
augmented with machine-learning techniques in order to make comparisons with stellar surface abundances,
interior abundance profiles inferred 
from stellar seismology, isotopic
abundances of stardust grains, 
the compact object distribution and properties inferred from gravitational wave observations,
isotopic compositions inferred from MeV $\gamma$-ray observations, and future neutrino detections (e.g., from Super-Kamiokande with Gadolinium with prospects to detect the relic supernova neutrino flux).

The dynamic evolution of massive stars is of particular importance as it sets the stage for supernova explosions \cite{sukhbold:16,2016ApJS..225...24P,Limongi2018,ritter:18,Andrassy2020}. It is therefore crucial to link advances in pre-supernova and supernova simulations and the underlying nuclear physics  
to observables such as light curves powered by radioactive isotopes,
compact object distributions, and multi-messenger signals.
To bridge this gap, computationally efficient frameworks, capable of large sets of simulations 
covering large domains for long stellar times, should be developed.

Distributions of the properties of compact stellar objects such as mass, radius and spin have recently gained importance as observables of stellar evolution endpoints owing to gravitational wave observations. The link between key nuclear reactions rates during stellar evolution such as the $^{12}$C($\alpha$,$\gamma$)$^{16}$O,
$^{12}$C+$^{12}$C, $^{12}$C+$^{16}$O, and $^{16}$O+$^{16}$O reaction rates \cite{deboer_2017_aa,Fang2017,Zickefoose2018,Jiang2018,Tumino:2018,Fruet2020,Tan2020} on compact object observables should be further investigated using transformative 1D and 3D simulations. Using this framework for updated predictions of compact object observables in light of the most recent progress in the understanding of key reaction rates should be expedited. 
Coupling this effort with the development of a 
framework to determine the underlying structure of the mass
distribution for compact objects that is informed by gravitational
waves and dense matter physics should also be a priority.

\subsection{What Do We Need?} \label{sec:stars:needs}
\begin{itemize}
    \item Underground accelerator facilities and other high sensitivity experimental techniques such as active targets or recoil separators that push stellar reaction-rate measurements closer to the energies encountered in stars.
    \item Neutron and $\gamma$-beam facilities with advanced capabilities.
    \item Capabilities for producing radioactive targets for reaction studies.
    Close collaboration with the DOE Isotope Program, with production
    facilities including the Isotope Production Facility at Los Alamos,
    the Brookhaven Linac Isotope Producer at Brookhaven, and the High-Flux
    Isotope Reactor at Oak Ridge, together with new opportunities afforded
    by isotope harvesting at FRIB will be essential.
    \item Capabilities to probe nuclear reactions in plasma environments. 
    \item Renewed efforts in nuclear reaction theory to understand and describe reaction mechanisms at extremely low energies. 
    \item Large-scale stellar spectroscopy surveys that can find and determine the composition of the most chemically primitive stars in the Galaxy, including the ones that preserve the elemental patterns created by the first stars after the Big Bang.
    \item Fully exploit new capabilities in asteroseismology and stardust analysis for nuclear astrophysics. 
    \item Advanced neutrino observation capabilities for detection of the relic supernova neutrino flux, improved measurements of the flux of CNO neutrinos from the Sun, and high statistics observation of a future galactic supernova. 
    \item An advanced MeV $\gamma$-ray mission that can detect decay radiation from radioactive nuclei produced in stars and supernovae. 
    \item Gravitational wave detections of binary mergers to constrain compact object mass distributions, which define the endpoints of stellar evolution.
    \item Advanced 3D models of phases of stellar evolution that impact element synthesis such as shell mergers, hydrogen ingestion, and core-collapse supernovae. 
    \item 3D models benchmarked to 1D models of stars that enable rapid calculation of element production and other observables for the full variety of stars.
    \item Stellar abundances based on improved atomic physics using 3D atmospheres and non-local thermodynamic equilibrium (non-LTE) computations.
    \item Re-factor and port the relevant software instruments to reflect the diversity in vendor architectures in next-generation GPU-based supercomputing.
    \item Broad support for sustaining core capabilities in key open-knowledge software instruments.
\end{itemize}

\vspace{0.05in}
\begin{tcolorbox}[colback=orange!5, colframe=orange!40!black, title=Sidebar: Building the Next Generation]
  \begin{center}
    \includegraphics[width=0.8\textwidth]{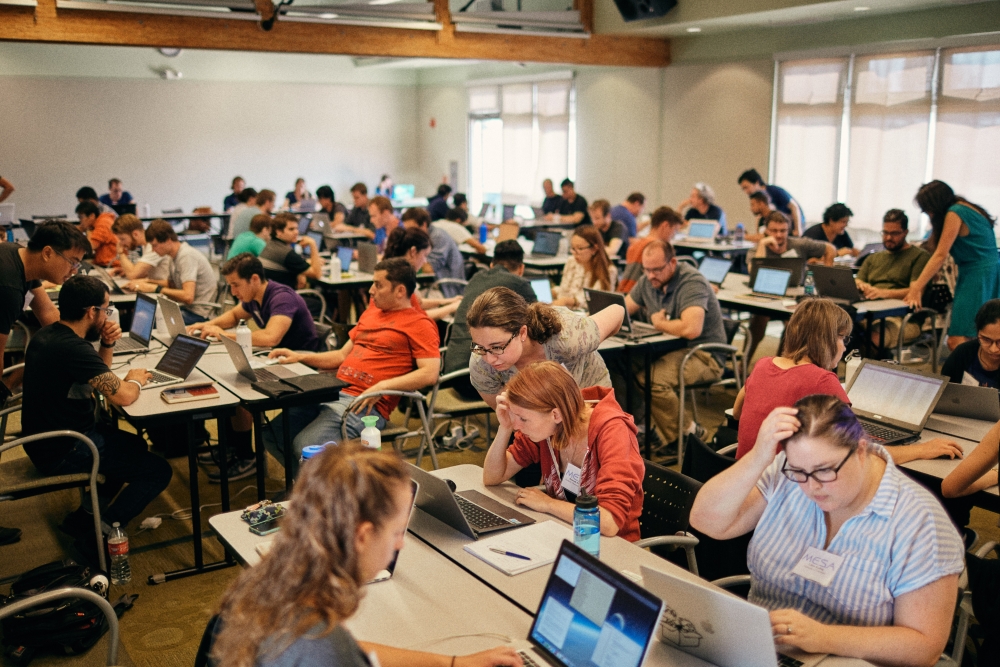}
  \end{center}
  \vspace{-0.20in} 
  \begin{center}
  {\footnotesize {\it Mastering a Software Instrument for Dynamic Nuclear Burning in Stars. \\ Credit: Jakub Ostrowski \url{www.jakubostrowski.com}.}} \label{fig:wg2:ss} \\  
  \end{center}
  \vspace{-0.08in} 
{\footnotesize
Summer schools have been in existence for hundreds of years.
They're hugely popular, especially for researchers at all levels (undergraduate, graduate, postdoctoral, faculty) 
who are looking to expand their mastery of specific software instruments. Though extensive hands-on labs, 
often lasting a week or more, participants 
gain familiarity and learn how to make better use of the software instrument in their own research. 
New science in nuclear astrophysics is driven by such training of the next-generation of researchers.
}
\label{sidebar:generation}
\end{tcolorbox}


\section{Origin of the Heavy Elements}
\label{sec:heavy}


\subsection{Introduction} \label{sec:elements:intro}

The elements heavier than iron (e.g., gallium, germanium, zirconium, silver, iodine, gold, uranium) cannot be made through nuclear fusion in stars. About half of these heavy elements found in our Solar System were synthesized via the slow neutron-capture process (s-process) mainly in evolved Asymptotic Giant Branch and massive stars \cite{Kaeppeler2011} as discussed in Section \ref{sec:stars}.  In this section we focus on the less understood origin of the other half of the heavy elements, traditionally attributed to the rapid neutron-capture process (r-process). Most likely sites of the r-process are considered to be the most extreme astrophysical environments, such as core-collapse supernovae, merging neutron stars, neutron stars merging with black holes, or black-hole accretion disks \cite{Horowitz2019,cowan:21}. These extreme events are very rare. This makes the direct astronomical observation of the freshly synthesized elements extremely challenging. Progress thus far has been mostly limited to detailed studies of the Solar System's isotopic composition using meteorites and stardust grains found on Earth, and of the buildup of heavy elements over cosmic time using the elemental abundances inferred from the spectra of old stars \cite{cowan:21}. A major breakthrough was the recent multi-messenger observation of the relatively close neutron star merger GW170817 enabled by the detection of gravitational waves \cite{GW170817}. Observations of a radioactively powered kilonova transient associated with GW170817 provided strong evidence for heavy element synthesis in neutron-star mergers \cite{Cowperthwaite:2017,2017ApJ...848L..19C,Drout2017}, though the only individual element identified was strontium \cite{Watson2019}. 

A major challenge in making progress are the short lifetimes of the rare isotopes involved in the nuclear reactions building up heavy elements in extreme astrophysical environments. These nuclei have been difficult to produce in the laboratory, and their properties and reactions are not well understood. An upcoming breakthrough is the development of a new generation of high-power 
rare-isotope accelerator facilities such as FRIB in the US, RIKEN/RIBF in Japan, TRIUMF/ARIEL in Canada, FAIR in Europe, SPIRAL2 in France, and SPES in Italy that promise to make most of the relevant nuclei accessible for laboratory studies. In addition,  simulations of heavy element nucleosynthesis in extreme astrophysical environments are computationally demanding, requiring large reaction networks as well as multi-scale-, and multi-physics calculations that push the limits of present-day computational facilities. In addition, important uncertainties in the nuclear and astrophysics of the s-process remain and need to be addressed in order to identify and quantify the contributions of the other heavy element nucleosynthesis processes (see discussion of the s-process in Section~\ref{sec:stars}).

Despite these challenges, a novel and much richer picture of heavy-element nucleosynthesis is now emerging: the primary process for the synthesis of heavy elements remains the neutron-capture reaction sequence followed by $\beta$-decays, which build up heavy nuclei from lighter seeds. However, instead of two distinct types of processes of slow and rapid neutron capture (s- and r-process, respectively), and a unique r-process site, there likely exists a continuum of processes: intermediate neutron-capture processes (i-processes) \cite{Cowan:77,Dar15,Ham16,roederer16ipro,denissenkov:17,clarkson:18,banerjee:18,Banerjee2019,SHC2020}, multiple r-processes of different strengths \cite{Horowitz2019,cowan:21}, and possibly an n-process \cite{Blak76,Truran1978,Rauscher2002,Pignatari2018}. What astrophysical sites enable these processes and how they combine to contribute to the inventory of elements in our Galaxy remain open questions. A fraction of the heavy neutron-deficient nuclides may also be synthesized by proton- and $\alpha$-particle captures (charged-particle capture process), photodisintegration (p-process or $\gamma$-process), or a mixture of proton capture and neutron-induced reactions  ($\nu$p-process) \citep{Schatz2001,frohlich06,rauscher:13,Pignatari2016,Tra18,battino:20}. 

In order for astrophysical simulations to make accurate predictions of what nuclei are synthesized, accurate nuclear data from both experiment and theory across a range of nuclei are required. Such accurate predictions are especially important to disentangle the very large number of different processes and sites now thought to contribute to heavy element synthesis. Model predictions can then be compared to astronomical observations of kilonovae (Fig.~\ref{fig:wg1:barnes20kn}), to measurements of the elemental compositions of ancient stars that recorded the compositions produced by a few or even individual nucleosynthesis events (Fig.~\ref{fig:wg1:roederer16iproc}), and to the isotopic composition of stardust grains that originate directly from a nucleosynthesis site. Validated model predictions are then incorporated into chemical evolution models that track the evolution of galactic compositions from their birth until today. These models can be confronted with the full set of astronomical data on stars in our Galaxy.  
Currently there are many discrepancies between the predictions and observations, emphasizing that both the astrophysical origins and the nuclear input require further investigation.

\begin{figure}
  \begin{center}
    \includegraphics[width=\textwidth]{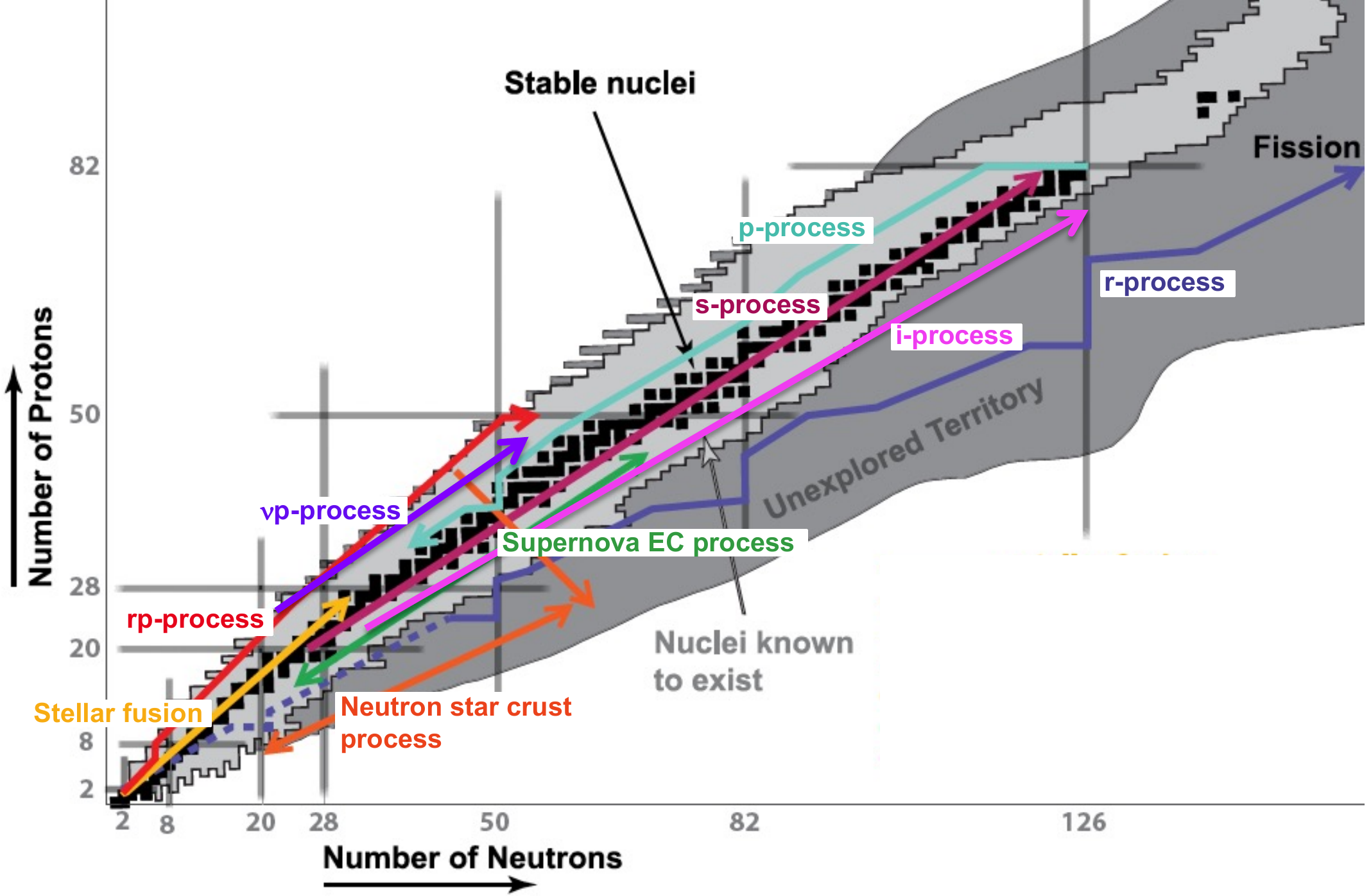}
  \end{center}
  \vspace{-0.20in} 
  \captionsetup{margin=0.2cm}
  \caption{Chart of nuclei: The various astrophysical processes are displayed schematically on the nuclear chart. Stellar fusion dominates nucleosynthesis of lighter elements up to the region of Fe. Heavier elements are formed predominantly by neutron-capture reactions via the s-processes, possibly a continuum of i-processes, and multiple r-processes including at least a weak and a strong r-process. A group of proton-rich isotopes, the p nuclei, are believed to be formed in the so-called p process, also known as $\gamma$ process, with possible contributions from the $\nu$p process. The rp-process and neutron star crust processes are not considered major contributors to the origin of the elements but play a role in interpreting observations of accreting neutron stars (Section \ref{sec:transients:intro}).  
}
  \label{fig:wg1:chartofnuclei}
\vspace{-0.2in}
\end{figure}

\subsection{Open Questions} \label{sec:elements:questions}

\begin{itemize}
\item What are the astrophysical sources of the heavy elements, what are their relative contributions, and how have these evolved over the history of the Milky Way and the Universe?
\item What are the properties of heavy radioactive isotopes and their reaction rates far from stability, how do they affect nucleosynthesis, and how do we push experimental technologies to access the full range of astrophysically relevant isotopes and their reactions?
\item How do we use the latest and rapidly improving experimental and computational developments to improve the accuracy and quantify the uncertainty of astrophysical isotopic yield predictions?
\item How do we distinguish between the multiple possible origins of light trans-Fe isotopes from germanium to cadmium? 
\item How does nucleosynthesis of the heaviest elements beyond lead and bismuth proceed, what is the role of nuclear fission, and how do these isotopes manifest in observations of astrophysical sites? 

\end{itemize}

\subsection{How Did We Get Here?} \label{sec:elements:how}

In nuclear physics, the past years have seen huge technological advances in the ability to produce and measure properties of nuclei far from stability. The development of radioactive ion beam facilities across the world has enabled the production of short-lived nuclei involved in all nucleosynthesis processes. In addition, new experimental techniques have been developed to investigate these exotic radioactive species and provide the necessary nuclear inputs for astrophysical models such as nuclear masses, decay properties, and reaction cross sections (see overview in \cite{Horowitz2019}). 

Most impressively, the extremely neutron-rich nuclei in the r-process are now coming within reach (Fig.~\ref{fig:wg1:chartofnuclei}). Building on early work at ISOL type radioactive beam facilities that reached r-process nuclei for the first time (e.g., \cite{Gill86}), pioneering work at NSCL \cite{Hosmer2010} and GSI \cite{caballero16} used fast radioactive isotope beams produced by fragmentation to cover significant parts of the r-process. A recent milestone was the BRIKEN campaign at RIKEN/RIBF that measured hundreds of neutron decay branches following $\beta$ decay between $^{75}$Co and $^{172}$Gd that are required for r-process models \cite{Hall21}. New techniques for the study of such $\beta$-delayed neutron emission  \citep[e.g.,][]{WILSON2021165806} build on these successes. Atomic physics-based approaches to measure atomic masses with ion traps, complemented by time-of-flight techniques using spectrometers and storage rings, have pushed precision measurements of nuclear masses well into the path of the r-process \cite{Hakala2012, Simon2012, Atanasov2015}. Novel ``reverse engineering'' techniques, coupled with state-of-the-art mass measurements \cite{ORFORD2020491}, are helping to bolster our understanding of the r-process conditions \cite{Orford2018}.

\begin{figure}
  \begin{center}
    \includegraphics[width=0.7\textwidth]{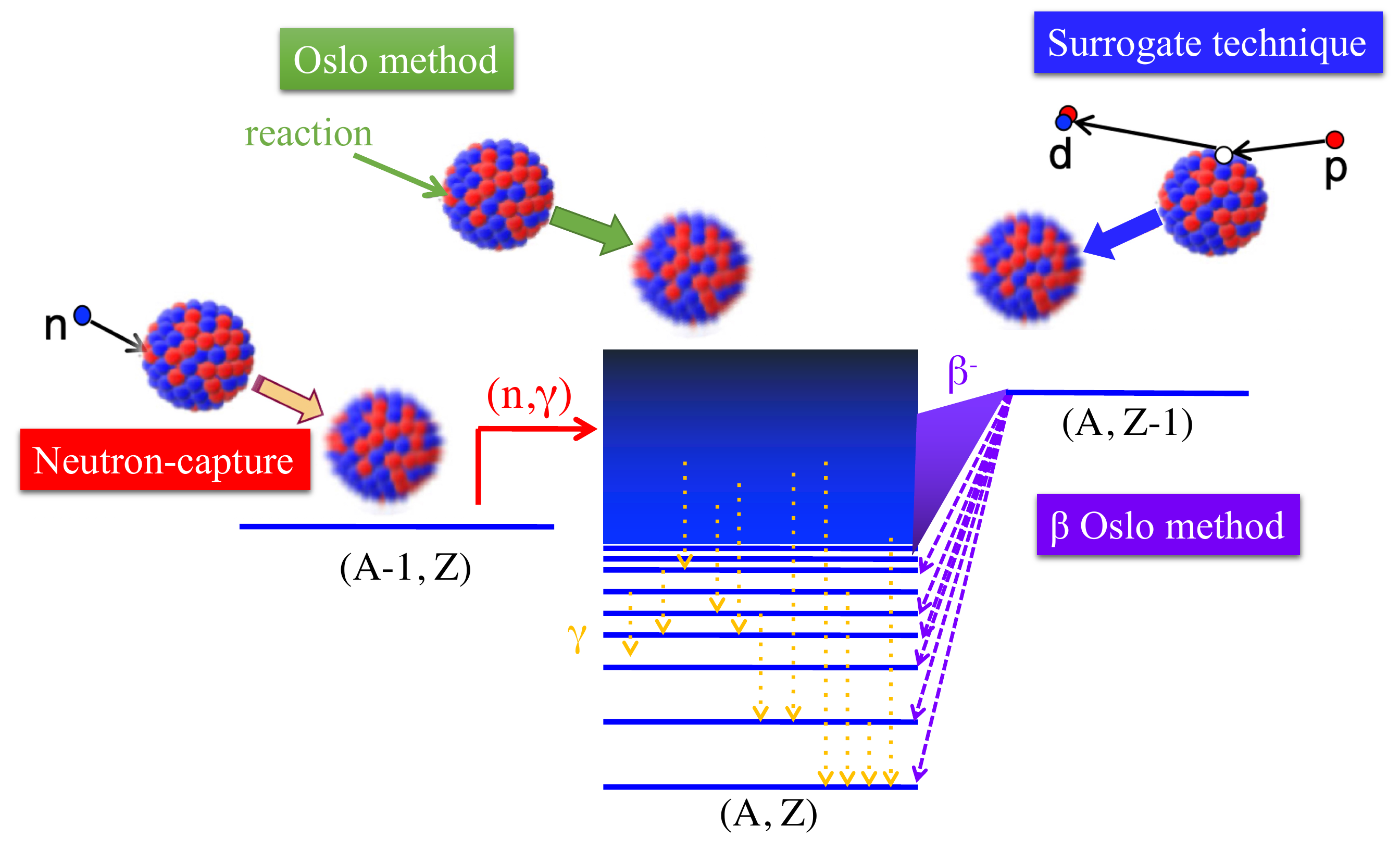}
  \end{center}
  \vspace{-0.20in} 
  \captionsetup{margin=0.2cm}
  \caption{Indirect techniques for constraining neutron-capture reactions. Where neutron-capture experiments are not feasible, alternative measurements can be used to determine crucial ingredients for constraining Hauser-Feshbach reaction calculations:  the surrogate reaction method uses inelastic scattering or transfer reactions to populate the compound nucleus and extract important nuclear decay information.  Alternatively, statistical properties (level densities, gamma strength functions) of the compound nucleus can be inferred from $\gamma$-emission following inelastic scattering or transfer reactions (the Oslo method), or $\beta$-decay (the $\beta$-Oslo method).
  }
  
  \label{fig:wg2:indirect}
\end{figure}

The recently recognized diversity of astrophysical neutron capture processes, including the s-process, the i-process, the r-process, and the n-process, further underscores the need to address the challenge of obtaining neutron capture rates for unstable nuclei. While direct measurements of neutron capture on very short-lived nuclides are presently not feasible, multiple indirect experimental techniques and advances in nuclear-reaction theory make it possible to obtain constraints for important reaction rates (Fig.~\ref{fig:wg2:indirect}): $\beta$-delayed neutron emission \cite{Kra83,Scielzo2020}, the $\beta$-Oslo \cite{Spyrou2014,Liddick2016} and inverse-Oslo \cite{Inge2020} methods, transfer reactions \cite{Thom2005,Jones2010,Mann2019}, the surrogate reaction method \cite{Ratk2019,Escher2012,Esch2018}, and the Trojan Horse Method~\cite{Tumino:2021} offer pathways to study neutron-capture reactions on short-lived isotopes.  In addition, measurements of evaporation spectra at stable-beam facilities  \cite{Voin21} provide complementary information for nuclei just off stability, e.g., for those relevant to i-process nucleosynthesis.

Successful direct reaction-rate measurements have been performed for charged particle reaction rates of importance to heavy element nucleosynthesis in the weak r-, p-, and $\nu$p-processes. For a few special cases, precision measurements with stable ion beams are possible, e.g., the activation cross section determinations of the weak r-process reactions $^{96}{\rm Zr}(\alpha,n)$ and $^{100}{\rm Mo}(\alpha,n)$ at ATOMKI~\cite{Kiss2021,Szeg2021}. For most cases, rare isotope beams are required. For example, the ($\alpha$,n) reactions on short-lived nuclides that synthesize heavy elements in the weak r-process are now being measured using a variety of complementary techniques such as the MUSIC detector at Argonne National Laboratory and NSCL~\cite{Avila2017}, along with the HABANERO neutron counter and SECAR recoil separator at NSCL~\cite{Berg2018}. Similarly, $(p,n)$ reactions are being measured at energies of interest for $\nu$p-process nucleosynthesis using the SECAR recoil separator with neutron coincidences~\cite{Gasti2021}.

Although the p-process consists of photodisintegration reactions starting on stable seed nuclei and moving off stability to neutron-deficient nuclei, these reactions are more often better probed with the inverse charged-particle reaction measurements~\cite{Rauscher2018}. Nonetheless, $\gamma$-beam facilities such as HI$\gamma$S can provide unique insights for special cases~\cite{Banu2019}. Optical-model potentials are of particular interest for these reactions. These potentials have been constrained by a broad range of measurements at (mostly stable beam) facilities across the world~\cite{Pignatari2016}. Direct measurements of the inverse p-process reactions with total absorption spectrometers at the University of Notre Dame have constrained properties of p-process nuclei near $A=100$~\cite{Quin2015,Kelm2020}, and measurements combining $\gamma$-spectroscopy and activation techniques performed at the University of Cologne and ATOMKI have focused on heavier p-process nuclei \cite{Scholz2016,Szucs2018}. Measurements on unstable nuclei involved in p-process nucleosynthesis have recently become possible with novel recoil separation techniques, such as the EMMA separator at TRIUMF \cite{Lotay2021} and the ESR storage ring at GSI~\cite{Glori2019}.

Nuclear theory has been and remains essential to fill in the gaps where nuclear experiments are not yet feasible. This is especially important for r-process nucleosynthesis. Significant progress has been made in theoretical predictions of nuclear masses using microscopic-macroscopic models, density functional models, and machine learning techniques~\cite{Neuf2020,Love2020}. A better theoretical understanding of fission at the endpoint of the r-process path (A$\sim$300) has shown that these processes affect r-process nucleosynthesis~\cite{Deva2020,Lemaitre2021}. Predictions of reaction rates for heavy nuclei rely on accurate optical potentials, and efforts are underway to calculate these from first principles~\cite{nunes_2020_aa,Whit2020}. The uncertainties of such calculations are still large, but statistical tools have been developed to identify experiments that can reduce these uncertainties at least for nucleon-nucleus potentials ~\cite{Lovell2021,Catacora2021}. For the important $\alpha$-nucleus optical potential, more theoretical and experimental work is needed to understand discrepancies between theoretical predictions and experimental data \citep{Mohr2020}.

\begin{figure}
  \begin{center}
    \includegraphics[width=0.73\textwidth]{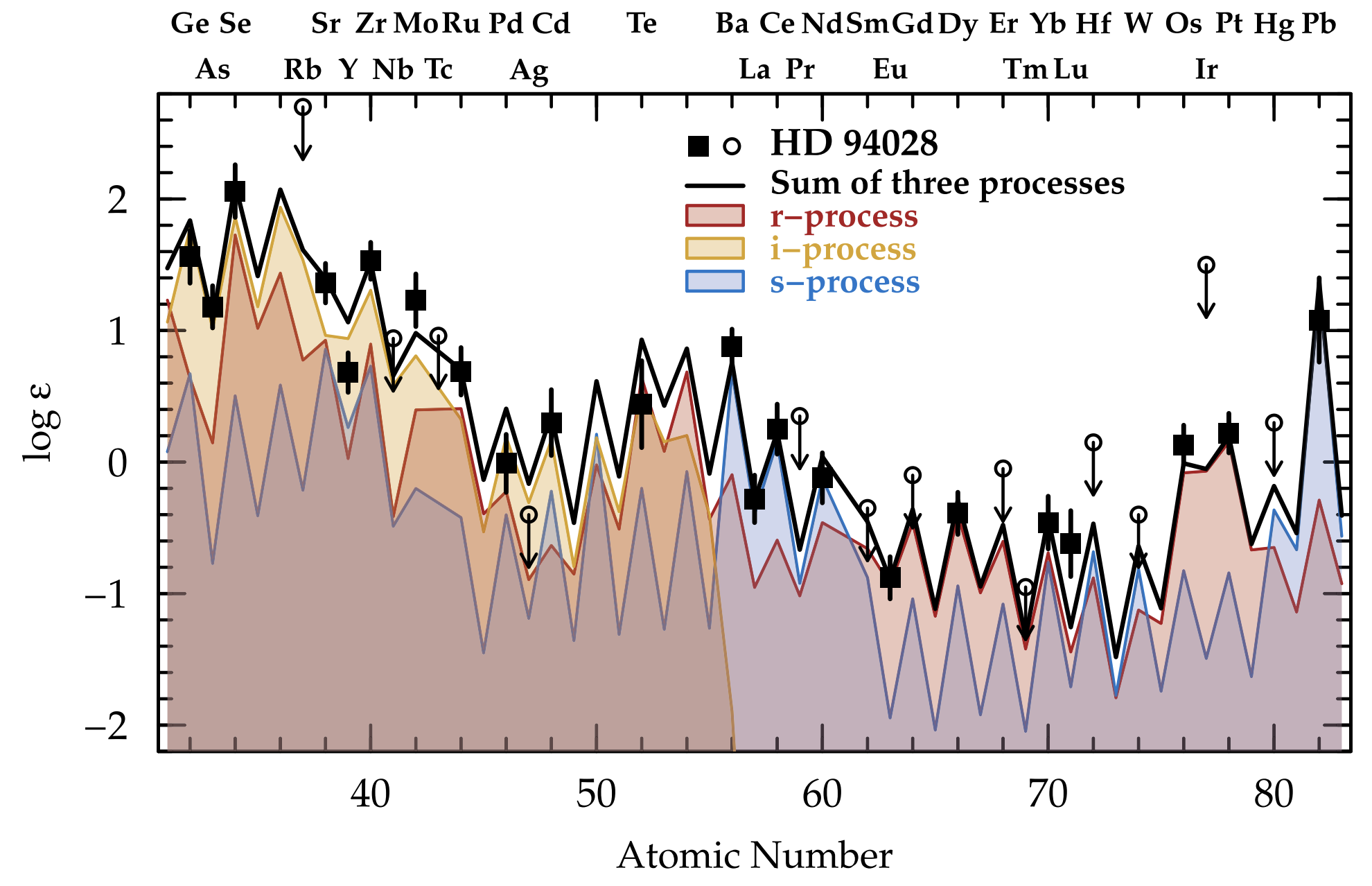}
  \end{center}
  \vspace{-0.20in} 
  \caption{Composition of an old metal-poor star (black points) with heavy elements from all of the $r-$, $s-$, and $i-$processes. All processes are needed to explain the observed element abundance pattern. Disentangling these processes requires reliable predictions of the individual element synthesis components and thus reliable nuclear data. It also requires observational data of a large number of elements, many of which are only accessible at ultraviolet wavelengths underlining the importance of an observational capability for UV spectroscopy \citep{roederer16ipro}.}
  \label{fig:wg1:roederer16iproc}
\end{figure}

Dramatic progress has also been made in heavy element nucleosynthesis observations. A spectacular accomplishment is the multi-messenger observation of the neutron star binary merger GW170817 with its associated kilonova (see sidebar on Pg. \pageref{sidebar:gw}). The observation unambiguously tied kilonovae and short $\gamma$-ray bursts to neutron star mergers, confirmed from the light curve evolution that lanthanides (and therefore heavy elements) are produced in neutron star mergers, and demonstrated that, within uncertainties, sufficient heavy elements may be ejected to make neutron star mergers an important r-process site. Furthermore, large scale stellar spectroscopy surveys of millions of stars and targeted smaller high-resolution followups with large telescopes, for example within the $R$-Process Alliance collaboration \cite{Hansen2018,Holmbeck2020}, have dramatically increased the sample of metal-poor stars observed to be strongly enriched in r-process elements to over a hundred. Together with higher metal content stars, these stars map the history and evolution of heavy element nucleosynthesis in the Galaxy and provide the detailed abundance patterns that any theory of the r-process ultimately must explain. Key results were 
\begin{enumerate*} 
\item the realization that multiple heavy element processes are responsible especially for the lighter r-process elements \cite{Travaglio.etal:2004, Montes.etal:2007, Qian.Wasserburg:2008, hansen12, Dar15, Ham16,Sku2020A},  
\item that on the one hand the r-process produces a quite robust abundance pattern in the Te - Pt element range \cite{Sneden.etal:2008,roederer14hst}, and that, on the other hand, there is an as-yet unexplained variation in the abundances of elements lighter than Te or heavier than Pt \cite{Hansen.etal:2014,ji:19,Hol19}, and 
\item that neutron star mergers are unlikely to be the sole source of the heavy r-process elements \cite{ Cote2019, Sim19a, Sku19, Sie19} 
\end{enumerate*}. 
A milestone was the spectroscopy of the star HD94028 \cite{roederer16ipro} for which abundances or limits on 56 elements were obtained, providing evidence for multiple contributions of various neutron-capture processes (see Fig.~\ref{fig:wg1:roederer16iproc}), albeit somewhat preliminary given the remaining large nuclear-physics uncertainties yet to be addressed. Traces of the i-process have been seen in all Galactic components, such as halo, bulge, disk, and smaller stellar systems \citep[e.g,][]{Koch2019}. Furthermore,
stardust grain analysis enables us to study the chemical evolution of the r- and p-processes by
directly probing relevant isotopes, e.g., isotopes of Mo \citep{hansen2014grains,Stephan2019}.

Computer models of astrophysical sites are essential to connect nuclear physics with observations, and to identify element synthesis sites through specific elemental signatures. Advances in computational capabilities and algorithms have enabled simulations in three spatial dimensions that have proven essential for the extremely dynamic environments responsible for heavy element synthesis \cite[e.g.,][]{Stephens2021}. At the same time, advances in understanding and implementation of key nuclear physics such as the dense matter equation of state or neutrino physics have been critical. For example, it has been shown that neutrino oscillations can have a significant impact on heavy element nucleosynthesis \cite{Duan2011,George2020,Li2021}. The most spectacular advances came in the modeling of neutron star mergers and their multi-messenger signals motivated by observations of GW170817. Sequences of models can follow the evolution of the merger event from 3D full general relativistic simulations of the actual merging \cite{Radice20, Foucart21}, through the development of multiple types of outflows \cite{Nishikawa2021} with their individual r-process nucleosynthesis contributions \cite{Just+15,Martin:2015hxa}, all the way to the atomic physics models to predict the electromagnetic kilonova signature \cite{Kasen:2017sxr, Even20, Gillanders21} and spectral features of r-process elements \citep{Watson2019}. Supernova r-process models involving collapsars \cite{Caballero12, Zenati20}, neutrino driven winds \cite{Witt21}, magnetic field driven jets \cite{Winteler+12, Nishimura.etal:2017, Yong21}, and hadron quark phase transitions \cite{Fischer2020} have also advanced considerably. Pioneering 3D mixing models of helium layers in stars \cite{Her14} and accreting white dwarfs \cite{Den19} have provided some of the first possible sites for the i-process. 

\begin{tcolorbox}[colback=gray!5, colframe=gray!40!black, title=Sidebar: Rare Isotope Beams]
  \begin{center}
    \includegraphics[width=0.8\textwidth]{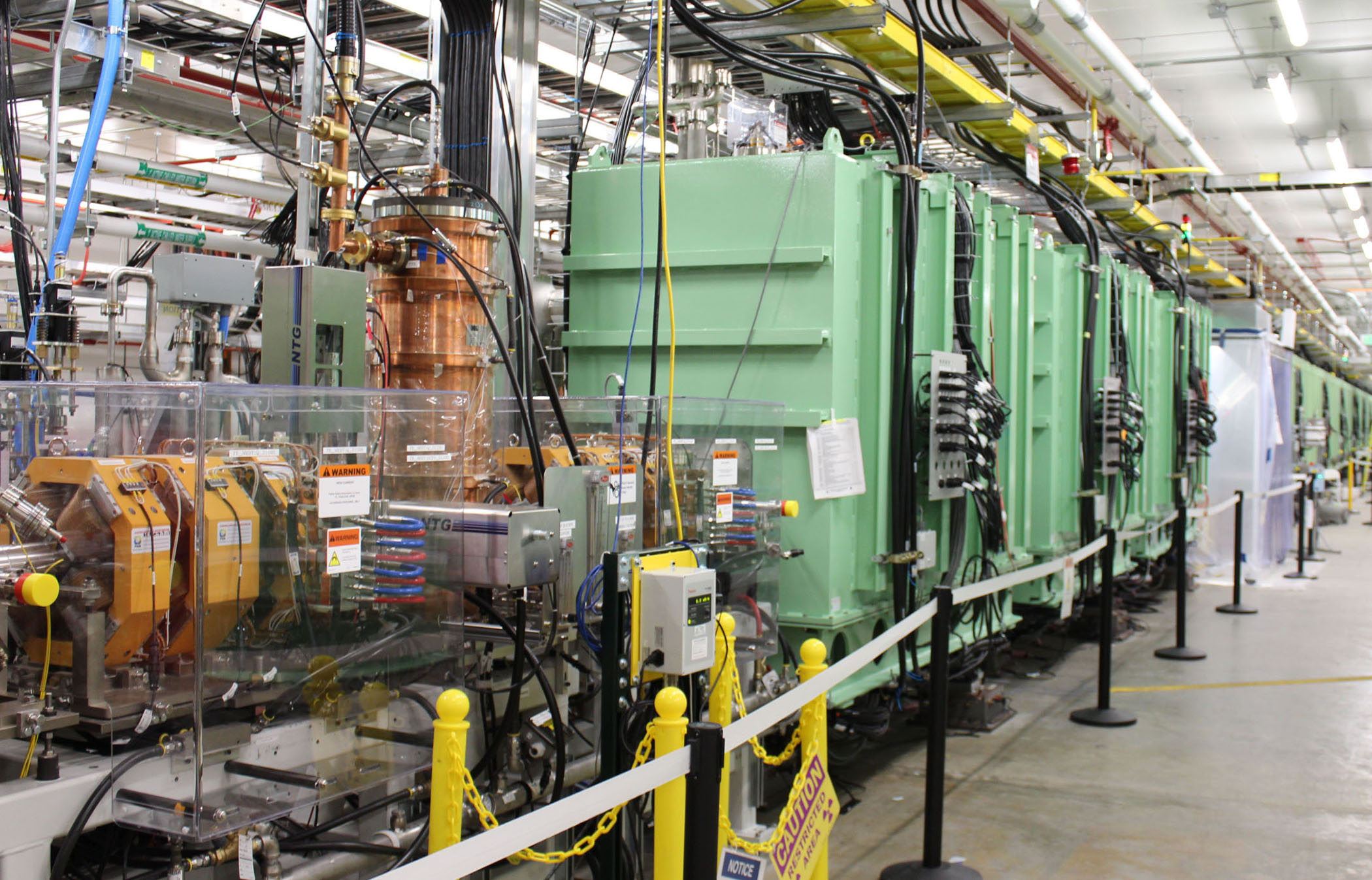}
  \end{center}
  \vspace{-0.20in} 
  \begin{center}
  {\footnotesize {\it A section of FRIB beamline. Credit: Facility for Rare Isotope Beams}} \label{fig:wg3:frib} \\  
  \end{center}
  \vspace{-0.08in} 
{\footnotesize

A new generation of rare-isotope accelerator facilities begins to address 
fundamental questions about the inner workings of nuclei and the formation 
of the heavy elements. These elements are likely forged in the extremely hot and dense astrophysical environments 
briefly encountered in supernova explosions and neutron star mergers. The new FRIB facility in the 
US will provide access to many of the rare isotopes serving 
as stepping stones for element creation in these sites. Other new or upgraded facilities
around the world, such as at Argonne National Laboratory in the US, RIKEN in Japan, FAIR in Germany,
ISOLDE in Switzerland, and ARIEL in Canada, provide complementary capabilities for 
the broad range of measurements and techniques needed to unravel 
the unknown properties of rare isotopes and their connection to the 
heavy element composition of the cosmos. 

}
\end{tcolorbox}

Implementing full nucleosynthesis models, which often involve thousands of nuclear reactions, into these advanced simulations has been a challenge due to the computational cost of including a complex reaction network. Despite this, major advances have been achieved in i-, r-, weak r-, $\nu$p-, and p-process model calculations, including Monte Carlo simulations propagating nuclear uncertainties to observables and identifying critical nuclear physics uncertainties \cite{Mumpower2016,Nishimura2018,Nishimura2019,Blis2020,Denissenkov21,Barnes21}. A remaining major challenge are the uncertainties in the nuclear physics input (masses, fission, $\beta$ decays, reactions) that prevent validation of models and inference from observations of the astrophysical conditions at the nucleosynthesis site. More experimental and theoretical nuclear data with quantified uncertainties are urgently needed, especially for the extreme neutron-rich nuclei involved in the r-process. 

\begin{tcolorbox}[colback=yellow!5, colframe=yellow!40!black, title=Sidebar: Gravitational Waves]
  \begin{center}
    \includegraphics[width=0.8\textwidth]{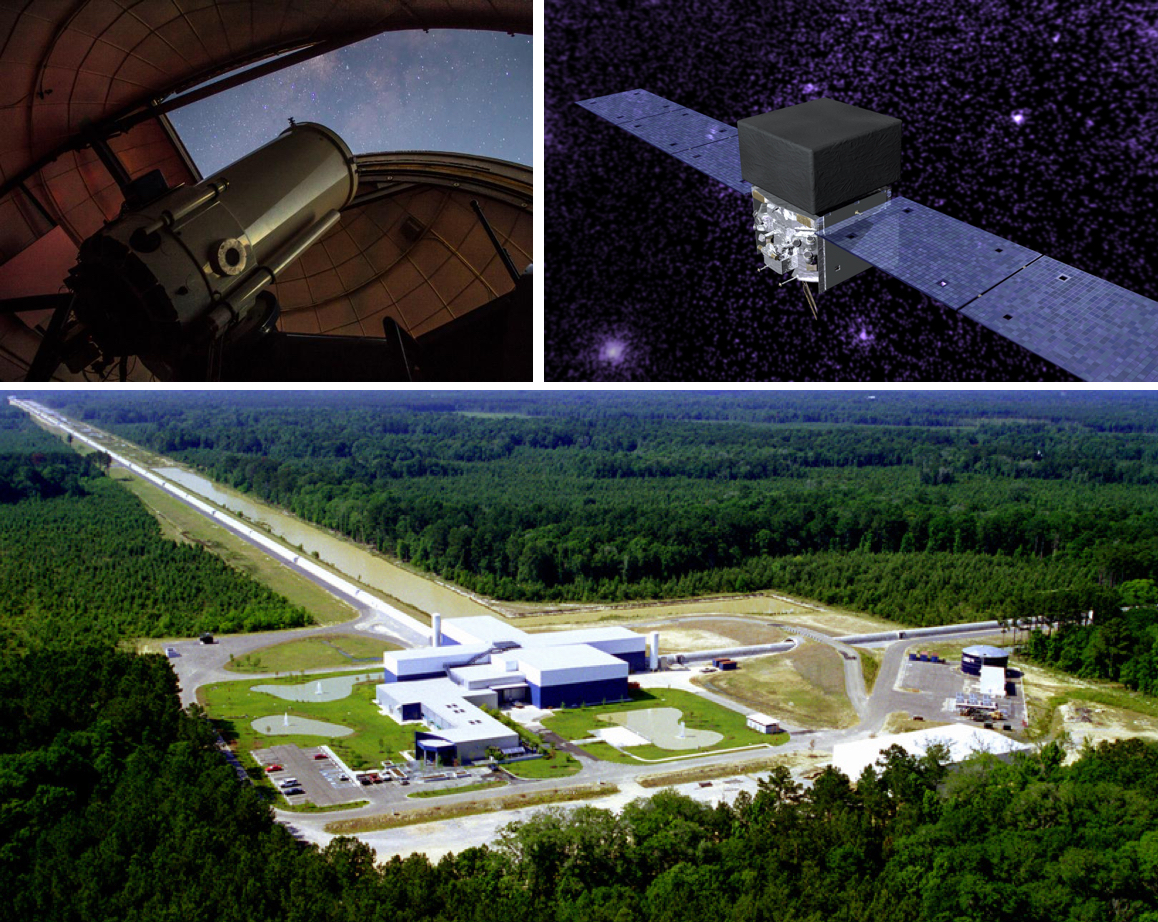}
  \end{center}
  \vspace{-0.20in} 
  \begin{center}
  {\footnotesize {\it Henrietta Swope Telescope, Fermi Gamma-ray Space Telescope, LIGO Livingston. \\Credit: Consuelo Gonz{\'a}lez Avila (Observatorio Las Campanas), NASA's Goddard Space Fligt Center, LIGO/NSF/Caltech/MIT }} \label{fig:wg3:ligo} \\  
  \end{center}
  \vspace{-0.08in} 
{\footnotesize
The first neutron star merger unambiguously detected across the multi-messenger spectrum, including the gravitational wave event GW170817, was a landmark moment in modern science, especially for nuclear astrophysics. This definitively connected neutron star mergers to short $\gamma$-ray bursts and proved that these events are important contributors to the nucleosynthesis of heavy elements. Tight sky localization with the LIGO and Virgo gravitational wave observatories and coincident detection in $\gamma$-rays enabled rapid follow-up across the electromagnetic spectrum and across the globe, revealing the kilonova afterglow over the following weeks. The kilonova transient harbored imprints of r-process nucleosynthesis, providing an observational handle for astrophysics theory to connect to experiments at facilities such as FRIB and nuclear theory calculations of extremely neutron-rich nuclides.

}
\label{sidebar:gw}
\end{tcolorbox}

Another critical advance for nuclear astrophysics has occurred in the development of Galactic chemical evolution (GCE) models that incorporate predictions from nucleosynthesis models \cite{Cot19a, Kobayashi20, Matteucci21}. Such GCE models now enable the propagation of new nuclear physics and their resulting abundance signatures to the compositional evolution of the Galaxy over time. Most importantly, this allows for the evaluation of the impact and contribution of nuclear processes in specific sites on the overall Galactic inventory of elements, taking fully into account the interplay of all nucleosynthesis processes. Important recent results were the demonstration that the neutron star merger ejecta properties inferred from GW170817 are consistent with neutron star mergers being a dominant source of r-process nuclei \cite{Cote17, Wanajo21} (though uncertainties are still large); the need for multiple r-process sources besides neutron star mergers \cite{Cote2019, Sim19a, Sku19, Sie19}; the demonstration that an i-process in rapidly accreting white dwarfs may contribute significantly to the origin of the elements in the Ge-Cd trans-iron range \cite{Cot18}; and the disentangling of multiple types of supernovae contributing to the origin of the p-nuclei \cite{Tra18}.



\subsection{What Needs To Be Done?} \label{sec:elements:todo}

Our understanding of heavy element synthesis is rapidly improving on all fronts, including new nuclear experiments, multi-messenger observations, spectroscopic surveys, and supercomputers. The next major steps involve realizing the potential of these new capabilities and making interdisciplinary connections between them as they evolve. Centers that enable interdisciplinary work and the effective exchange of ideas will play an important role in both, generating scientific breakthroughs, and training the next-generation workforce (Section~\ref{sec:centers}).

In nuclear physics,  advances in experimental capabilities are expected from the next generation of radioactive ion beam facilities coming online \cite{Horowitz2019}. In the US, the Facility for Rare Isotope Beams (FRIB) at Michigan State University will provide access to a large fraction of the exotic nuclei relevant for heavy element nucleosynthesis processes. The nuCARIBU facility at Argonne National Laboratory will offer intense beams around the fission fragment peaks, and the N=126 Factory, also at Argonne, will provide beams of nuclei critical to our understanding of the r-process rare-earth peak. Other facilities around the world, like FAIR in Europe, ISAC-ARIEL in Canada, and RIKEN/RIBF in Japan are also producing nuclei far from stability in complementary ways. A more niche but no less important path to heavy neutron-rich nuclides will be offered by multi-nucleon transfer reactions at JYFL and FAIR \cite{watanabe2015i,Deva2020}. The experimental techniques and devices required to take advantage of the new beams have mostly been developed. The main goal for the near future is to fully exploit these new capabilities to advance the understanding of heavy element nucleosynthesis. This will require close collaboration with nuclear theorists to interpret experimental data and reveal the deeper connections between the structure of exotic nuclei and nucleosynthesis. In light of the large number of isotopes involved in nucleosynthesis processes, it will be beneficial to combine nuclear theory with machine learning techniques \citep{Boehnlein2021} to interpolate sparse experimental data as they become available, thereby maximizing the impact of early experiments with the new facilities. It will also be critical that prior to experiments astrophysical model calculations, together with nuclear theory considerations, provide strong guidance to identify the most important measurements and map out their potential impacts on the open questions in the field. 

\begin{figure}
  \begin{center}
    \includegraphics[width=0.73\textwidth]{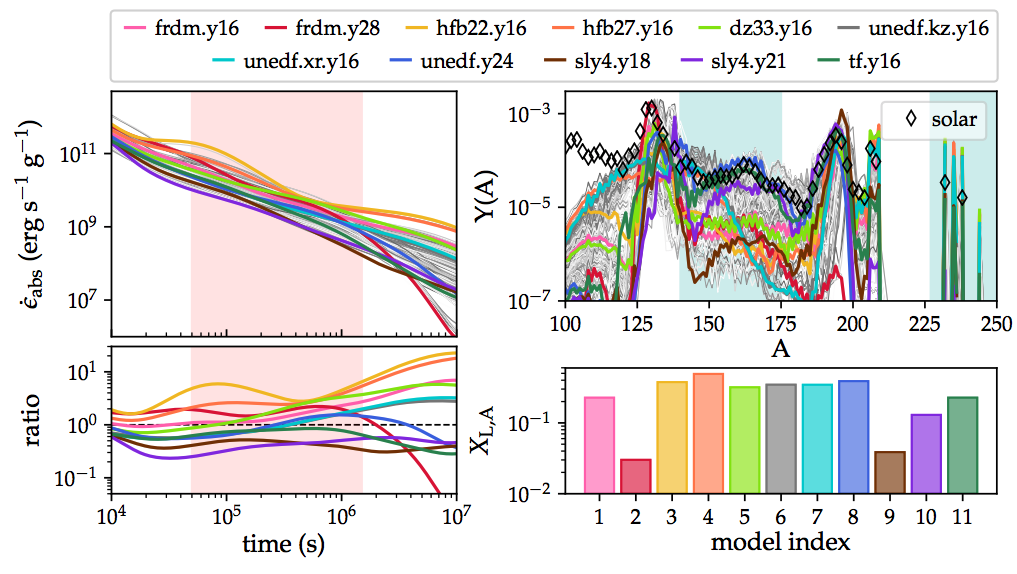}
  \end{center}
  \vspace{-0.20in} 
  \caption{Impact of nuclear physics uncertainties on r-process nucleosynthesis and kilonovae \citep{Barnes20}. As described in Table 1 of \cite{Zhu21} and Table 2 of \cite{Barnes20}, the labels provide short-hand notation for the different mass models / fission barrier models applied (FRDM2012 \cite{Moller16FRDM} / FRLDM \cite{Moller15BH}, HFB22 \cite{Goriely13HFB22} / HFB14\cite{HFBBH}, HFB27 \cite{Goriely13HFB27} / HFB14, Duflo-Zuker \cite{DufloZuker95} / FRLDM, UNEDF1 \cite{Kortelainen12UNEDF1} / FRLDM, SLY4 \cite{Chabanat98SLY4} / FRLDM, and Thomas Fermi \cite{TFmass} / Thomas Fermi \cite{TFBH}) as well as the neutron-richness considered for each case (for example .y16 implies $Y_e=0.16$). Absolute nuclear heating rate is shown (top left) as well as a ratio comparison to a power law assumption (bottom left) for all cases considered. The variances in the predicted nucleosynthetic abundances are shown as a function of mass number (top right) with a comparison of their total mass fractions (bottom right) of lanthanides + actinides (which are important contributors to the opacity).}
  \label{fig:wg1:barnes20kn}
\end{figure}

Despite the major developments undertaken so far, further advances in experiment and theory are needed. The FRIB400 upgrade of FRIB will further increase the reach towards the most neutron-rich nuclei in the r-process \cite{FRIB400}. The developments of the SECAR recoil separator at FRIB, the DRAGON and EMMA separators at TRIUMF, and the CRYRING storage ring at GSI/FAIR enable direct measurements of proton and $\alpha$-induced reactions in the weak r, p- and $\nu$p-processes. The future ISLA separator at FRIB will further strengthen these capabilities. Experimentally constraining neutron-capture rates on unstable nuclei for the s-, i-, n-, and r-processes remains another long standing challenge that must be addressed. With the promising advances in indirect techniques, systematic errors need to be better understood, including uncertainties from reaction theory. Examples include the optical potential, both from a phenomenological and a microscopic standpoint~\cite{nunes_2020_aa,Esch2018}, as well as the $\gamma$-ray strength function. At the same time, there are promising techniques for direct measurements: In the near term, the harvesting of radioactive isotopes, for example at FRIB, may offer opportunities for direct measurements of neutron capture on longer-lived isotopes, for example for the s-process. The LANSCE spallation target upgrade will significantly increase the keV neutron flux available, making
measurements on smaller radioactive samples possible \cite{ZMK18}. Further into the future, storage rings are a very promising tool to collide circulating short-lived radioactive ion beams with neutron beams~\cite{Reifarth2014a,Reifarth2017}. In North America, three projects are presently being discussed (at FRIB, at LANL, and at TRIUMF) and a collaboration has started to identify synergies and common projects \cite{MCC21}. Another challenge is understanding nuclear fission at the endpoint of the r-process. Fission rates and fragment distributions influence the abundances of lighter elements produced in the r-process \cite{Deva2020,Lemaitre2021} and will remain out of reach of experimental facilities for the foreseeable future. Further advances of the nuclear theory of fission and benchmarking fission models with experiments will be important as well as the development of new approaches to synthesize a broader range of actinide isotopes in the laboratory \cite{Devaraja2015,Horowitz2019}. An improved understanding of some nuclear reactions on light elements, as discussed in Section~\ref{sec:stars:needs}, will also be critical for heavy element nucleosynthesis such as the helium burning reactions that provide the neutrons for the i- and n-processes. An example of how nuclear physics uncertainties propagate to the interpretation of important observables is shown in Figure~\ref{fig:wg1:barnes20kn}, which illustrates the large variations in theoretical predictions for the nuclear heating rate and lanthanide plus actinide mass fractions, which go into kilonova light curve calculations. In that context, improved atomic physics measurements of cross sections, opacities, and oscillator strengths would help advance kilonova light curve modeling and aid the more accurate extraction of elemental abundances from stellar spectroscopy.

In addition to better nuclear and atomic physics, further developments of astrophysical models are needed. The sites of heavy element nucleosynthesis all invoke a wide range of physics. Three-dimensional (magneto-)hydrodynamics, general relativity, photon and neutrino transport, neutrino physics including flavour oscillations, thermonuclear kinetics, and the wide range of timescales involved in these explosive astrophysical events all conspire to make state-of-the-art simulations computationally expensive, but have all been shown to be critical for reliable nucleosynthesis predictions~\cite{Mart2017,Radi2018,OCon2018,Reic2021}. It will be important to take advantage of increasing computational capabilities, including exascale computing, and algorithm development to further advance these models. A particularly important limitation of most present models that needs to be overcome is the limited number of nuclear reactions included, which requires post-processing and tracer-particle approximations to be applied to predict the element patterns created. Overcoming this limitation is especially important for scenarios where such approximations are known to be inadequate, for example when the nuclear burning and convective mixing time scales are comparable. 

Such model improvements will enable meaningful quantitative comparisons between models and observations, further increasing the need for precise nuclear physics. Thus, nuclear uncertainties will have to be quantified and, taking advantage of increased computing power, propagated through suites of sophisticated astrophysical simulations that span the full range and variety of conditions encountered in nature. This will provide guidance for both nuclear experiment and nuclear theory, identifying the largest uncertainties hampering observational comparisons. At the same time, comparisons of models with observations that account for uncertainties will enable extraction of quantitative information on the astrophysical sites, their nuclear reactions, and their physical environment.

To interpret the large incoming amount of stellar spectra and isotopic abundance data, and ultimately to answer the question of the origin of the elements, it is crucial to accurately model the processes by which elements are dispersed throughout the Galaxy and incorporated into stars. The study of chemical evolution increasingly requires sophisticated hydrodynamic galaxy formation simulations with a focus on inhomogeneous transport of the heavy elements. It will therefore be important to take advantage of progress in nuclear physics and models of sites. This will require to produce extended sets of nucleosynthesis yields for a broad range of scenarios that can be used in chemical evolution models. At the same time, it will be important to quickly test new sites or new nuclear physics for its impact on chemical evolution as a whole.  



The stellar spectroscopy, stardust grains, and gravitational wave communities should investigate avenues for collaboration to resolve open questions in heavy element synthesis, especially for the trans-iron elements, short-lived radionuclides, n-process, and r-process.
Millions of stellar spectra will be obtained in the coming years, so stellar abundances now need to be accurately, precisely, and homogeneously extracted on an massive scale, especially for heavy elements like Ba, Eu, Th, and U. Ultraviolet spectroscopy is needed to measure crucial elements like Ru, Ag, Mo, Te, Ag, Pb, Th, U, which break degeneracies between nucleosynthesis processes. More observational capability needs to be developed in this area. Measurements of the ratio of heavy element abundance to the abundances of lighter elements, including the so called $\alpha$ elements such as Mg, can be particularly powerful to trace the history of r-process nucleosynthesis in various Galactic components \cite{guiglion2018,Sku19,Cote2019,Reichert2020,Tautvaisiene2021}. These needs are well aligned with the priorities of the ASTRO2020 Decadal Survey \cite{ASTRO2020}: ``Industrial scale spectroscopy'' is the discovery area, and an IR/O/UV space telescope with spectroscopy capability is the priority recommendation in the space frontier.  

The laboratory analysis of stardust is another important observational area, where new capabilities will have a large impact on nuclear astrophysics (see sidebar on Pg.~\pageref{sidebar:lion}). Most importantly this approach provides the isotopic composition of astrophysical sites, which is critical, especially for the disentangling of production sites of p-nuclides and other elements in the trans-iron region (A $\sim$ 60-120, Ge to Cd) where multiple proton or neutron-capture processes may contribute. Another important observational development is the expected rapid growth of direct observations of element production in transient events like supernovae and neutron star mergers, due to surveys like the Vera Rubin Observatory's Legacy Survey of Space and Time and gravitational wave observations from LIGO and Virgo.  Another multi-messenger observation of a relatively close neutron star merger such as GW170817 would be of particular importance for the field (see ASTRO2020 Priority Area: New Windows on the Dynamic Universe \cite{ASTRO2020}). In particular, better late-time observations of the associated kilonovae may contain detectable heavy-element signatures, for example from actinides. An advanced MeV $\gamma$-ray observation capability would greatly enhance direct observations of heavy element nucleosynthesis sites in our galaxy and enable the direct identification of specific synthesized isotopes~\cite{Timm2019,andrews:20}. 

\subsection{What Do We Need?} \label{sec:elements:needs}

\begin{itemize}
    \item Advanced rare isotope beam facilities that push isotope production to as neutron-rich nuclei as possible. This includes FRIB, nuCARIBU, N=126 Factory, ARIEL, FAIR as well as the FRIB400 upgrade for FRIB. At FRIB a full range of instrumentation is needed to maximize sensitivity for the broad range of reactions and nuclear properties involved in heavy element nucleosynthesis, including SECAR, the FRIB decay station, the HRS, SOLARIS, GRETA, and ISLA. 
    
    \item Storage ring and isotope harvesting developments for direct neutron-capture measurements.

    \item Intense rare isotope beams near stability for the charged-particle reactions of the weak r-, $\nu$p-, and p-processes. 
    
    \item Developments in nuclear theory to reliably predict the properties and reactions of extremely neutron-rich or heavy isotopes that are beyond the reach of experiment. 
    
    \item Access to leadership-class computing resources and improvements to model codes. 
    
    \item Large spectroscopic surveys of stars (2-4m class) (e.g., SDSS-V, DESI, 4MOST, WEAVE, GALAH) and accompanying improvements in large-scale homogeneous abundance determinations; high resolution follow-up studies at large telescopes (10m class or more, including the future ELT, GMT, and possibly TMT (Thirty Meter Telescope)) with near UV spectroscopy capabilities and public access to these capabilities; a space-based high resolution UV spectroscopy capability such as LUVOIR (see ASTRO2020 decadal survey \cite{ASTRO2020}) to derive abundances of key heavy elements. Advances in the theoretical modeling of stellar atmospheres (3D) to extract more accurate information from stellar spectroscopy (e.g., non-LTE) also need to be incorporated. 
    
    \item An advanced MeV $\gamma$-ray mission to detect decay radiation from rare isotopes in space such as COSI or AMEGO. 
    
    \item Continued Gravitational wave observations with increased sensitivity with Advanced LIGO/VIRGO, and in the future with Cosmic Explorer and Einstein Telescope, as well as multi-messenger followup of neutron star mergers and their associated kilonovae. 
    
    \item Increased effort in isotopic composition measurements of stardust grains. 

\end{itemize}

\clearpage


\section{Understanding the Transient Sky}
\label{sec:transient}

\subsection{Introduction} \label{sec:transients:intro}

While for the naked-eye observer the night sky appears as static and immutable, for deep-sky surveys and X-ray observatories there are numerous transient events that dominate the night. These transient events offer unique windows into physics at extreme temperatures and densities. In many cases they represent major sources of new elements and provide unique diagnostics of matter in extreme conditions. They are, therefore, of key importance for nuclear astrophysics. It has been challenging to observe transient events with high fidelity due to the limited field of view of instruments with high sensitivity. This is now rapidly changing as astronomy enters an era of time-domain observations. The ASTRO2020 Decadal Survey identifies Time-Domain Astrophysics (the study of how astrophysical objects change with time) as the highest priority sustaining activity for space \cite{ASTRO2020}. New developments include dedicated ground-based surveys for transient searches with broad sky coverage and near-daily cadence~\cite{Ivezic2019}. Another key development is the advent of multi-messenger astronomy, where, for example, a gravitational wave signal will be used as a trigger for rapid followup with a multitude of other instruments, with multi-wavelength coverage but a more limited field of view~\cite{Bailes2021}. These new capabilities have already revolutionized nuclear astrophysics and will continue to do so in the coming decade. 

\begin{wrapfigure}{r}{0.5\textwidth}
\vspace{-0.30in}   
  \begin{center}
    \includegraphics[width=0.48\textwidth]{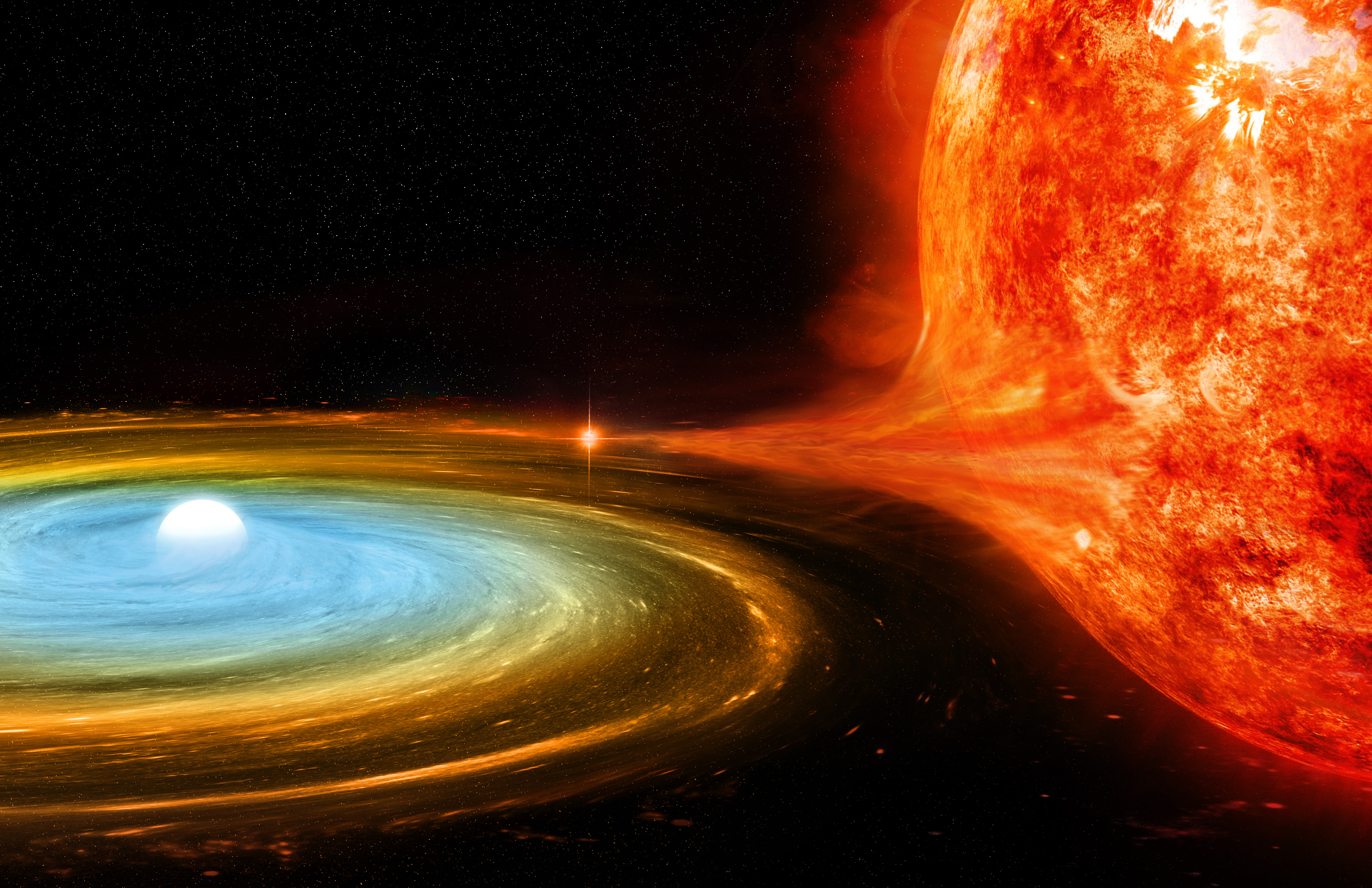}
  \end{center}
  \vspace{-0.20in} 
  \captionsetup{margin=0.2cm}
  \caption{Artist's rendition of the binary system ASASSN-16oh, a rapidly accreting white dwarf star recently observed by NASA's Chandra X-ray Observatory.  This system has an unusually high accretion rate and may be the progenitor system for a thermonuclear supernova. Image credit: NASA/CXC/M.Weiss}
  \label{fig_binary}
\vspace{-0.20in} 
\end{wrapfigure}

There are already a broad range of known transient stellar phenomena of interest for nuclear astrophysics. Neutron-star mergers and the associated kilonovae are discussed in Section~\ref{sec:heavy}. Here we focus on events associated with accreting compact objects (e.g. Fig. \ref{fig_binary}): classical novae, Type-I X-ray bursts, accreted neutron star crust cooling, and thermonuclear (Type Ia) supernovae.  X-ray bursts and classical novae are driven by explosive hydrogen and helium burning on the surface of compact stellar objects, neutron stars or white dwarfs that accrete hydrogen and helium-rich material from a companion star~\cite{Jose2007,Meisel18a} (see Section~\ref{sec:densematter} for more details on neutron stars). X-ray bursts are associated with neutron stars and are the most commonly observed stellar explosions in the Galaxy. They synthesize proton-rich isotopes up to mass numbers of approximately 100~\cite{Schatz2001,Woosley2004,Jose2010}. These relatively short bursts ($\rng[--]{10}{100}{\usk\second}$) recur on typical timescales of hours to days, reaching peak temperatures in the burning region of up to \val{2}{\Giga\K}.  Classical novae reach comparatively lower peak temperatures (up to and potentially above \val{0.4}{\Giga\K}) due to the lower surface gravity of the more diffuse underlying white dwarf. Novae are observed to produce elements up to about \calcium[40]. What the heaviest elements are that can be synthesized in novae remains an open question though~\cite{Jose2007b}. Similar to X-ray bursts, classical novae are recurrent but the time between explosions is typically of the order of 100,000 years, with the exception of recurrent novae, which have outburst recurrence times of 1-100 years. Thermonuclear supernovae also occur in accreting white dwarf systems. However, unlike novae, the explosion is not powered by nuclear reactions in the accreted layer, but instead the entire white dwarf serves as nuclear fuel. As a consequence, in a typical luminosity event the more massive white dwarf in the binary system is completely disrupted, and typical luminosity thermonuclear supernovae do not recur. Neutron stars undergoing periodic accretion outbursts result in a non-explosive X-ray transient~\cite{Wijnands2017}. This is due to a gradual cooling of the crust (see Fig.~\ref{fig:NS_cross_sec}) over months or years after the end of an accretion episode, which can last many years. Accretion often resumes after some time and the entire process may repeat, resulting in multiple observed cooling events for some systems. Several crust cooling systems are also known to host X-ray bursts during the accretion phase~\cite{Galloway2021}.

X-ray bursts, cooling neutron star crusts, novae, and thermonuclear supernovae are important to understand for different reasons. While it is unclear whether significant amounts of the synthesized nuclei in X-ray bursts are ejected, the burst ashes define the composition of the neutron star crust and shape a variety of observable signatures that can be used to constrain the properties of neutron stars~\cite{Deibel2016,Meisel2017}. Cooling crust observations provide a direct signature of matter properties at extreme density and neutron-richness. Novae are thought to be important sources of specific light element isotopes such as $^{7}$Li, $^{13}$C, $^{15}$N, and $^{17}$O, as well as $\gamma$-ray emitters such as $^{26}$Al and $^{22}$Na \cite{Gomez1998,Starrfield2020}. Thermonuclear supernovae are major contributors to nucleosynthesis, providing about half of the iron-group elements Cr, Mn, Fe, and Ni, in the Galaxy \cite{Seitenzahl2017}, and, potentially, also contribute to proton-rich isotopes of heavy elements beyond iron through the p-process (Section ~\ref{sec:heavy}). Furthermore, the use of these objects as so-called standard candles to inform, e.g., cosmological expansion, necessitates an improved understanding of such transients~\cite{Hosseinzadeh2017}.

\subsection{Open Questions} \label{sec:transients:questions}
There are many open questions of specific interest to nuclear astrophysics related to X-ray bursts, novae, and thermonuclear supernovae:

\begin{itemize}
\item What do observations of X-ray bursts and crust cooling tell us about neutron stars? 
\item Are there observational signatures of freshly synthesized elements in X-ray bursts?
\item What creates the diversity of observed X-ray bursts and associated phenomena such as burst oscillations? 
\item What is the role of multi-dimensional effects such as burning front propagation across the neutron star surface?
\item What mechanism is responsible for the unexplained strong shallow crustal heating inferred from observations of accreting neutron stars?  
\item How is classical novae white dwarf material mixed into the accreted envelope, how much of galactic $^{7}$Li and $^{26}$Al are produced, and what are the heaviest elements synthesized?
\item What unique isotopic signatures can be used to identify stardust from novae? 
\item What are the isotopic abundances of iron-group elements produced in thermonuclear supernovae, and do thermonuclear supernovae contribute to the origin of the p-nuclei? 
\item What do nucleosynthesis signatures tell us about the nature of the progenitor systems and explosion mechanism of  thermonuclear supernovae? 
\end{itemize}

\begin{tcolorbox}[colback=blue!5, colframe=blue!40!black, title=Sidebar: New Views of the Cosmos]
  \begin{center}
    \includegraphics[width=0.8\textwidth]{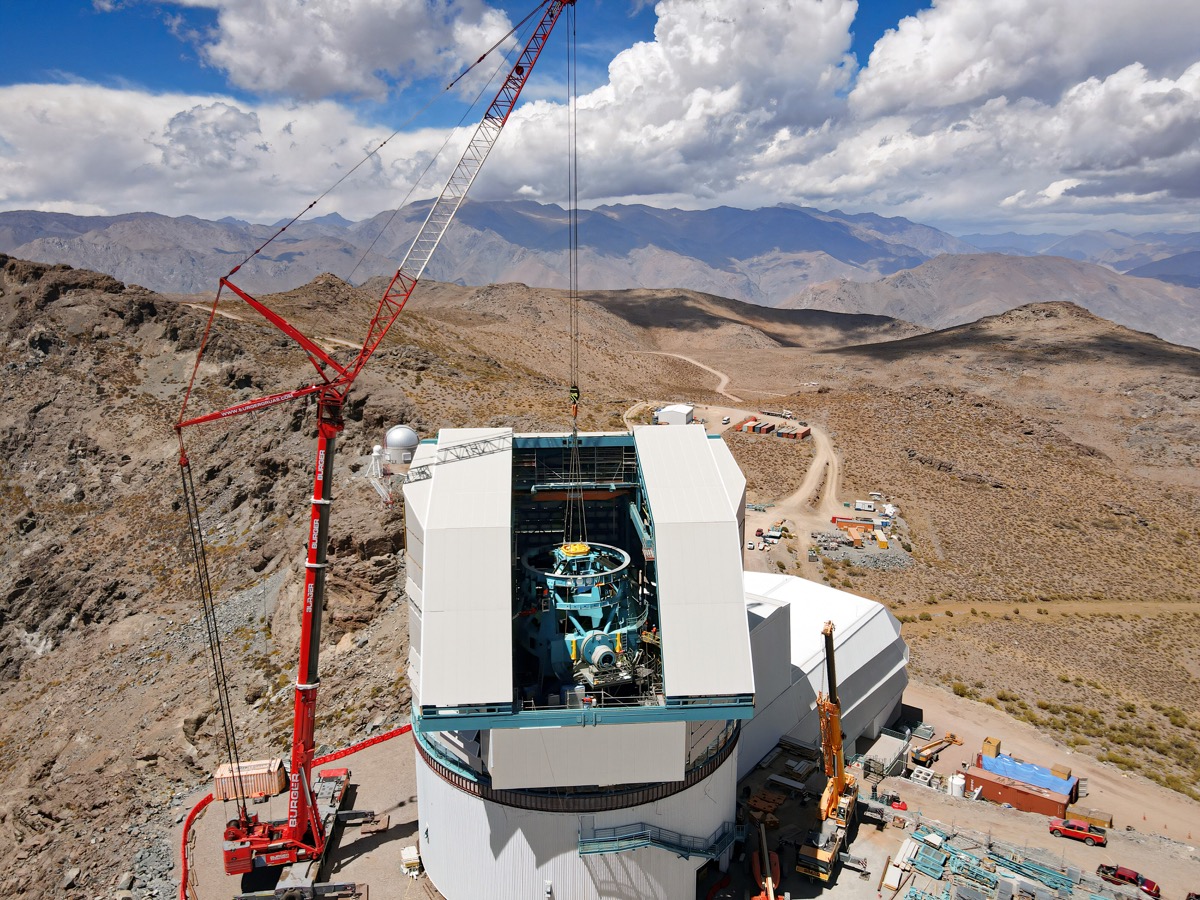}
  \end{center}
  \vspace{-0.20in} 
  \begin{center}
  {\footnotesize {\it Construction at the Vera C. Rubin Observatory. Credit: ESA/Hubble, Rubin Observatory/NSF/AURA.}} \label{fig:wg2:vro} \\  
  \end{center}
  \vspace{-0.08in} 
{\footnotesize
We stand on the precipice of a revolution.  This next era of our
investigation of the cosmos is about to be kick-started by a dramatic
leap in telescopes.  During the next several years, several
transformative ground- and space-based instruments will start scanning the skies.  For example, the Vera C. Rubin Observatory shown above, is a ground-based telescope currently under construction on the El Pe\~{n}\'{o}n peak of Cerro Pach\'{o}n in northern Chile.  New science will be driven by the largest digital camera ever created
and using it, for example, to investigate many kinds of stellar
variability producing real-time movies of transient objects in the night sky in the ultraviolet, visible, and infrared.
}
\label{sidebar:verarubin}
\end{tcolorbox}

\subsection{How Did We Get Here?} \label{sec:transiens:how}
More than 100 X-ray bursters are known in the Galaxy and most of them show bright bursts several times a day. Consequently, X-ray observatories such as RXTE have obtained data on thousands of bursts that are available in databases such as MinBAR~\cite{Galloway2020}. More recently, NICER has provided a large amount of new data with unprecedented X-ray spectral coverage~\cite{Keek2018}. Observations of the response of long-term burst behavior to changes in accretion rate are available as well~\cite{Galloway2017}. These observations reveal a wide diversity in burst phenomenology from source to source and among bursts of the same source. These include single, double, and triple bursts; short bursts ($\sim$10 s), long bursts ($\sim$100 s), intermediate long bursts (hours), and superbursts (hours to days); short recurrence time bursts; radius expansion bursts; and bursts with millisecond period oscillations~\cite{Galloway2021}. 

Experiments with beams of radioactive isotopes along the proton drip line have identified waiting points in the rapid proton-capture process (rp-process) in X-ray bursts~\cite[e.g.,][]{schatz:98,Lang14,Ong17,Tu11,Lam16,Roge11,DelS14,Bazi08,Clark2004}. This has led to long bursts being recognized as rp-process signatures, while short bursts are either helium flashes or powered by a limited rp-process~\cite{Heger2007,intZand2017}. With these advances recent one-dimensional models were also able to reproduce observed burst sequences and their response to changes in accretion rate~\cite{Heger2007,Meis18b,John20,Lam2022}. However, such successes are limited to the so called ``textbook burster," a single system with unusually regular bursts~\cite{Bildsten2000}. Models can also qualitatively explain intermediate long bursts and superbursts as explosions of thick layers of helium and carbon, respectively, though there are open questions regarding how the required conditions for ignition and the amount of required fuel can be reached~\cite{Strohmayer2002,Cumming2006,Keek2012}. 

Advances in nuclear physics have solidified this picture. Most of the masses and ground-state half-lives of nuclei involved in the rp-process are now experimentally known ~\cite{Meisel18a}. Recent mass measurements were performed with Penning traps~\cite{Clark2004,Fallis2011,Dilling2018,Valverde18}, MR-TOF~\cite[e.g.][]{Wolf2013}, and TOF-B$\rho$ facilities~\cite[e.g.][]{Chartier1998,Estrade2011,Meisel2020}. Recent decay studies 
have focused on $\beta$-delayed particle emission probabilities~\cite[e.g.,][]{Lorusso2012,DelS14,Roge11b,Suzuki2017,Hoff2020,Ciem20}. Modifications of $\beta$-decay rates in the astrophysical environment, for example due to continuum electron capture, have been calculated and are significant in some cases \cite{Sarriguren2011}. The remaining nuclear-physics challenges are the large uncertainties in nuclear reaction rates affecting ignition and light curves. A broad range of techniques have been developed to address this challenge and constraints have been possible for a number of reaction rates \cite{Meisel18a}. Broadly, these techniques fall into the two categories of direct and indirect measurements. Direct measurements of explosive hydrogen and helium burning reactions at or near the astrophysical energies where they occur have been performed with rare isotope beams using recoil separators (DRAGON) \cite{Lotay2016,Williams2020,Lennarz2020,Lovely2021}, active targets (MUSIC, AT-TPC, ANASEN, and others) \cite{Avila2017,Schmidt2018,Randhawa2020,Anastasiou2021}, or gas targets (JENSA) \cite{CHIPPS2014553,CHIPPS2017297,SCHMIDT20181,Schmidt2018}. In many cases direct measurements have not been feasible due to limited radioactive ion beam intensities and the extremely small cross sections involved. As such, indirect measurements have played a major role in our understanding of the nuclear physics of transients. These include $\beta$-decay techniques to locate and characterize resonances of interest~\cite{Grawe2007,Glassman2018}, excitation energy and spin measurements via $\gamma$ spectroscopy (e.g. SeGA, GRETINA (and GRETA in the future), and  Gammasphere)~\cite{Clement2004,Jenkins2004,Seweryniak2005,Lang14,Ong17,Wolf2019}, $\gamma$-particle coincidence techniques (e.g. GODDESS)~\cite{Hall2019,Hall2020}, the Trojan Horse reaction
Method~\cite{Tumino:2021}, the Asymptotic Normalization Coefficient method \cite{AlAbdullah2014}, elastic scattering \cite{Bardayan2000,Zhang2014}, constraints on statistical nuclear properties (e.g. with $(^{3}{\rm He},n)$ neutron time-of-flight measurements at Ohio University and the University of Notre Dame)~\cite{Alma12,Solt21}, as well as charged-particle spectroscopy with high-resolution charged particle detector arrays \cite{Pain2015,BARDAYAN2015311} and the recently refurbished Split-Pole spectrographs at TUNL~\cite{Setoodehnia2021,Marshall2021}, Florida State University~\cite{Good2021}, and University of Notre Dame~\cite{Bardayan2019}. 

Over the quarter-century since their discovery, around a dozen transiently accreting neutron-star systems have been observed to undergo prolonged cooling periods~\cite{Wijnands2017}. The steep density gradient in the outer layers of the neutron star means that exterior layers cool far quicker than interior layers, such that the surface steadily comes into thermal equilibrium with deeper regions as time progresses~\cite{Shternin2007,Brown2009}. As such, the cooling light curve provides a tomographic picture of the accreted neutron-star crust~\cite{Page2013,Turlione2015,Lalit2019}. The light curve may be used to constrain bulk properties of the neutron star and microscopic properties of the crust~\cite{Deibel2015}. However, the cooling light curve is sensitive to the crust composition, which is influenced by surface nuclear burning such as X-ray bursts. The light curve also depends on nuclear reactions in the crust that heat (during the accretion phase) and cool via neutrino emission. The relevant nuclear reaction sequences have only recently been fully identified \cite{Haensel2003,Schatz2014,Lau2018,Chugunov2019,Wang2021} and include electron capture, $\beta$-decay, fusion, and neutron transfer.  Nuclear masses play a key role in determining which processes may occur and how much energy they can generate, and significant experimental and theoretical progress has been achieved. In contrast, predicting the rates of the reactions has been a challenge \cite{Yakovlev2010,Chugunov2019} and only a few dedicated experiments have been performed to date \cite{Ong2020,Vada2018}. 

Unlike X-ray bursts, nova light curves are powered by the cooling of ejected material and are less sensitive to the details of the nuclear physics. However, novae eject their nuclear ashes into space, whereas X-ray bursts do not. Observations of spectral features have provided detailed composition information that can be directly compared with nucleosynthesis models~\cite{Schwarz2007,Downen2013,Denissenkov2014}. Analysis of observational data requires detailed radiation transport models to account for the fact that only part of the gas is visible at a given time and that this part evolves depending on ejecta morphology \cite{Schwarz2007,Helton2012}. 

The nuclear reactions synthesizing the ejected elements are hydrogen-burning reactions involving a mix of somewhat neutron-deficient rare isotopes and stable nuclei. Generally, relatively intense beams of the relevant nuclei have been available and the high temperatures compared to traditional stellar burning result in reasonably large cross sections that are within reach of current experimental techniques \cite{Ruiz2006}. For these reasons, despite their explosive character, novae are among the nucleosynthesis sites with the most complete nuclear-physics knowledge. This has been achieved using the same techniques applied to X-ray burst reactions for unstable nuclei (see above and e.g. \cite{Bardayan2002, Bishop2003, Jenkins2004, Peplowski2009, Sallaska2010, Bennett2016, Friedman2020}), and the same techniques applied to stellar-burning reactions  (Section~\ref{sec:stars}, e.g. \citep{Lennarz2020}). Despite this impressive progress, there remain a few key nuclear uncertainties that need to be addressed in the future requiring beams that have been difficult to produce, such as $^{30}$P \cite{Iliadis2002,Wrede2014}. 

Models that employ up-to-date nuclear reaction information have been successful in reproducing observations (which have large uncertainties) by assuming a fraction of the white dwarf surface is mixed into the burning zone~\cite{Jose2020}. In the case of a neon nova, this material is rich in neon, which can serve as a seed for a mild rp-process producing elements up to, and possibly beyond, Ca \cite{Andrea1994}.  Progress has also been made in understanding this mixing process by using advanced three-dimensional simulations of accretion and ignition indicating that mixing occurs as part of the explosion~\cite{Jose2020}. Despite this mixing, recent models predict that the white dwarf is growing in mass. Such nova systems are therefore expected to evolve towards a thermonuclear supernova explosion once the Chandrasekhar mass limit is reached~\cite{Starrfield2005,Chomiuk2012}. 

Among the three explosion scenarios discussed in this section, only novae are expected to be significant producers of dust~\cite{Gehrz1998}. This opens the possibility to obtain detailed information on nova nucleosynthesis from stardust grains found in meteorites. Indeed, grains have been found with some of the isotopic signatures expected for novae~\cite{Amari2001}. However, this identification relies on nucleosynthesis predictions which require more accurate nuclear reaction rates~\cite{Pepin2011,Iliadis2018,Bose2019,Kennington2020}. 

Thermonuclear supernovae are modeled as thermonuclear explosions of white dwarfs. The evolution of the progenitor system towards explosion, the ignition of the thermonuclear burning, and details of the explosion mechanism are still uncertain~\cite{Seitenzahl2017}. This is of particular importance because of the prominent role these events play in observational cosmology and in the cosmic cycle of matter~\cite{Perlmutter1999,Abbott2019,Palla2021}. A wealth of observational data has been collected over the past decade (and will grow with future surveys such as the Vera Rubin Observatory - see sidebar on Pg.~\pageref{sidebar:verarubin}). The interpretation of these observational data in a consistent theoretical framework poses one of the most important challenges in stellar astrophysics~\cite{Narayan2016}.  Recent results from observations as well as predictions of observables from explosion models and nucleosynthesis considerations challenge the canonical view of thermonuclear supernovae as arising exclusively from the explosions of carbon-oxygen white dwarfs near the Chandrasekhar mass limit~\cite{Livio2018,Wang2018}. The most recent calculations explore three-dimensional explosion models~\cite[e.g.,][]{Fink2013,Sato2015}. The postprocessing tracer particle technique has been successfully used to overcome the challenge of accurately following the nuclear reactions of the roughly 400 isotopes that are relevant during a thermonuclear supernova explosion \cite[e.g.,][]{Lach2020}. Such postprocessing has enabled the calculation of large model grids that explore different explosions, metallicity effects, and the impact of nuclear uncertainties. Of particular relevance for nucleosynthesis in thermonuclear supernovae are electron-capture reactions. They directly impact the elements produced, and affect the dynamic evolution of the explosion from the initial mass accretion of the white dwarf, through carbon simmering, and through to the actual thermonuclear explosion. As such, electron-capture rates are critical in linking observables to the nature of the progenitor system, one of the key open questions. The sensitivity of thermonuclear supernovae nucleosynthesis to electron-capture rates has been analysed in a few works with conflicting results~\cite{Parikh2013,Mori2018,Bravo2019}. Charge-exchange reaction measurements have been used to constrain these electron-capture reactions by benchmarking theoretical shell-model calculations and quantifying nuclear uncertainties \cite{Langanke2021}. These data available in a dedicated database \cite{0004-637X-816-1-44,0954-3899-45-1-014004,weakrlib}. Of particular importance are new techniques that have been developed to perform these measurements far from stability, such as the use of the active target AT-TPC with the S800, and in the future the HRS, spectrometer at FRIB \cite{AYYAD2020161341,GIRAUD2021}. This will also be important for neutron star crust reactions and the prediction and interpretation of neutrino signals from core-collapse supernovae (Section \ref{sec:stars}).

\begin{tcolorbox}[colback=pink!5, colframe=pink!40!black, title=Sidebar: Peering Into Neutron Stars]
  \begin{center}
    \includegraphics[width=0.8\textwidth]{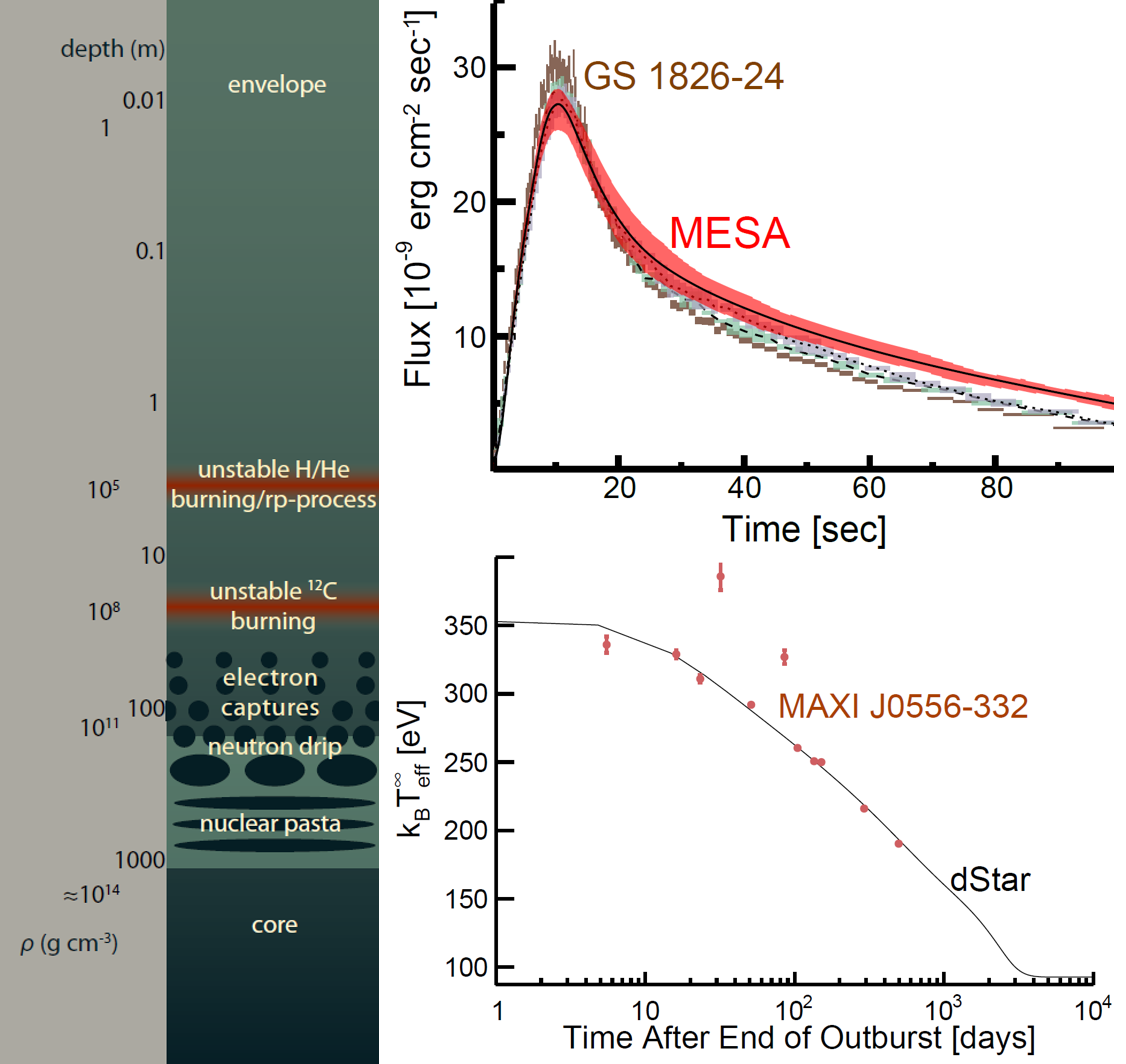}
  \end{center}
  \vspace{-0.20in} 
  \begin{center}
  {\footnotesize {\it  Schematic cross section of an accreting neutron star (left) and model-observation comparisons for associated X-ray transients, X-ray bursts (upper-right, showing X-ray flux as a function of time) and crust cooling (lower-right, showing the surface temperature inferred from X-ray observations as a function of time).}} \label{fig:wg4:side} \\
  \end{center}
  \vspace{-0.08in} 
{\footnotesize
Neutron stars siphoning gas from a nearby companion star are nature's ultradense matter laboratory. The gas is ultimately deposited onto the neutron-star surface, heating up, and generating a number of unique transient events such as X-ray bursts and cooling stellar surfaces, for instance from the sources GS 1826-24 and MAXI J0556-332. Over the past 20 years, major advances in nuclear experiment, nuclear theory, and astrophysics computations have enabled the first model-observation comparisons, as shown above for X-ray bursts calculated with MESA and crust cooling calculated with dStar. Experiments at facilities such as FRIB will fill in the gaps in nuclear data required for higher-fidelity model calculations. Comparison of these models with continued observations of the X-ray sky, including observations with advanced X-ray telescopes, promise to reveal the high density matter phenomena that may occur inside neutron stars. Advances in computation will enable model calculations to move beyond the one-dimensional approximations presently used.
}
\label{sidebar:XRB}
\end{tcolorbox}

\subsection{What Needs To Be Done?} \label{sec:transients:todo}

There are tremendous opportunities in the near future to significantly advance the understanding of thermonuclear transients, determine their contribution to the origin of the elements, and use them as laboratories for extreme environments that cannot be replicated on Earth. This is primarily due to the unprecedented advance in rare-isotope beam production capabilities, as well as advances in computational modeling. 

With the availability of large-scale X-ray burst observational data sets, the focus in the near future will be on advancing the modeling of X-ray bursts to explain these observations, address open questions, and extract information on both the binary system and the underlying neutron star. This will require much improved nuclear reaction rates, including reactions on relatively long-lived nuclear isomers \cite{Lotay2022}, and the development of realistic multi-dimensional models. The hydrogen- and helium-burning nuclear reactions in X-ray bursts involve unstable nuclei that have been historically challenging to produce as sufficiently intense beams to perform experimental measurements. New facilities like FRIB and RAISOR at Argonne National Laboratory will change this. For example, the SECAR recoil separator and the AT-TPC active target will enable new direct measurements taking advantage of the unique FRIB radioactive ion beam production capabilities~\cite{Berg2018,Randhawa2020}. The broad range of indirect techniques and associated major equipment developments at a large number of accelerator laboratories will come to full fruition and guide as well as complement direct measurements. Better calculations of modification of reaction and decay rates due to the rather extreme temperatures and densities during X-ray bursts may also be needed in some cases, and can be guided by targeted radioactive beam experiments.

The development of multi-dimensional X-ray burst models that capture burning front propagation across the neutron star surface and other important effects have been a long-standing challenge due to the extremely high speed of sound near the surface of a neutron star. Such burning front propagation effects are directly linked to observations of burst oscillations, which are connected to luminosity variations across the neutron-star surface \citep{Watts2012,Bilous2019}. Burst oscillation models combined with improved burst nuclear physics \cite{Chambers2019} may be used to constrain neutron star properties. With advances in computing power and algorithms, significant progress on multi-dimensional X-ray burst models has been made recently~\cite{Eiden2020,Harpole2021} but much remains to be done. 

Observationally, a major challenge yet to be addressed is the unambiguous detection of spectral features produced by elements synthesized in X-ray bursts, either by directly observing the convective surface or by observing potentially accumulated ejected material around the bursting system. Such observations would provide direct information about the nucleosynthesis and, through the extracted red-shift, about the neutron star compactness and the nuclear equation of state  (Section~\ref{sec:densematter}). An advanced X-ray telescope with time resolved spectroscopic capabilities such as STROBE-X \citep{STROBE-X} will be needed. 

In addition, it is now clear that burst physics are tightly coupled to the physics of neutron star crusts. Both problems need to be addressed simultaneously and consistently ~\cite{Keek2017,Meisel18a,Meis18b}. Cooling models need to use surface burning ashes produced from X-ray burst models that are consistent with the system, while experiment and theory need to substantially improve the precision and coverage of nuclear physics input that such models require. New rare isotope facilities like FRIB will provide unprecedented access to study the properties and reactions of the relevant rare neutron-rich isotopes in the crust of neutron stars. This will enable precision determination of ground-state properties, including nuclear masses using Penning traps, storage rings, and MR-TOF. Measurements of weak transition strengths using total absorption spectroscopy, neutron counting, and charge-exchange reaction measurements will also become possible. To reach all the relevant nuclei out to the limit of neutron stability will require the FRIB400 energy upgrade of the new FRIB facility. Fusion measurements involving neutron-rich nuclides will provide experimental grounding of nuclear theory calculations needed for pycnonuclear fusion reaction rate estimations. Meanwhile expanding the set of observed crust cooling sources will improve the fidelity of crust cooling models by providing more data with which to test theory. Early and regular coverage of the observed light curve are particularly crucial to breaking model degeneracies~\cite{Wijnands2017}.

Improved nuclear physics and multi-dimensional modeling are priorities for advancing the understanding of novae. Only a few reactions remain with uncertainties that limit our understanding of nova nucleosynthesis, including $\sodium[22](\proton,\gamma)\magnesium[23]$, $\aluminum[25](\proton,\gamma)\silicon[26]$, and $\phosphorus[30](\proton,\gamma)\sulfur[31]$~\cite{Denissenkov2014}. Direct measurements and more indirect studies of these reactions are needed. Furthermore, the endpoint of nova nucleosynthesis is still unclear.  Therefore, for a complete accounting of the nuclear physics involved in these explosions, targeted direct and indirect measurements, especially the reactions involving nuclei with mass numbers around 30-40, are required. We expect that with new facilities and equipment available now all relevant reactions are within reach of experiments. 

While progress has been made in realistic multi-dimensional hydrodynamic modeling of the mechanism that mixes white dwarf material into the ejecta~\cite{Jose2020}, advances are needed to address nucleosynthesis and whether white dwarfs in novae increase in mass over repeated outbursts~\cite{Chomiuk2021}, thus becoming potential progenitors for thermonuclear supernovae. 
A concerted effort in nuclear physics, modeling, and observations will be needed to understand the full range of nova nucleosynthesis. Observationally, more data on ejecta compositions of novae spanning the full range of elements are needed. Analysis of broader characteristic abundance patterns in stardust grains may provide an alternative avenue to determine the composition of nova ejecta~\cite{Iliadis2018}. 

Accurate modeling of the small-scale dynamics of thermonuclear flames is essential to understand the physics of thermonuclear supernovae. There has been significant progress in developments of computational methods and physical models for supernova simulations. While these advances have been encouraging, the current state of the art is far from satisfactory~\cite{Roepke2018}. Open issues include a proper representation of the instabilities for propagation of deflagrations that directly affect ejecta properties and the distribution of the synthesized isotopes, the transition from deflagration to detonation, though significant progress has been made recently \citep[e.g.,][]{Poludenko2019}, and the initial conditions of the explosion, since the progenitor structure and the ignition process cannot be directly observed and must be modeled. A good characterization of thermonuclear supernova progenitor systems is yet to be established~\cite{Lach2020}. The astronomical identification of a progenitor can help with suggesting potential scenarios, but it cannot resolve all of the initialization problems. Extensive simulations using higher accuracy models and connected nucleosynthesis calculations with improved nuclear physics may help discriminate among possible known scenarios and also explore alternate possibilities. Another important challenge is an accurate description of the radiation processes responsible for spectrum formation in the evolving ejecta, for example by including non-equilibrium effects \cite{Shen21}.

The carbon fusion reaction is key to understand thermonuclear supernovae and also drives the ignition of superbursts. While recent advances at stable beam facilities have provided new data, the cross section at the lowest relevant energies remains unknown and extrapolations are complicated by the interplay of unknown resonances and poorly understood possible hindrance effects~\cite{Beck2020}. Thermonuclear supernova nucleosynthesis is mostly governed by nuclear equilibrium at high temperatures and densities reducing the dependence on individual nuclear reaction rates. Only a few proton-capture reactions have been shown to impact nucleosynthesis, including the production of $\gamma$-ray emitters, and can be studied at FRIB \cite{Bravo2012}. However, electron capture rates are not in equilibrium and affect observables significantly. New rare isotope beam facilities such as FRIB will enable the first charge-exchange measurements with rare isotopes to probe electron-capture processes. Thermonuclear  supernovae have also been suggested as a site for the p-process, the related nuclear physics needs of which are discussed in Section~\ref{sec:heavy}.

\subsection{What Do We Need?} \label{sec:transients:needs}

\begin{itemize}
    \item Advanced rare isotope facilities such as FRIB that produce (1) the most neutron-rich nuclei and map the neutron drip line up to mass numbers of around 100 for neutron star crust studies requiring the FRIB400 energy upgrade (2) high intensity rare isotope beams of neutron deficient isotopes to directly measure X-ray burst, neutron star crust, and nova reactions and the associated experimental equipment such as recoil separators (e.g. SECAR at FRIB) and storage rings. 
    
    \item Advanced stable beam facilities and detector systems that can constrain carbon fusion rates and rp-process reactions indirectly.
    
    \item Improved theoretical understanding of nuclear reactions at the lowest energies, in particular very low-energy fusion and the role of cluster structure.
    
    \item An advanced X-ray observatory with time-resolved X-ray spectroscopy capability to observe X-ray binaries and novae, such as STROBE-X (see ASTRO2020 \cite{ASTRO2020} priority for Probe-class mission). 
    
    \item A curated database of X-ray burst and transiently accreting neutron stars observations suitable for comparison with nuclear physics-based models. 
    
    \item Advanced 3D computational models of X-ray bursts and thermonuclear supernovae. 
    
    \item Improved theoretical understanding of the physics of accreted neutron star crusts.
\end{itemize}



\section{Neutron Stars and Dense Matter}
\label{sec:densematter}


\subsection{Introduction}

\begin{figure}[ht]
    \centering
    \includegraphics[width=0.7\columnwidth]{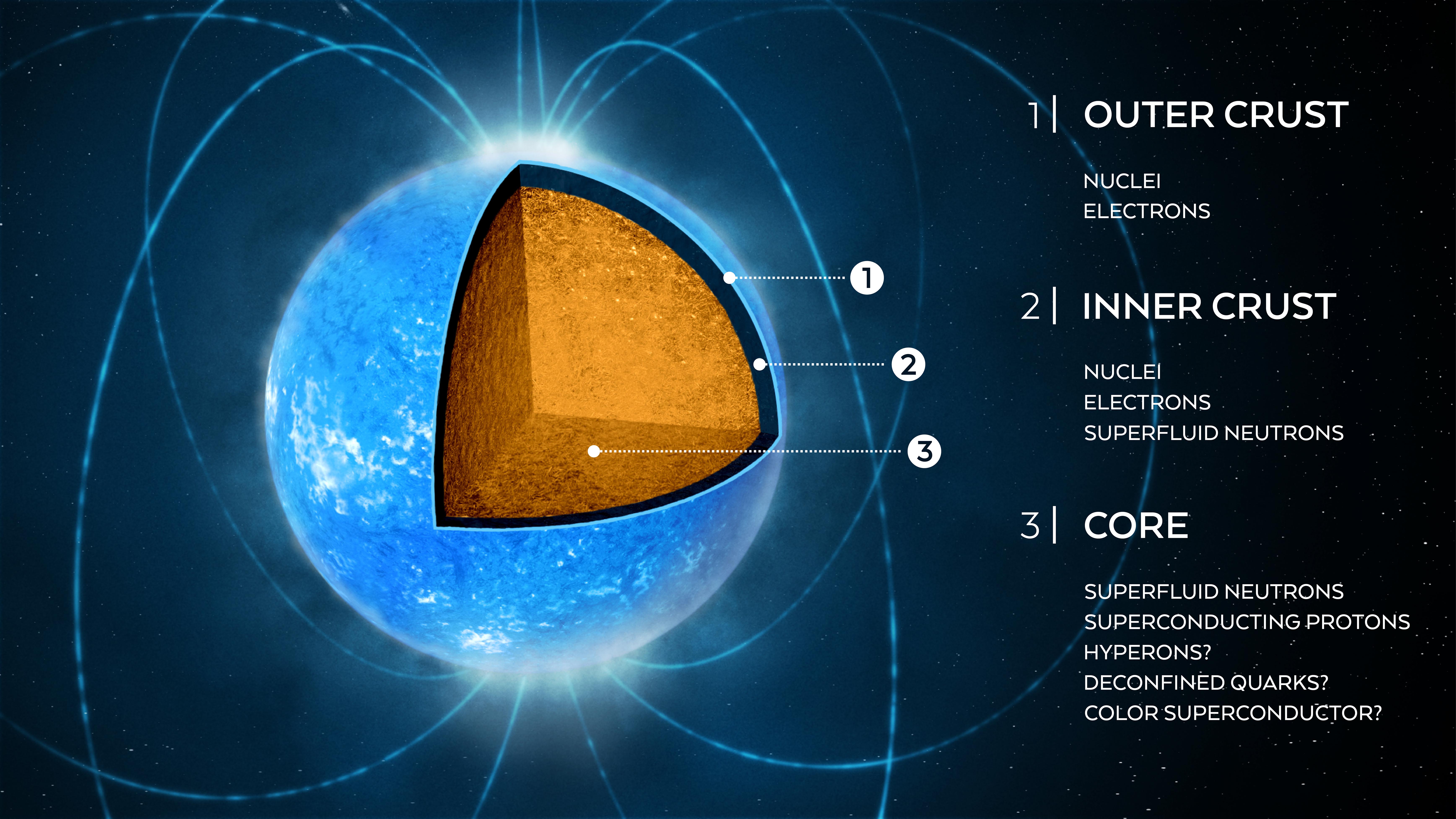}
    \caption{
        Cross section of a neutron star, showing the composition for each of its layers \cite{Watts:2016uzu}.
    }
    \label{fig:NS_cross_sec}
\end{figure}

Neutron stars are the densest non-singular objects in the present Universe.
Most are remnants of core-collapse supernovae with masses $1-2$ times that of our Sun contained in a radius of about 12~km~\cite{Ozel:2016oaf}.
The extreme conditions encountered in their interiors are the outcome of the interplay of all four fundamental forces, with nuclear physics and gravity taking center stage~\cite{Lattimer:2015nhk,Baym:2017whm,Oertel:2016bki}. 
Without strong nuclear interactions among the particles in their core (Fig.~\ref{fig:NS_cross_sec}), neutron stars would be overcome by the crushing force of gravity and could not exist.

Despite abundant observations of neutron stars in our Galaxy and beyond, their properties are still uncertain, posing a puzzle that attracts the attention of scientists in many disciplines, including nuclear theorists and experimentalists, astronomers and astrophysicists, relativists, 
 and computational physicists. 
 Studying neutron stars requires microscopic models of the dense matter in the stellar interior 
that must describe matter over a wide range of densities and temperatures. These models have to account for potentially exotic, unknown forms of matter in the neutron-star cores while remaining consistent with the results of terrestrial nuclear experiments.
The models can be tested against, and informed by: \begin{enumerate*}
\item experiments with atomic nuclei and astronomical observations, including neutron star mass measurements using radio signals,
\item combined neutron star mass-radius measurements from X-ray observations and observation of gravitational waves emitted during neutron star collisions,
\item observations of surface temperature, spin, and magnetic field evolution, and
\item the detection of neutrinos from core-collapse supernovae.
\end{enumerate*}
Current and future observations and experiments will provide strong tests of theories for nuclear matter and will help us to unravel the nature of dense matter.


\subsection{Open Questions}

\begin{itemize}
    \item How robust are different nuclear-physics models in describing the interiors of neutron stars? At what densities are they applicable?
    \item What particles are present in neutron-star interiors and which states of matter do they form?
    \item How can we best connect experiments with atomic nuclei to the properties of neutron-rich matter in the crust and core of neutron stars?
    \item Do we fully understand the systematic uncertainties in the analyses of radio, X-ray, and gravitational-wave data from neutron stars, and of experimental nuclear structure and heavy-ion collision data?
    \item How can we robustly combine this multitude of constraints spanning widely different scales 
    in the era of informative observations?
\end{itemize}


\subsection{How did we get here?}

Nuclear experiments provide a wealth of information about strongly interacting matter.  Such experiments include measurements of nuclear properties across the nuclear chart, e.g., the thickness of the neutron skin of heavy atomic nuclei (e.g., PREX~\cite{PREX:2021umo}), and heavy-ion collisions that create matter resembling that encountered in the cores of neutron stars \cite{Danielewicz:2002_Science,Liu2021,Fuchs:2006_PPNP,Lynch:2009_PPNP,Lynch2021,Russotto:2016ucm,Estee2021}.
These experimental results provide key input and benchmarks to construct theoretical models of dense matter.
Over the past decade, there have been significant advances in the theoretical modeling of neutron-star matter, based on chiral effective field theory~\cite{Epelbaum:2008ga,Hebeler:2015hla,Lynn:2019rdt,Keller:2020qhx,Drischler2021} in combination with developments in quantum many-body methods.
Importantly, these advances have provided more accurate descriptions of neutron-star matter with theoretical uncertainty estimates, but they are limited to nuclear densities where theories are relatively certain.
%
At higher densities, where information about the relevant degrees of freedom in dense matter is scarce, neutron-star observations provide the greatest potential to determine the nature of dense matter. Heavy-ion collision experiments provide probes at densities below twice the normal nuclear matter density. They also  have the ability to isolate the symmetry energy contributions to the equation of state \cite{Danielewicz:2002_Science,Fuchs:2006_PPNP,Lynch:2009_PPNP,Russotto:2016ucm} and the effective nucleon masses ~\cite{Tsang2009,Morfouace:2019_PLB}, which play important roles in the structure and dynamics of dense matter.

A wealth of observational data on neutron stars was provided in past years, including
mass measurements using radio timing~\cite{Demorest:2010bx, Antoniadis:2013pzd,Cromartie:2019kug,Fonseca:2021wxt}, 
X-ray observations of two neutron stars by NASA's Neutron-Star Interior Composition Explorer (NICER)~\cite{Riley:2019yda, Miller:2019cac, Riley:2021pdl, Miller:2021qha}, 
and detections of gravitational waves from coalescing neutron-star binaries, like the groundbreaking detection of GW170817~\cite{GW170817} 
by the Advanced LIGO and Virgo interferometers~\cite{TheLIGOScientific:2014jea,TheVirgo:2014hva} 
together with the electromagnetic counterpart~\cite{LIGOScientific:2017ync}. These observations
allowed us to measure masses and radii of neutron stars, although sizable statistical and systematic uncertainties remain and need to be understood. 
While NICER data seem to rule out neutron stars with relatively small radii at a given mass, gravitational-wave observations rule out the largest, least compact neutron stars with high confidence (Sidebar on Pg.~\pageref{sidebar:densematter}).
Combined inference of the masses and radii of neutron stars using a variety of neutron-star observations, theoretical modeling, and experimental information about dense matter has recently become the standard approach~\cite{GW170817,Bauswein:2017vtn,Radice:2017lry,Annala:2017llu,Fattoyev:2017jql,Coughlin:2018fis,LIGOScientific:2018cki,De:2018uhw,Most:2018hfd,Lim:2018bkq,Tews:2018iwm,Landry:2020vaw,Dietrich:2020efo,Essi2021,Legred:2021hdx,Raaijmakers:2021uju,Miller:2021qha,Pang2021,Lim2021,AlMa2021,Biswas:2021yge,Annala2021,Huth:2021bsp}. 
This approach provides the strongest constraints on neutron stars to date (Sidebar  Pg.~\pageref{sidebar:densematter}).

Additional constraints on neutron-star properties come from many sources. Continued radio timing provides an increasingly precise measurement of the moment of inertia of pulsar A of the double pulsar system J0737-3039~\cite{Kramer:2021_PRX}, and has recently enabled resolution of important details of a pulsar glitch~\cite{Ashton:2019_Nat}. X-ray observations of accreting neutron stars (Section~\ref{sec:transient}) and the cooling of isolated neutron stars, where thermal radiation from their surfaces is observable in X-rays~\cite{Yakovlev-etal2001, YakovlevPethick2004, Page-etal2006, Page-etal2009, Potekhin-etal2015}, provide information about the thermal properties of the interior. In particular, the young neutron star in the supernova remnant Cas A is of interest because its surface temperature decreases faster than expected within the standard neutrino cooling scenario. This anomalous temperature drop is naturally explained within a minimal cooling framework by the onset of core superfluidity~\cite{Page-etal2011, Shternin-etal2011}. Radio and X-ray observations also provide information about the long-term rotational and magnetic field evolution of the star.

%
%

\begin{tcolorbox}[colback=red!5, colframe=red!70!black, title=Sidebar: Comparing Observations with Dense Matter Experiment and Theory]
  \begin{center}
    \includegraphics[width=\textwidth]{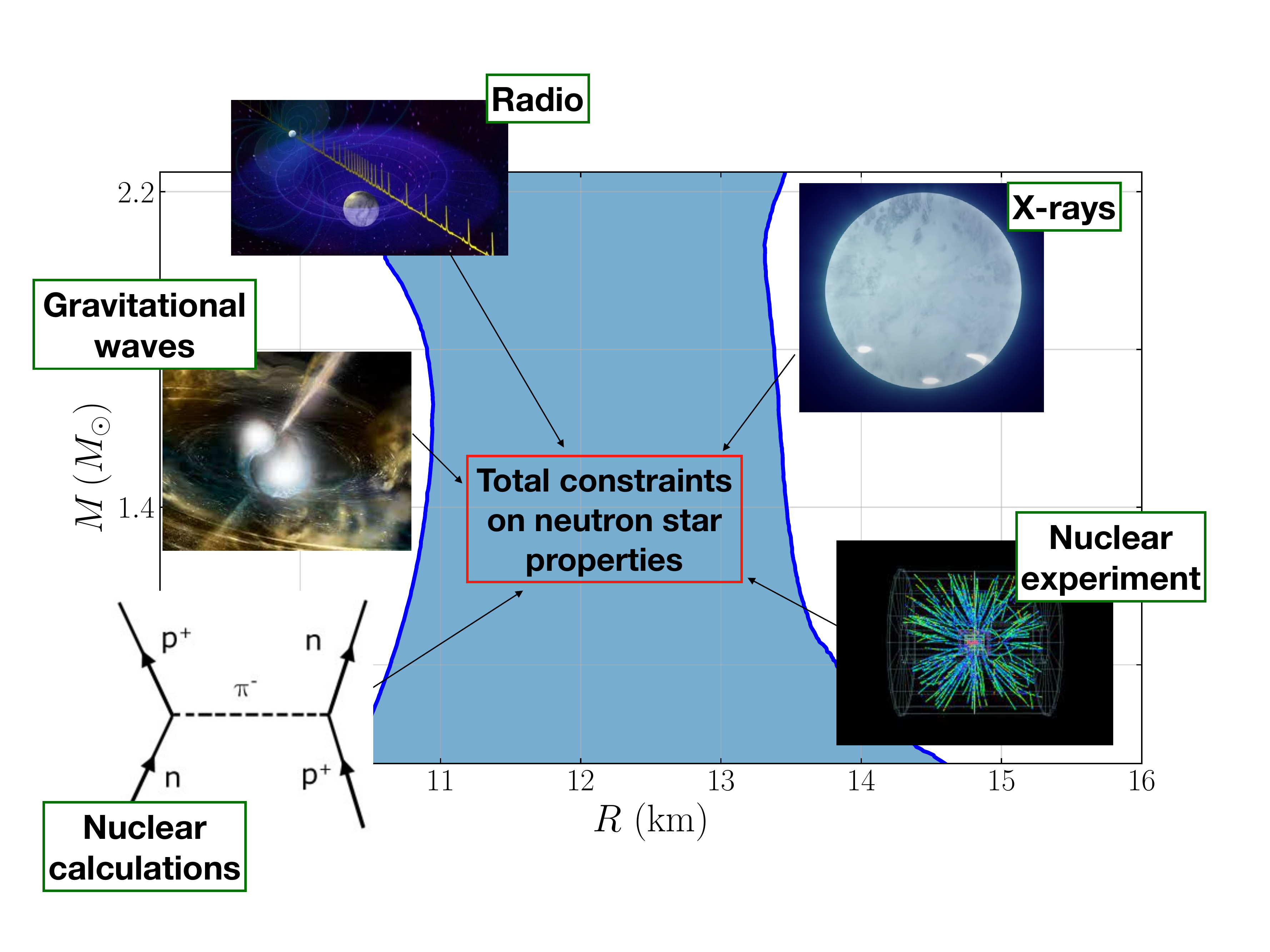}
  \end{center}
  \vspace{-0.20in} 
  \begin{center}
  {\footnotesize {\it  The mass-radius relation of neutron stars is constrained by a multitude of observations across the electromagnetic and gravitational bands, terrestrial experiments, and nuclear calculations.} Credits: Center \cite{Legred:2021hdx}, Radio: B. Saxton NRAO/AUI/NSF, Gravitational Waves: NSF/LIGO/Sonoma State University/A. Simonnet, X-rays: NASA's Goddard Space Flight Center, Nuclear experiment: ALICE/CERN } \label{fig:wg3:multimess} \\  
  \end{center}
  \vspace{-0.08in} 
{\footnotesize
A wealth of observational, experimental, and theoretical information contributes to the emerging picture of the properties of neutron stars. Gravitational waves and X-rays are inconsistent with very large and very small radii, respectively, resulting in a preferred radius of about $12$~km. Radio observations reveal neutron stars with up to twice the mass of our Sun demonstrating the existence of relatively massive neutron stars. These constraints are complemented by observations of light emitted as part of heavy-element production during neutron-star mergers. 
Terrestrial nuclear experiments and nuclear-theory calculations anchor the properties of nuclear matter at densities lower than those explored in the center of neutron stars. The emerging constraints will be further tightened by upcoming observations and experiments.
}
\label{sidebar:densematter}
\end{tcolorbox}

\noindent
Crucial input to these analyses comes also from simulations of core-collapse supernovae and neutron-star mergers that require dense-matter properties as input and can connect them to a broad range of observables.
Core-collapse supernova models were among the very first simulations ever undertaken in
computational astrophysics. This now 60-year history of development has led to modern simulations that
have the requisite physical fidelity to provide predictive power over a wide array of observables, including
explosion energy, gravitational wave signatures, and neutrino signals~\cite{Lentz2015,Janka2016,Ebinger2020,couch:20,Burrows2021,Bauswein2022}. Fully understanding the progenitor
dependence and the impact of neutrino flavor oscillations on these observables requires significant
increases in computational power, both in hardware and software. Many of these same advances will
be essential in making possible high-fidelity simulations of neutron-star mergers, as they are the only
avenue available to simultaneously predict gravitational waves,  electromagnetic waves, and nucleosynthesis yields from
these events for a given set of  nuclear-physics input that includes the properties of dense nuclear matter~\cite{Sekiguchi2011,Hotokezaka2011, Foucart:2015, Radice:2016dwd,Radice20,Baiotti2017,Bauswein2020}.


\subsection{What needs to be done?}

Upcoming observations of neutron stars and new data on neutron-rich matter from experiments will provide  information with improved uncertainties, allowing us to obtain a more refined picture of neutron stars and tighter constraints on their properties. 
It is crucial that we can fully exploit this new data without being dominated by systematic uncertainties.
Furthermore, it is key to prepare for the possibility of discovering ``new" physics, like phase transitions to exotic forms of matter, from the data. 
This requires the development of robust and flexible statistical tools rooted in nuclear physics theory and experiment. 

Experimental information on properties of atomic nuclei sensitive to interactions among neutrons, for example the neutron-skin thickness in heavy nuclei~\cite{PREX:2021umo}, 
will help us to obtain key information on nuclear matter probed in the outer core of neutron stars. 
A frontier related to neutron-star science is the nature of very neutron-rich nuclei at the limits of existence, as measured at powerful rare isotope facilities such as FRIB in the USA, RIBF in Japan, and FAIR in Germany.
Additionally, heavy-ion collision experiments at these and other facilities such as NICA in Russia will directly probe dense matter to higher densities and larger neutron-proton asymmetries. 
It is crucial to better understand  systematic uncertainties in the analysis of such experiments and to improve transport model simulations of heavy ion collisions \cite{Wolter2022}.
These experimental constraints, together with progress in theoretical quantum many-body calculations of heavier nuclei, will allow us to improve and benchmark current and future models of nuclear interactions. 
Because these models break down at high densities, where a transition between hadrons and quarks likely occurs, a consistent description of such a transition may need to be included.
 
On the observational side, many new data are expected. The ASTRO2020 Decadal Survey has identified as a priority area ``New Windows on the Dynamic Universe'', including gravitational-wave astronomy and multi-messenger observations of high energy astrophysical sites \cite{ASTRO2020}. 
New X-ray telescopes such as the Imaging X-ray Polarimetry Explorer (IXPE)~\cite{Weisskopf-etal2016} successfully launched in December 2021, XRISM, to be launched in 2023, or Athena, the European Space Agency’s soft X-ray observatory scheduled for launch in 2031, will enable further advances and improve modeling of X-ray sources.
Meanwhile, other new X-ray telescopes are in development: eXTP~\citep{eXTP:2018anb,Watts:2018iom},
which could launch in the late 2020s, and Strobe-X~\citep{STROBEX2019}.
These missions will enable X-ray modeling of pulsars, accreting neutron stars, and classical novae (Section~\ref{sec:transients:intro}). 
Accreting neutron stars offer alternative independent avenues to determine neutron-star properties  because the stellar response to the accretion of matter generates additional observables  \cite{Watts:2019lbs} and higher neutron star masses enhance relativistic effects that are advantageous \cite{Psaltis2014}.

In the next decade, sensitivity improvements to the LIGO and Virgo detectors will increase the prospects for dense-matter science with gravitational waves.
The next LIGO-Virgo observing run, to be held jointly with KAGRA, is scheduled for late-2022 and is expected to return about ten binary neutron-star detections.
The subsequent observing run, planned for 2025 with LIGO Advanced+ detectors, upgraded beyond twice their original design sensitivity, will be even more sensitive to neutron-star mergers.
The potential of gravitational-wave astronomy will be fully realized with the advent of proposed third-generation gravitational wave detectors like the Einstein Telescope~\citep{Maggiore:2019uih} and the Cosmic Explorer~\citep{Reitze:2019iox} in the 2030s, with ten times the LIGO design sensitivity. 
A first detection of a post-merger gravitational wave signal could be made by the end of the 2020s, 
teaching us about matter at the most extreme conditions in the Universe.

These future observations of neutron stars across the mass spectrum 
will constrain dense matter in neutron stars across different density scales. 
Resolving more of the merging neutron-star population will also enable a search for outliers -- for example, neutron stars of similar mass but very different radius -- that could reveal the existence of phase transitions. 
A large-scale effort to explore the vast parameter space of neutron stars and their mergers using numerical simulations is needed in preparation for the next generation of multi-messenger observations.
The fact that the electromagnetic counterparts depend on the dynamics of the system over very different timescales, ranging from milliseconds after the merger to years, poses a significant challenge. 
At the moment, different groups are attacking this problem with independent codes, each optimized for a particular phase of evolution of the binary.
However, these efforts will need to be combined to develop a consistent coherent picture of the multi-messenger emissions from neutron star mergers. 

Finite-temperature effects are expected to be quantitatively important for the outcome of neutron-star mergers and will be imprinted in the post merger gravitational-wave signal and its electromagnetic counterparts. 
There are still only a few studies of finite-temperature and composition effects in mergers. 
These studies show that phase transitions can dramatically impact the merger outcome and leave an imprint on its gravitational-wave and electromagnetic signals  \citep{Most2018,Bauswein2019}. 
However, a variety of physics can drive phase transitions and their impact is only understood qualitatively.
More studies are needed to understand how certain microscopic effects lead to observable features in the dynamics of mergers.

Complementary information about neutron stars can be obtained from analyzing the temperature evolution of isolated neutron stars.
Future observations of young neutron stars such as Cas A and the potential neutron star in the remnant of SN 1987A~\cite{Page-etal2020} will constrain theoretical cooling models and could potentially set constraints on the energy gaps of superfluid matter calculations~\cite{LombardoSchulze2001, Dean2003, Gezerlis-etal2014}. 
Superfluid characteristics impact a number of neutron star phenomena~\cite{HaskellSedrakian2018} such as pulsar glitches~\cite{Anderson:1975_Nature,Link:1999_PRL} and neutron star oscillations~\cite{Andersson:2001_PRL}. Density-dependent gap models of the entire neutron star are needed to accurately model, e.g., magneto-thermal evolution~\cite{Potekhin-etal2015, PonsVigano2019} or determine superconducting magnetic flux distributions~\cite{Wood-etal2020}. Neutron star crust physics also influences thermal and magnetic field observables. 
As the modeling of such phenomena not only requires knowledge of superfluid gaps but also information about the neutron star equation of state and composition, including the crust, theoretical approaches that provide consistent information about various aspects of dense matter would be particularly useful for such tasks.

Finally, we need to prepare for the next Galactic supernova observation. A multi-messenger observation of such an event would provide transformational data for the understanding of dense matter. 
Recently, supernova simulations have used more modern treatments of dense matter that better match the constraints provided by astronomical observations and laboratory measurements, but the robust analysis of future observations requires three-dimensional simulations based on detailed knowledge of the properties of nuclear matter and neutrino-matter interactions. Accurate nuclear physics, for example the weak interactions of neutron rich rare isotopes that can be constrained with FRIB experiments, will also be needed to interpret such observations. 

Capitalizing on these opportunities for dense-matter science will require a concerted effort in theory, experiment, simulations, and modeling to keep up with detector sensitivity improvements and the expected wealth of data.
Studies spanning the full range of nuclear-physics models, including exotica like phase transitions, are critical to lay the theoretical groundwork needed for the correct interpretation of future observations of neutron stars and to maximize  scientific return.

\subsection{What do we need?} 

To reach these goals, we need:
\begin{itemize}
    \item Analysis of systematic uncertainties in models and data analyses from nuclear experiments and observations of neutron stars.
    \item FRIB400 upgrade of FRIB to compress neutron-rich matter in heavy-ion collisions to twice the normal nuclear density.
    \item Exploration of wider and more detailed  microphysical models in simulations of supernovae and neutron star mergers.
    \item Consistent theoretical models, e.g., nuclear interaction models or energy-density functionals, that can be applied to nuclear systems ranging from atomic nuclei to dense matter with systematic uncertainties and at finite temperatures.
    \item A consistent theoretical description of crust and core physics including transport and superfluid properties to allow observational data on neutron star dynamics and cooling to constrain dense-matter properties. Where possible, ensembles of models should be created for statistical inference as with core models in the past decade.
    \item Experimental data on masses and weak interactions of neutron-rich, rare isotopes.
    \item Increased sensitivity of gravitational-wave detectors to observe more neutron-star mergers with increased sensitivity to  late-time signals (see ASTRO2020 Decadal Survey \cite{ASTRO2020}).
    \item Precision X-ray observations of neutron stars, especially high-quality, large photon number,  high resolution spectral-timing-polarimetric data (ASTRO2020 Decadal Survey \cite{ASTRO2020}). 
    \item Continued radio timing of pulsars to improve the precision of moment of inertia and mass measurements, and to resolve more pulsar glitches at shorter timescales.
\end{itemize}


\section{Diversity in Nuclear Astrophysics} \label{sec:diversity}

\subsection{Introduction} \label{sec:diversity:intro}
Nuclear astrophysics integrates a wide range of research areas and brings together subfields in nuclear physics, nuclear and cosmochemistry, astronomy and astrophysics, gravitational physics, accelerator physics, and computational science.  Complementary activities across the globe are in constant exchange. Nuclear astrophysics, as a field, therefore benefits strongly when ideas and contributions from the broadest possible range of participants are shared for the joint advancement of the community. Nuclear astrophysics requires an environment where diverse scientific communities with different communication cultures and demographic compositions  comfortably and safely interact in productive ways. The main challenge to achieve this goal is creation of a safe environment in which people from marginalized communities are invited to participate actively and are supported. The field of nuclear astrophysics therefore benefits greatly from advancing diversity, equity, inclusion, and accessibility (DEIA) goals and, at the same time, is well suited to spearhead such efforts.

As we look at who is participating in the field, we recognize that many groups are not present in proportion to their representation in the population.  If we truly believe that the ability, drive, and interest to succeed in nuclear astrophysics is distributed in the population without regard to sex, gender, sexuality, ethnic or socio-economic background, or whether a person is disabled or neurodivergent, then we must acknowledge that such groups are systemically marginalized. Dismantling the mechanisms of marginalization must be part of how we reach the DEIA goals that will advance the field.

As we look to eliminate barriers in the field, we must seek the advice of those who understand and have experienced these barriers to identify where and when they occur, how they manifest, and their true impact on researchers. As we learn from colleagues and experts, within and outside of the field, about these obstacles, we must use both informal (mainly individual, voluntary) actions and formal (institutional, structural changes) actions  to vanquish them.   Institutional and individual changes to our culture should happen in parallel to reinforce one another as we obtain broad buy-in from the community.  It is crucial that we provide incentives for including practices to advance DEIA goals as well as disincentives to ignore those efforts.  

Advancing the goals of nuclear astrophysics as a more welcoming and hospitable field requires sustained engagement that starts with recognition of the challenges that need to be overcome. Eliminating these challenges will require broad adoption of DEIA goals and a willingness to make necessary changes to our institutional and cultural structures, as emphatically noted in Section 3 of the recently released Astro2020 Decadal Survey \cite{ASTRO2020}.
\subsection{Open Questions} \label{sec:diversity:questions}
Open questions and challenges that need to be addressed in this area are: 

\begin{itemize}
\item How can we achieve appropriate representation in nuclear astrophysics? 
\item How can we create a welcoming and inclusive field that supports and nurtures all young scientists?
\item How can we take advantage of the unique mixture of diverse subfields and international communities that comprise nuclear astrophysics to achieve these goals?
\item What can individual researchers do to further DEIA goals? 
\item What institutional and cultural changes are needed to achieve DEIA goals? 
\item How can we incentivize inclusive practices and disincentivize oppressive behaviors?
\item Institutional DEIA officers play a key role as facilitators of change. How can the group that actively facilitates change be broadened so that community members from marginalized groups are not the only ones expected to be active in this work? 
\item The COVID pandemic has transformed our use of digital communication. These new practices and innovative online interaction platforms are powerful tools for advancing DEIA goals. However, the same tools can also be isolating. What lessons can be learned from recent experience, how can we address shortcomings and can we identify useful practices that should be made permanent?

\end{itemize}

\subsection{How Did We Get Here?} \label{sec:diversity:how}
Considerable effort in recent years has been undertaken by various groups to identify challenges, to develop pathways and recommendations to overcome them, and to start to address the DEIA problem in physics. These include the AIP TEAM-UP initiative \cite{TEAM-UP}, the APS STEP-UP program \cite{STEP-UP}, and the International Science Council's Gender Gap in Science Project \cite{GenderGap}. 
Centers such as JINA have advanced the DEIA agenda by broadly adopting a Code of Conduct and implementing it at all JINA-supported meetings.  Centers like JINA, and its European counterparts, have also spearheaded the inclusion of DEIA topics in workshops and conferences, including invited plenary talks, dedicated sessions, and novel discussion formats. JINA-CEE now provides a wide range of resources related to DEIA efforts as a resource to the field \cite{jinadiversity}. Centers and networks aggregate DEIA metrics.

Thanks to the significant efforts made, particularly in the past couple of decades, by various international networks in nuclear astrophysics (JINA, ChETEC, IReNA, etc.), the nuclear-astrophysics community has made strides towards awareness of the importance of gender diversity. For example, attention has been paid to inviting women speakers 
%
to present their work at conferences, and to making sure participation by women is encouraged and welcomed. As a consequence, recorded attendance, presentations, and inclusion of women in committees at conferences in the field has reached typical values of around 30\% (41\% of the speakers at the JINA Horizon meeting that originated this white paper were women). 
This is, however, only the start of the journey. As detailed below, the field's progress in including other marginalized communities is poor. It is also vital that we realise that people who belong to more than one marginalized community face
more barriers than those who belong to only one such community. 

DEIA priorities vary somewhat from region to region and country to country, and nuanced approaches are required at regional levels. An example of a successful, larger-scale, regional initiative to advance gender and socio-economic diversity in the field is the Changing Face of Physics campaign in 2018 by the E.A. Milne Centre for Astrophysics at the University of Hull, UK (also part of JINA, ChETEC, and IReNA). Key elements of this initiative include close connections to more than 50 local schools to identify capable students interested in physics that would be excluded when using traditional entrance selection criteria, mentorship by undergraduate and post graduate students from similar backgrounds, and research internships. As a result, Physics at Hull has seen a doubling of women
into Physics in only three years, though more work remains.

Many important challenges related to gender equality and inclusion of other marginalized groups remain, e.g, 
\begin{enumerate*} 
\item  Unconscious biases, for example when women and people of color are overlooked as potential invited speakers, 
\item the gender pay gap and differences in approach (and success) of
minorities in applying for jobs or promotions \cite{Ivie20,Bishu17}, 
\item the need to inspire younger generations to be involved in the community, 
\item making it easier in practice for everyone to participate in activities such as workshops or meetings, and 
\item achieving a healthy work-life balance. 
\end{enumerate*}
Family and other caring responsibilities often impact women disproportionally. Women in physics continue to experience sexual harassment in large numbers \cite{Libarkin2019} correlating with a feeling of not belonging. Similar challenges are encountered by members of the LBGT+ communities (lesbian, gay, bisexual, transgender and other sexual and gender minorities). APS and IOP climate surveys indicate that 16-20\% of LBGT+ members in physics recently experienced exclusionary behavior and a third considered leaving Physics \cite{workplacereport,LBGT_APS}. The international nature of collaborations poses particular challenges as it increases exposure of LBGT+ scientists to cultures with low levels of inclusion \cite{workplacereport}.  

%
%
The track record of the physics community in terms of participation by race remains dismal. In the US the fraction of Black students among physics bachelor degrees between 1999 and 2020 dropped from 4.8\% to 3.1\% \cite{Marvis2022} despite robust increases of participation of Black students in other STEM disciplines. While these numbers are low, they are hiding an even bleaker reality as the vast majority of Physics degrees were awarded by a small number of Historically Black Colleges and Universities (HBCUs).
At most other US universities numbers of Black students graduated range from zero to two over this time period.  Furthermore, the fraction of physics doctorates awarded to Black students has dropped since and is now below 1\%.

\begin{tcolorbox}[colback=black!5, colframe=black!40!black, title=Sidebar: Women who Made Nuclear Astrophysics]
  \begin{center}
    \includegraphics[width=0.8\textwidth]{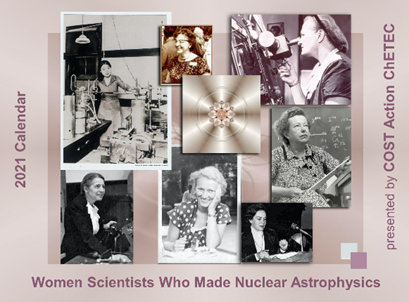}
  \end{center}
  \vspace{-0.20in} 
  \begin{center}
  {\footnotesize {\it  Cover of the calendar ``Women Scientists Who Made Nuclear Astrophysics".} Credit: \cite{women_calendar}, and clockwise 1) Yuasa: courtesy Ochanomizu University, History Museum, 2) Müller: courtesy R. J. Rutten, Utrecht University 3) Payne: courtesy AIP Emilio Segr{`e} Visual Archives, Physics Today Collection 4) Mayer: U.S. Department of Energy (Public Domain), 5) Eryurt: Courtesy METU Physics Department 6) B{\"o}hm-Vitense Courtesy U. of Washington via Julie Lutz 7) Meitner: Smithsonian Institution Archives, Wikimedia } \label{fig:dei:side} \\\
  \end{center}
  \vspace{-0.3in} 
{\footnotesize
Many women scientists were critical to the development of nuclear astrophysics since its inception. 
 The nuclear astrophysics community has worked on promoting these women as role models and to remember and celebrate them. A poster and a calendar celebrating 12 historical figures of women who contributed to developments in the field have been created in an effort led by M. Lugaro and C.~V. Hampton \cite{women_calendar}. The poster was presented at the Nuclei in the Cosmos conference in 2018 and can be downloaded from the ChETEC website \cite{womenposter}. The calendar was translated into 25 languages, and distributed for the years 2021 and 2022 in several countries including the United States, Japan and Europe \cite{calendarPDFs}.
The main aim of the project is to reach out to young students and present them an image of scientists that counters prevailing stereotypes. Carefully chosen photographs portray scientists both at an early age and advanced career phase, so young girls and boys can start recognizing outstanding scientists also as young women. 
The calendar was presented at secondary schools and social media events to specifically reach young students. Moving from historical figures to current women scientists would be a natural extension of the above outreach, anti-bias projects. While some efforts have been made in this direction, for example, the ChETEC website lists women researchers who recently secured prestigious grants from the European Research Council \cite{ChETEC_womenERC}, a more coordinated, international effort is needed to ensure that the young public is reached. Using infrastructures such as current projects that involve high school students, e.g., within JINA and ChETEC-Infra, and visitor centers associated with observatories, laboratories, or planetaria is a successful direction in which to expand the reach.
}
\label{sidebar:women}
\end{tcolorbox}

\subsection{What Needs To Be Done?} \label{sec:diversity:todo}
Action is needed, not only at individual and institutional levels, but also by scientific collaborations, conference organizing committees, and funding agencies. Recruitment is a challenge, but until we improve the retention of students, we cannot expect recruitment to be the sole answer.
As we improve our ability to retain students in the field, the ability to recruit will also improve. Research has shown that the proportion of undergraduate students from marginalized communities interested in science and engineering is much greater than the proportion of degrees awarded to minorities in those fields \cite{ASAI2020754}. We need to identify how cultural practices in introductory physics and astronomy classes and early encounters with research experiences have deterred students from persisting in areas where they have interest. This must lead to restructuring of recruitment and retention efforts at the undergraduate level \cite{Malcom17}.  Bringing more interested high-school students into a field that fails to nurture their sense of belonging and identity as a scientist prevents achievement of our DEIA goals.

Large collaborations and centers (Section~\ref{sec:centers:intro}) have an opportunity for particularly impactful actions. Most importantly, just like science, advancing DEIA should be a collaborative endeavor for the field as a whole, with opportunities to exchange ideas and observations, develop new directions, and share best practices, successes, and failures. In the following we summarize specific actions that the nuclear astrophysics community recommends to implement for progress towards a diverse, equitable, inclusive, and accessible field of nuclear astrophysics:

{\bf Treat diversity goals on an equal footing with scientific goals:} Diversity goals are as important as the scientific goals of   the field. They should be valued as part of how we execute our research and thus be part of the scientific merit of our work.  Therefore, diversity goals  should be integrated accordingly in conferences,   workshops, review and promotion criteria, criteria for prizes and   awards, publications, and public dissemination of achievements. We   advocate for dedicated sessions on DEIA at conferences and workshops   that are treated like scientific sessions. This has been shown to foster awareness across multiple generations of scientists, and stimulate discussion. 

{\bf Provide incentives to promote DEIA goals:} Incentives are a key agent for advancing DEIA goals. Recognizing DEIA achievements on equal footing as science   achievements (see above) will go a long way in creating such   incentives. Another example would be recognition of clearly developed DEIA plans and activities in a project at the same level as   scientific and technical components. Funding agencies and institutions need to provide incentives through grants and fellowships, e.g.,~requiring some meaningful discussion of DEIA in impact statements and giving such statements equal weight to research descriptions, especially for established researchers. 

{\bf Address barriers:} There are many kinds of barriers to marginalized-community success in science that many scientists are unaware of. 
Barriers to marginalized ethnic minorities  (See AIP TEAM-UP report \cite{TEAM-UP} for details) include financial and family concerns, or the lack of resources at small or under-resourced universities. Barriers to disabled scientists need to be addressed by targeting resources to explore innovative approaches such as sign  language for science, sonified data and content for blind people, as well as measures that help alleviate the impact of disabilities that are not immediately visible. An increasingly important type of disability limiting   participation in nuclear astrophysics, as in other science fields,   concerns mental health, which constitutes a significant percentage  in the spectrum of DEIA barriers. Institutions and leaders need to bring this inclusion barrier into clear focus and address it (Section~\ref{sec:education:intro}).

{\bf Address bias:} Policies that seek to mitigate biases must be both intentionally implemented and regularly reviewed for effectiveness for achieving DEIA goals. For example, double-blind reviews reduce unconscious bias by focusing on the science rather than the identity of the proposer. Bias in recruiting and hiring should be reduced by following best practices in writing job advertisements, in streamlining interviews, and in using interview panels with pre-selected questions and scoring criteria.  

{\bf Retention:}  The retention of researchers from marginalized communities is key to achieving DEIA goals.  A commitment to affirm an early or late career researcher's  identity as a scientist, and to reinforce a sense of belonging in a department or collaboration, should be the responsibility of all faculty and collaboration members with whom the researcher engages \cite{TEAM-UP}. Retention and recruitment can be helped by policies and practices that acknowledge and are designed to address the family concerns of professionals, including support for solving the two-body problem in which married or partner couples seek meaningful professional careers in the same location. In addition, practices of cluster hiring that allow new staff to identify a supportive cohort can be helpful if implemented correctly. 

A healthy work-life balance must be promoted and normalized: the current culture that reinforces the notion that to be successful one must be working significantly more that 40 hours/week is one mechanism by which people are marginalized and contributes significantly to mental health problems. Both the structure of our programs and those that mentor early-career researchers must send a clear message that one can be successful in the field without working more than standard working hours. As a community we must also highlight and support flexible working hours to accommodate the needs of those with other life commitments.

At the student level, valuing the experience and insights of each student as they seek a career path in science, requires changes to our evaluation of readiness to undertake graduate work. Career paths of each student differ and our DEIA goals should reflect values and community norms that support achievement of individual student’s career goals.  Retaining a diverse pool of graduate students requires investing resources in these efforts. It is also important that students see people from marginalized communities as leaders in the field. 

{\bf Open access to data, resources, knowledge, and ideas:} This includes digital access, for example open access to  data and research software, as well as shared access to experimental, observational, and computational facilities. The community must also organize open access to knowledge and promote early sharing of ideas through open networks that explicitly strive to include participants from groups currently marginalized in the field. Centers and networks play a key role in providing access to a broad range of   researchers at different levels of seniority and from a broad range   of institutions. They should strive to create   structures and processes specifically designed to improve   access. For example, JINA schools have focused on introducing a diverse group of students to open-access computational research tools and to experts in the field who provide guidance and collaboration. The NuGrid collaboration has  adopted a mandatory open-access policy in which all collaboration   projects have to be open to all participants. This could be extended to the center or international network-level.

{\bf Buy-in:} The buy-in of all stakeholders and the community as a   whole is critical for achieving DEIA goals. Centers and networks must facilitate the establishment of discussion spaces for people to engage in the   conversation and form their opinions. Attention needs to be paid to making these discussion spaces safe for affected marginalized communities to avoid negative impacts. 
 
{\bf Committed leadership:} To advance the DEIA goals both bottom-up and top-down approaches are needed. Leadership in the community and high-level policy groups need to engage with clear and visible actions. An example would be the implementation of DEIA metrics in evaluation and assessment. Overall, a shared leadership model needs to be adopted that incorporates the broad community’s range of career status, expertise, and perspective.
 
{\bf Codes of Conduct:} Collaborations and conferences should develop, ideally with outside experts, and implement codes of conduct, also referred to as community agreements, and effectively communicate them. These should include an aspirational DEIA statement advanced through policy, an outline of institutional responses to violations such as harassment and bullying, as well as best practices for bystanders such as reporting and   providing support to the people affected. Codes of conduct should be reviewed regularly. 

{\bf Mentoring:} Effective mentoring is recognized as a  particularly important element for creating inclusive   communities. Institutions and collaborations should implement  mentoring frameworks that go beyond the single mentor approach and  include mentor training \cite[e.g.,][]{nap2019}). One possibility would be the formation of multi-institutional support groups for minorities, potentially coordinated by centers, with support from professional coaches. 
 
 {\bf Outreach:} Public outreach provides an opportunity to reach marginalized communities and dispel racial and gendered misconceptions by presenting diverse role models. Owing to new developments especially during the pandemic, where in-person outreach has largely been substituted by online outreach, it's now possible to reach new audiences that are not in the immediate vicinity of the research institutions. Outreach programs should be adapted to ensure marginalized communities are reached and there are no barriers to participation, including technical barriers for online programs. 
 
{\bf Training:} Collaborations should consider training activities that promote DEIA topics. Such training should be tailored to the needs of the field. Centers should facilitate the collection and sharing of experiences in the field with such trainings.

{\bf Centers and Networks:} Centers and networks (Section~\ref{sec:centers}) have the opportunity to apply particularly impactful action on all the points in this section. To further   broaden the impact and lower access barriers it is important to expand networks to smaller, less-connected institutions both   nationally and internationally. As they are broad-reaching and multi-institutional, centers have an excellent opportunity to provide a support network for those who might not otherwise have access to interdisciplinary training, research opportunities, and resources. This could include community-wide mentorship and networking programs, a centralized location for job notices and workshop announcements, mental health and wellness resources, assistance with translation or data accessibility issues, and a diverse database of potential speakers for conference and seminar planning. Centers also offer the opportunity to coordinate responses to code of conduct violations across institutions. 
 
{\bf Online Access to Meetings:} While recognizing that online participants miss important elements of the scientific meeting experience, we have noted that moving to online meetings during the COVID pandemic has supported women to attend more meetings and that the percent of female participation has increased -- on top of the meetings becoming in general more inclusive. Online access may also be beneficial to groups with limited ability to travel, for example because of health issues, disabilities, or economic reasons. We need to continue monitoring this potential effect, and suggest that meeting organizers consider continuing to provide the opportunity to attend meetings both online and in person, and keep track of the attendance levels. 

{\bf Collect and Monitor Metrics:} The monitoring of statistics on participation is important to understand if progress is being made. Nuclear astrophysics must continue to collect information that demonstrates that DEIA goals are being met.  Centers can help to normalize and curate this information. Collaborations should track participation, for example by analyzing co-authorship, to identify disparities.

On the personal level, a major step that each of us can take is to examine ourselves for unconscious bias. Tools exist (for example, the tests developed by \cite{ProjectImplicit}) to enable individuals to help expose their unconscious biases, which can lead to positive changes in an individual's approach to decision-making that might otherwise negatively impact diversity in the sciences. Unconscious bias happens below the level of an individual's or institution's awareness, however while drawing attention to it can be helpful in some cases, there is evidence that this is not sufficient. Policies and procedures must still be in place that curtail the opportunity to act on bias. Individuals play an important role  in creating a more inclusive and respectful workplace. Even small steps -- such as taking the time to learn the correct pronunciation of someone's name, or ensuring that someone is given appropriate credit for an idea in a meeting -- over time  make a difference. 

\subsection{What Do We Need?} \label{sec:diversity:needs}
\begin{itemize}
    \item Broad community buy-in and a concerted effort to make DEIA goals a priority for all members of the field, on equal footing with the scientific goals. 
    
    \item Centers are important in bringing DEIA efforts center stage, fostering new ideas, and implementing policies and procedures that drive broad change across all of nuclear astrophysics and its various subfields and interdisciplinary partners. 
    
    \item Transition from identifying what should be done to making and sustaining actual changes  through  equity-minded  approaches  that  acknowledge  the  differing  needs  of individuals. All members of the community have a role to play. Individual actions  in these roles may range  from  individual  and  personal  to using influence to establish inclusive policies and procedures in their collaborations, departments and universities.
    
\end{itemize}

\section{Career Development: Perspective of Early Career Researchers}
\subsection{Introduction} \label{sec:education:intro}

For an early career researcher, the task may seem clear: do research and publish. This may make sense at face value. However early career scientists require holistic mentorship to develop skills in more than just research in order to navigate their way to a permanent job and be successful, be it in academia, industry, public service, or other sectors. Here we summarize some thoughts on career development shared by early career researchers following the JINA Horizon's Junior Workshop, which included participants from the global IReNA network (see \cite{JINAhorizons} for links to talks and other career related resources including a ``hiring survey" of short answer responses from senior scientists, which highlights the diverse criteria considered during the hiring process).
The general topics for which early career researchers expressed concern were: mentoring, understanding career options, networking and community, resources for topics outside research, navigating the job market, family and work life balance, mental health, and imposter syndrome. We note that each of these issues impact different types of early career scientists in distinct ways and in this sense are related to the discussions of equity, diversity, and inclusion in Section~\ref{sec:diversity}; however, we do not discuss such connections here.

\subsection{How Did We Get Here?} \label{sec:education:how}
Nuclear astrophysics, with its need to integrate broad sets of skills and knowledge across diverse disciplines and with its collaborative and international character, is well suited for training of skills that are transferable to a broad range of careers. Many skills learned are practical and overlap with practices in the private sector (e.g. coding, exposure to lab equipment). Summer schools, workshops, and programs are often noted by early career researchers to provide exposure to a variety of topics and tools that then become integrated into their skill sets. An education in physics also provides opportunities to travel or work anywhere in the world thereby meeting a large spectrum of people. Additionally, early career scientists have the freedom to follow their interests, and can make discoveries at the edge of knowledge, thereby contributing to the advancement of science and society. 

While there are clearly many benefits to an education in nuclear astrophysics, it can be challenging for early career scientists to understand how their experience translates to the next stage of their career. In recent years, there has been significant progress with raising awareness of the importance of professional development  and the preparation for successful careers. Physics centers like JINA and IReNA (Section~\ref{sec:centers}) have introduced career development activities as part of their programs, and as major components of meetings. Nevertheless, there remain challenges that need to be addressed, and it is important to try to identify the best practices for preparing the next generation of scientists for the workforce.

\begin{figure}[!h]
\begin{center}
\includegraphics[scale=0.23]{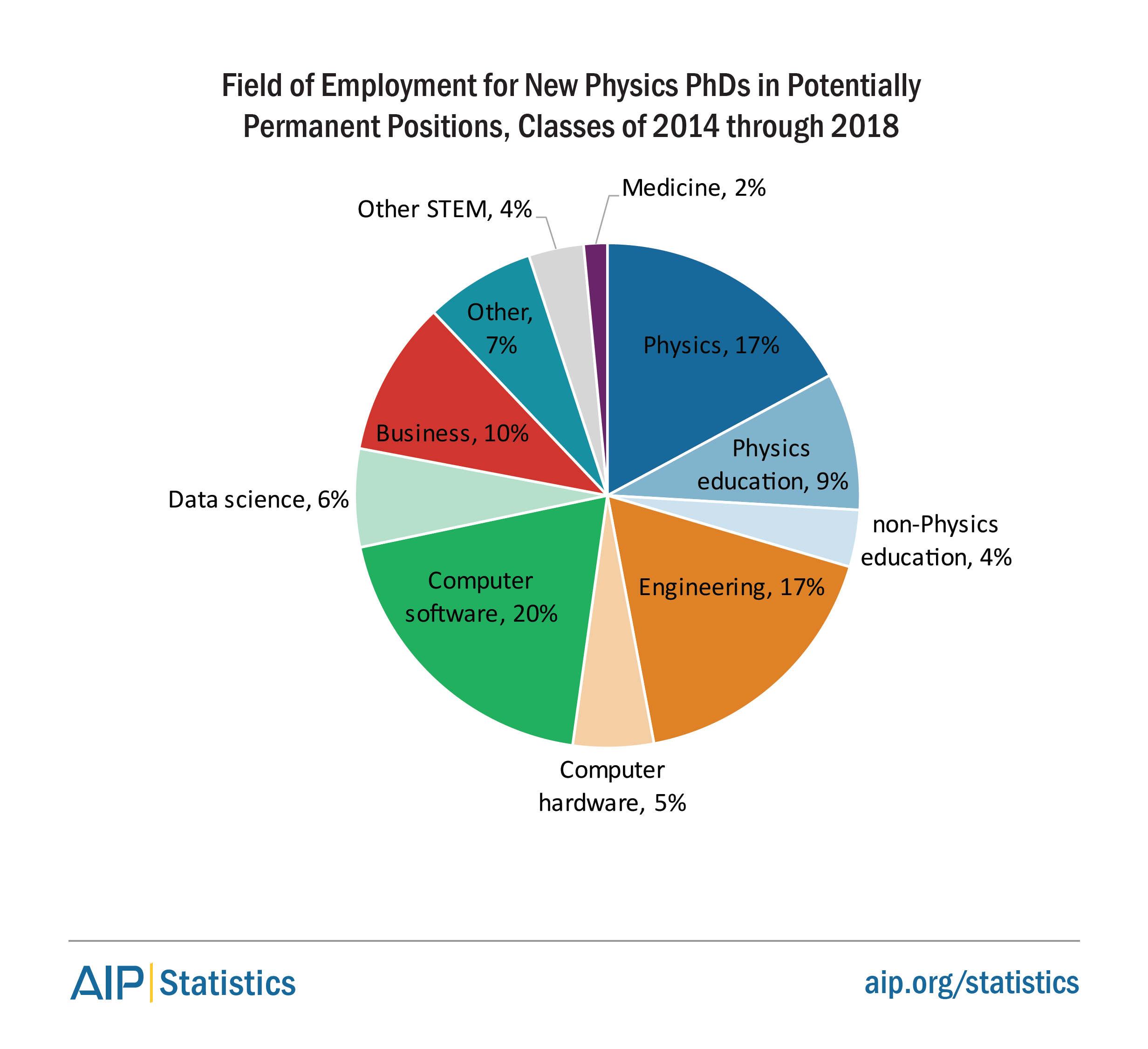}
\end{center}
\caption{Chart showing fields of potentially permanent employment for physics PhDs awarded between 2014 and 2018, inclusive. The different fields are displayed corresponding to the percentage of the PhD awardees going into that field immediately on graduation. Credit: AIP Statistical Research Center \cite{AIP_SRC}}
\label{fig:physPhDcareers}
\end{figure}

Ultimately, physics PhDs have a wide range of exciting job prospects, both inside and outside of academia. In fact, most physics PhDs will not end up in academia, as is demonstrated in Fig.~\ref{fig:physPhDcareers}. This is the elephant in the room of any discussion about career paths for physicists that early career scientists are largely encouraged toward pursuit of an academic job but may be only vaguely aware of the competitiveness to obtain such a job. Despite bright job prospects in areas ranging from computer software to finance and business, early career scientists noted that discussions of career opportunities in national labs and in industry are rarely discussed in their university settings. Mentors who themselves have landed in academia may even unintentionally display bias about what should be the next stage of the mentee's career. Often early career scientists feel that looking into careers outside of academia comes with a feeling of failure. 

Mentors and institutions have a responsibility to prepare students for a broad range of careers inside and outside of academia. However, many early career scientists feel mentors focus too exclusively on the research development, and they lack important guidance and skills needed to build a successful career. As noted in \cite{McPherson2017}, key challenges for early career scientists do not necessarily relate to the higher learning materials, but the lack of clarity around what their time during studies should entail, which can be helped by greater clarity of expectations and earlier feedback. Some areas of deficiency relate to the development of important transferable skills such as communication skills or a better understanding of the job application and interview process and required preparations.  

Imposter syndrome and mental health are other challenges that are often not sufficiently addressed. Indeed recent work in the social sciences has noted that transitions from undergraduate to postgraduate education are not straightforward and can be characterized, at least initially, by anxiety, self-doubt and disorientation \cite{McPherson2017}. Imposter syndrome is experienced by a large part of the scientific community at all career stages and often prevents early career scientists from integrating in the scientific community. Another challenge is the lack of advice on work-life balance, or how to balance the pursuit of a career while planning or growing a family. Many early career researchers are also confronted with the so-called ``two-body problem", facing the prospect of having to choose between advancing the career of one partner at the expense of the other. Although many face such challenges, early career scientists may not have or may not be aware of
resources outside their personal mentor such as counseling, mentorship programs,
or support groups.

All mentors need training on how to recognize the signs of depression and anxiety in early career researchers, while not over-interpreting the behavior. The mentors should be aware of the full range of resources available to early career researchers, and the best approaches to helping their mentee reach out to counseling and psychiatric centers. Mental health training for advisors is essential, and incentive to provide such training within a
research group should be provided.

An additional issue early career researchers are currently experiencing is to navigate networking in the remote conference era brought about by the COVID pandemic. Many early career scientists are concerned about diminishing opportunities to get to know more senior researchers from different institutions and a lack of conference visibility. Even when attending a meeting, virtually or in person, many early career scientists feel intimidated to approach more senior researchers, thus missing out on important opportunities.

\subsection{What Needs to be Done?} \label{sec:education:todo}

\noindent{\bf Professional development:} Communication skills are important no matter where an early career researcher's career path leads. A communication skill which is widely recognized to be important is the ability to present scientific results effectively (in fact the need to give a great talk was a recurrent point mentioned in the hiring survey by those giving insights on how hiring committees evaluate candidates). Here early career researchers noted that it would be helpful if there was a more central resource for learning good talk practices, practicing talks, or presenting among peers. Additionally, effective communication skills are needed for a diverse number of settings, not just when standing in the front of the room. For instance, involvement in multiple collaborative efforts can provide some needed exposure to relaying and synergizing distinct viewpoints. The ability to clearly outline science goals and results is also needed to teach effectively as well as to be successful when approaching funding agencies. Scientific writing is another important skill that needs to be learned. This spans a broad range from scientific publications to research and grant proposals. Common questions of early career researchers are how to write effective proposals, abstracts, figure captions, or conclusions and how to be a courteous and fair referee. 

A pre-requisite for effective training towards successful careers is being comfortable in one's community. Mentors should consider group dynamics and how it may affect individual mentees since this can influence an early career researcher's view of their value. This can translate into their perceived place in the field and their ability to compete for positions. Early career researchers also need a space to be able to express their anxieties, perceived short-comings, and fears about the uncertainties that come with building a science career, particularly because of the common fear that mentors or other more senior researchers could interpret their concerns as foolish. 

Bringing imposter syndrome out in the open and having senior researchers acknowledge it would help to reduce its power since this would highlight that: (1) if you are experiencing imposter syndrome you are not alone and (2) you can navigate through such feelings to a successful career. Senior researchers can help to change the environment for the better with acts as simple as paying attention to the tone struck when teaching courses, being mindful of the manner in which they address other senior researchers during conferences, and speaking up to address any inappropriate behavior of colleagues. Although science benefits from a vigorous back and forth between experts, if the tone is dismissive, it will likely deter an early career researcher from feeling adequate to join the conversation. This is just one example of why codes of conduct such as that implemented by JINA serve an important function since they define and enforce what is considered a productive and comfortable academic environment.

Senior researchers also have personal experience of imposter syndrome, mental health problems and rejection: it is important they help normalize sharing these experiences and that support is provided when students and early career researchers face the universal academic experience of rejected papers and grant proposals.

Students would also benefit from a more well-rounded approach to mentoring where PhD goals, career options, and lifestyle preferences are discussed early and often. Research projects should be tailored to the ultimate career goals of a mentee, for instance a machine learning project would help to develop skills needed in a software development career. Having regular meetings with long-term mentors who are familiar with the evolution of the personal goals of the mentee could be of great value toward helping an early career researcher evaluate their progress. Indeed multiple mentors with unique perspectives and experiences are crucial for early career researchers to fully consider their career options.

{\bf Training for Careers Outside of Academia:} Although mentors may worry that emphasizing the competitiveness of the academic job market will discourage their budding mentee, mentors and the community must find a way to be more up front about career prospects. Tunnel vision toward an academic career can make sense if a researcher is certain that they would like to work toward this end. However, early career scientists, busy with their research tasks, may develop such a tunnel vision not purely out of interest but rather due to a lack of exposure to other options. Mentors should help students consider their choices more completely, for instance by encouraging them to attend career fairs. Additionally early career scientists should be made more aware of career related resources, some of which have been provided by organizations like CIRTL \cite{CIRTL} and AIP \cite{AIPcareers}, including discussions of careers outside academia (e.g. 
\cite{toolbox}).

In particular, senior researchers can help to mitigate the perceived stigma of industry jobs by discussing non-academic career options early on and by being mindful of the attitude that they convey about their mentees obtaining jobs outside of academia. An obvious difficulty is that academic mentors themselves likely do not know much about industry career paths so outside help is needed. Mentorship training programs and courses could help to fill these and other gaps. For instance, organizations such as the CIMER Project at the University of Wisconsin-Madison  \cite{CIMER} aim to provide resources and training for improving mentoring relationships at all stages of a mentee's career. Although such special courses should likely be a mandatory part of becoming a mentor, mentorship programs are often not readily available.  

Many early career researchers noted a lack of perspective as to the possible sectors of industry and job titles that align with the skills learned during a physics PhD. University programs could help with information on non-academic career outcomes, mentorship programs, or keeping a record of up-to-date contact information for past students who have landed in industry. Physics centers like JINA could play a role in further aggregating such information.

Physics programs could also benefit from more workshops, industry visitors, invitations for seminars from industry professionals, and even industry internship programs. Having such exposure built in to the options available during their PhD studies would not only help students make a more educated decision on their career goals, but would also help physics PhD students stay competitive for industry jobs when compared to other disciplines such as engineering and computer science whose degree programs may already incorporate industry partnerships. Additionally, this could help to the address ``gaps" highlighted in recent work (e.g. \cite{Groeneveld2022}) between what students learn and the skills that are expected and needed in the private sector.

{\bf Training for Careers in Academia:} If an early career researcher decides to pursue an academic job, then a postdoc is likely their next step after graduate school. To get to this next stage, graduate students need to be preparing during their PhD to be an attractive candidate through networking and collaborating. Through such interactions the student can gain additional mentors, letter writers, and allies. Students also need assistance in navigating the market of postdoc positions, and can benefit from recommendations on parties to send application materials to privately, rather than solely responding to job ads.

When moving on to a postdoc position, early career researchers are often unclear about strategies to make their short time as a postdoctoral researcher effective and aligned with their career goals. Mentors could help mentees develop such strategies prior to starting a position. Additionally, postdocs reported feeling that there was less support available to them compared to graduate students. Increased support programs for postdocs in their local departments could go a long way toward navigating their careers at such a crucial stage. An example is MIT's semester long class on "Leadership and Professional Strategies and Skills (LEAPS)" for STEM interested grad students and postdocs. LEAPS focuses on self-awareness, tools and tips for how to chart one's own career, and how to lead and guide students and group members within academia and industry. 

Another crucial step towards an academic career is the transition from postdoc to a permanent position. Early career researchers would like there to be a more open and active dialogue with their mentor early in their careers regarding how their research, publications, community involvement, etc. may be viewed by the community and future hiring committees as well as guidance on how to build a successful resume (e.g. drawbacks of being on long author lists, the importance of first author papers, or whether to branch out to gain expertise in several topics versus building expertise in a more narrowly focused topic). Guidance on application strategies is also important. It would be helpful for early career researchers to learn more about search committee thought processes and participate in training activities related to applying for faculty positions. In many cases, early career researchers are unclear about the role of the cover letter, how to self-market effectively, if and how to tailor applications to different jobs, and how to choose where to apply. It would be helpful if senior researchers in the department, who have recently been on a hiring committee, could provide feedback to grad students and postdocs on application materials and hold mock interviews. Valuable resources such as application materials training may be available from services outside one's department such as the graduate school, office of postdoctoral affairs, or from dedicated career development workshops such as those organized by APS and JINA. However, such resources can be overlooked by grad students and postdocs who are most actively focused on research progress. Graduate students could also benefit from having a postdoc mentor keep them informed of what they have learned while navigating the job market, which also serves to give postdocs much needed mentoring experience. 
Mentors should make early career researchers aware of the full range of academic jobs. Students could go to an institution with a high level of research activity; however,
these universities account for a minority of universities in
the US \cite{NCES}. The majority of students in the US are trained at generally smaller, regional universities with higher teaching loads, and that is where many faculty positions will be available. Many senior faculty are unaware of what these sorts of jobs entail and advisors can give the impression that working at such institutions are beneath them. However, many of our colleagues successfully combine research with the different demands of such an institution, and the advantages to working in smaller departments include having fewer barriers to making positive reforms in the program, greater connection to students, and opportunities to collaborate with local high schools, community colleges, and local private organizations on outreach and education (the APS STEP-UP program \cite{STEP-UP} is a great example of the latter). A faculty member can make significant positive impacts in ways that would be much more difficult to achieve at a large, research-heavy, university.

It is unfortunately rare for people trying to navigate the academic job market to get feedback from hiring parties. All too often interviewees learn that they did not get the job either by never hearing back or from a generic email with no personalization. This is a lost opportunity to provide community mentorship to early career researchers. Regular, brief feedback as to how the committee felt the candidate fit and did not fit the search criteria could go a long way to help early career researchers understand how hiring committees think and the uniqueness aspect of each job search. 

Cultural differences between countries can also be difficult to navigate as administration, funding, and the expectations of a hiring committee are all subject to cultural factors. Due to the international nature of the field, postdocs often seek permanent positions in countries that are different from the country they were trained in. International networks such as IReNA could help address such challenges.  This is just one example of why access to distinct viewpoints and example approaches from multiple mentors are crucial for early career researchers to successfully navigate the job market. Here early career researchers should be made aware of valuable programs designed for them to find more support such as JINA mentoring lunches or the APS Career Mentoring Fellows program.

\subsection{What is needed?} \label{sec:education:needs}

\begin{itemize}
    \item  Preparing students for careers beyond academia should not be an afterthought, but a cultural norm. Mentors and academic institutions should make sure to broaden the career prospects presented to PhD students by, for example, integrating career fairs into graduate programs or inviting outside speakers from industry. It should be made clear to early career scientists that moving on to a career in industry is not only okay, but common and full of exciting opportunities.

     \item Professional development for a broad range of careers should be available to all graduate students and postdocs in nuclear astrophysics, and should be integrated in research mentoring and training. Exposure and access to industry partners could help students and academic institutions better align their training to be compatible with both industry and academia.
     \item Early career researchers should be made more aware of what employers, both in industry and academia, are looking for. Centralized resources to assist with job searches, self-marketing, application materials, and interview preparation would help ensure all candidates get a fair chance.
    \item Early career researchers need to be able to build their confidence and discuss the challenges they are facing. They should be provided with and made aware of resources outside their personal mentor such as counseling, mentorship programs, or support groups. The community and mentors must also work to ensure a healthy working environment with an inclusive atmosphere.
    \item International experiences are of particular importance for early career researchers and should be encouraged and supported. 
\end{itemize}

\section{The Role of Centers}
\label{sec:centers}
\subsection{Introduction} \label{sec:centers:intro}
Nuclear astrophysics is an interdisciplinary field at the intersection of nuclear physics and astrophysics with its own compelling fundamental scientific questions as described in the previous sections of this white paper. Addressing these questions requires simultaneous work using the most advanced nuclear accelerator facilities, the most advanced telescopes, space-based observatories, and ground based detectors of cosmic messengers such as gravitational waves or neutrinos, as well as cutting edge computational models and theory. Some of the major capabilities and facilities have been developed with nuclear astrophysics in mind, others not. Nuclear astrophysics centers maximize the scientific impact of all these major investments, exploit their full discovery potential, and ensure a large and diverse community takes advantage of them. 

Furthermore, the scientific questions in nuclear astrophysics are all interconnected. It is the same stellar populations that are responsible for nucleosynthesis that are producing the compact objects probed by LIGO and give rise to transients. Stellar evolution sets the stage for stellar explosions and compact object formation. Multi-messenger astronomy provides constraints across all types of stellar objects. The same nuclear physics, from reaction rates to dense matter properties, plays a role in multiple types of scenarios and questions. Instead of investigating phenomena in isolation, it is therefore important for the field to take into account the full range of nuclear physics and observational constraints across interconnected astrophysical sites towards a consistent descriptions of all nuclear processes in the cosmos. 

Centers are essential for nuclear astrophysics to build the interdisciplinary communities of scientists necessary to carry out such coordinated work, to facilitate the exchange of ideas and data across fields, and to trigger the necessary developments in each subfield required to address the open questions. Nuclear astrophysics requires an extraordinarily diverse range of capabilities and expertise to come together, and centers provide the framework to trigger the necessary large-scale networks of collaborations and to evolve communication and coordination over extended periods of time. Centers also play a key role in interdisciplinary training of new generations of young scientists able to navigate a rapidly evolving research environment by transcending traditional field boundaries. Last but not least, centers bring together the scientists from different fields to develop and define the open questions and scientific opportunity – this white paper with its working groups of nuclear scientists and astronomers is just one example.

\subsection{How Did We Get Here?} \label{sec:centers:how}
The particular need of centers in nuclear astrophysics is reflected in the history of the field. In the early stages of nuclear astrophysics around the middle of the last century, the pioneering research groups that defined the new field served that purpose, most importantly the Caltech group around Nobel prize winner William A Fowler. The Nuclei in the Cosmos conference series initiated by Heinz Oberhummer and Claus Rolfs in Austria in 1990, and since rotating internationally, started to provide a touch point for a growing, broader, and international nuclear astrophysics community. It served as the primary forum that brought together nuclear scientists and astronomers and helped define the modern field of nuclear astrophysics. In 1999 the Joint Institute for Nuclear Astrophysics (JINA) was founded. It started as a small network of institutions primarily located in the US Midwest. In response to the growth and developing needs in the field, JINA grew rapidly and includes today 350 scientists from 27 institutions in 12 countries. JINA developed the modern, broad center-based network approach to nuclear astrophysics. It serves a dual role as interdisciplinary research center driving key developments in experiment, observation, and theory, while at the same time serving as a center for the entire field of nuclear astrophysics, providing a forum for exchange, coordination, and community building. The approach has been successful and effective in building a broad interdisciplinary nuclear astrophysics community and accelerating scientific progress. Subsequently a similar approach has been adopted in other countries, starting with NAVI in Germany, UKAKUREN in Japan, and BRIDGCE in the UK. Since 2010 a number of collaborative networks have been initiated at the European level, including Eurogenesis, ENSAR1, ENSAR2, ChETEC, and most recently in 2021 ChETEC-INFRA. In Canada the CanPAN nuclear astrophysics network was initiated in 2021. The JINA-led US initiative of the International Research network for Nuclear Astrophysics (IReNA) took the approach to the next level, connecting initially 6 (and now 9) international research networks to form a network of networks. IReNA involves scientists from 17 countries and enables communities to take advantage of complementary international capabilities in nuclear astrophysics research and education. 

All these initiatives include strong educational components and foster active involvement of young scientists, including students and postdocs with particular attention to equal access (see Sidebar: Building the Next Generation on Pg.~\pageref{sidebar:generation}). In the US JINA has developed, often in collaboration with international partners, a novel school concept that focuses on hands-on activities on specific topics and brings together students and postdocs to work together in interdisciplinary teams. In Europe, three summer schools focusing on nuclear astrophysics were developed in the early 2000s: The European Summer School on Experimental Nuclear Astrophysics in Italy, the Russbach School on Nuclear Astrophysics in Austria, and the Carpathian Summer School of Physics in Romania. These schools are complemented by the Nuclei in the Cosmos School, which has been established as a permanent addition to the Nuclei in the Cosmos conference. Recently, the Chetec:INFRA framework established an online school series SNAQs (School on nuclear astrophysics questions) to provide educational and networking opportunities to students during the COVID-19 pandemic. 


\begin{tcolorbox}[colback=white!5, colframe=gray!90!black, title=Sidebar: International Research Network for Nuclear Astrophysics]
  \begin{center}
    \includegraphics[width=\textwidth]{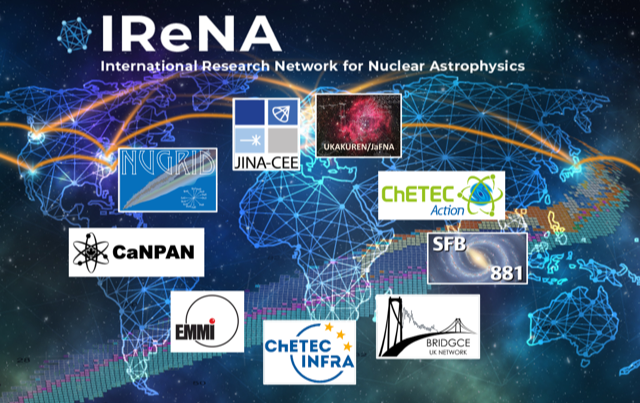}
  \end{center}
  \vspace{-0.20in} 
  \begin{center}
  {\footnotesize {\it IReNA is a National Science Foundation AccelNet Network of Networks.}} \label{fig:wg5:irena} \\  
  \end{center}
  \vspace{-0.08in} 
{\footnotesize
Success in nuclear astrophysics requires international collaboration. Research networks pool regional expertise and resources, kickstart and strengthen collaborations, and help establish common goals in a research community. The International Research Network for Nuclear Astrophysics (IReNA) is a network of networks, joining research networks in nuclear physics, astronomy, and computational science in North America (JINA-CEE, CanPAN), Europe (BRIDGCE, ChETEC, ChETEC-INFRA, EMMI, SFB-881), Asia (UKAKUREN, JaFNA), and virtual space (NuGRID). Since it was founded in 2019, IReNA has improved communication across countries and disciplines to take advantage of the latest developments in astronomy, nuclear experiments, and theory. Enabled by NSF’s AccelNet program, IReNA has employed novel mechanisms for connecting regional research networks across the world into a global community. Such international frameworks will be key to maximizing progress in nuclear astrophysics research in the coming years by opening opportunities to exchange multidisciplinary knowledge and expertise, and enhance training of the next generation of researchers. 
}
\label{sidebar:IReNA}
\end{tcolorbox}

\subsection{What Needs To Be Done?} \label{sec:centers:todo}
The need for centers and larger scale collaboration in nuclear astrophysics is stronger than ever before, and essential for realizing the vision for the field outlined in this white paper.  The development of such centers has been driven by the extraordinary diverse range of expertise and capability across different disciplines that needs to come together to address nuclear astrophysics questions, and the distributed character of this expertise and capabilities, often with single research groups at a given institution. This diversity is now undergoing a phase of significant further growth, with rare isotope science entering a new stage with a new generation of accelerator facilities coming online, with the advent of multi-messenger astronomy in particular the detection of gravitational waves, new underground accelerators for nuclear astrophysics, and new areas in astronomy with strong potential impact on nuclear astrophysics such as asteroseismology or time domain astronomy. At the same time, the number of research groups working in nuclear astrophysics, and the number of facilities and instruments focusing on this field are growing as well. 

Successful nuclear astrophysics centers combine traditional center and network aspects – they are centers in that they serve as attractors and focal points for the broader community, and they form networks by connecting multiple institutions and islands of expertise to achieve scientific goals beyond the capability of individual research groups. The most important roles and features of centers in nuclear astrophysics will continue to be:  

{\bf Connecting islands of expertise across disciplines to advance the science:} such connections are essential to achieve the scientific goals. In nuclear astrophysics this is of particular importance as the expertise and capabilities are exceptionally broad. Centers address the challenge of these islands of expertise being highly distributed across institutions and countries, sometimes as single isolated research groups, or even a single researcher. In addition, to address the overarching science themes nuclear astrophysics ultimately relies on relatively sparse astronomical, planetary,  and cosmochemical observations – a rapid connection between nuclear physics work, modeling, and the latest observations at every stage is therefore especially important. This requires a range of expertise that is not typically available in a given research group. 

{\bf Sustain and grow an inclusive, interdisciplinary nuclear astrophysics community:} centers play a critical role in forming the interdisciplinary research community needed to seize the scientific opportunities in nuclear astrophysics. Successful centers have an explicit focus on initiating and fostering communication between different areas, and overcome cultural and scientific language divisions across fields. It is important for centers to be open, nimble, and flexible to be able to respond to new developments and reach out to new communities in a rapidly evolving field. For researchers at small or remote institutions, or in countries without a strong nuclear physics research effort, networks are an invaluable way to interact with a broader community and for the community to take advantage of their expertise. 

{\bf Foster a diverse scientific community:} With their multi-institutional nature large scale centers can be effective agents in advancing goals towards full representation of marginalized communities (Section~\ref{sec:diversity}). Centers can draw on the combined expertise and best practices in multiple institutions, countries, and scientific fields, and new insights and actions can have a broader impact. An important role of centers is also to connect researchers from a broader range of institutions such as large research universities, small colleges, national laboratories, and minority serving institutions. 

{\bf Provide interdisciplinary educational opportunities for early career researchers:} Centers provide resources and mechanisms for training and knowledge transfer across disciplines that do not otherwise exist in educational systems. Centers are therefore essential for training the nuclear astrophysics workforce needed to take advantage of the new interdisciplinary opportunities in the future. At the same time, the same interdisciplinary skills that make nuclear astrophysics researchers successful, apply to the demands of modern work places and thus provide students with professional development opportunities for a broad range of careers. Centers help early career researchers to see broader context and the big picture of their work and help overcome isolation in small research groups. Interdisciplinary teaching resources help early career scientists prepare for teaching roles that were not in their particular area of training.

{\bf Trigger new directions and new scientific opportunities:} Centers bring together different research areas, foster the exchange of new ideas, and connect new capabilities and developments in multiple fields. This leads to new developments and directions in nuclear astrophysics, new scientific opportunities, and new discoveries that would not otherwise happen. 

{\bf Facilitate transfer of data and knowledge across subfields:}  Centers facilitate exchange of data and knowledge across field boundaries. This includes infrastructure for sharing data and results, addressing the particular challenge that often researchers in one field have to be able to access data in another field.

{\bf Leverage resources and enable broad access and sharing:} Centers play an important role in facilitating access to instruments, equipment, methods, and other resources, as well as to knowledge of how to use these efficiently. In many cases centers also drive the development of resources with particular importance for nuclear astrophysics that are subsequently becoming available to the community.

{\bf Sustain networks of collaborations:} Centers can provide continued interaction points for networks of collaborations, such as small focused workshop series or long-term visitor programs. Such frameworks are critical to maintain research momentum across disciplines over extended periods of time. 

{\bf Facilitate effective dissemination of results and outreach:} Centers can coordinate communication of scientific advances to experts via targeted messaging networks and to the general public via center-supported outreach mechanisms.

{\bf Combine good practices from different fields:} Centers enable the community to combine best practices from different fields and create an environment where different fields can inspire each other to improve. Examples include best practices related to Diversity, Equity, Inclusion and Accessibility (DEIA) (Section~\ref{sec:diversity}), data sharing culture, and open source computer codes.

{\bf Give the field a voice:} Centers play an important role in the community bringing together the researchers from different fields to define the frontiers of nuclear astrophysics and give the field a voice. This is an important complement to existing field specific mechanisms such as the Nuclear Physics Long Range Plan and the Decadal Surveys in nuclear physics and astronomy, which operate predominantly within their discipline. 

{\bf Foster partnerships and collaboration:} Centers foster multi-institutional and international partnerships and collaboration. They also provide opportunities for early career searchers to develop the skills needed to navigate and thrive in such collaborations.

\subsection{What Do We Need?} \label{sec:centers:needs}
\begin{itemize}
\item Centers and center-based networks are essential to achieve the scientific goals of the field, and to continue to define the frontiers of the field as new discoveries are made and new capabilities emerge. Sustained support for nuclear astrophysics centers is necessary to maintain the rapid pace of progress seen in the field in recent years.
\item Larger scale collaboration, fostered by centers, is essential in nuclear astrophysics. Funding agencies should support such collaboration that cuts across typical single investigator and group funding structures, institutional leaders should embrace and appreciate collaborative breakthroughs and successes, and science policy should encourage and facilitate international collaboration. 
\item Closer connections between the various communities important for nuclear astrophysics in the multi-messenger era. These include a broad range of nuclear science, nuclear data, astrophysics, neutrino physics, gravitational wave physics, cosmic-ray, neutrino observation, cosmochemistry, and planetary science communities as well as a broad range of institutions from small universities, minority serving institutions, research universities, national laboratories, NASA, and international institutions. 
\end{itemize}

\ack
We thank all participants of the 2020 JINA Horizons meeting for their active participation that formed the basis of this paper. The organization of the meeting and the writing of this paper have been supported by the US National Science Foundation through awards PHY-1430152 (Joint Institute for Nuclear Astrophysics JINA-CEE) and OISE-1927130 (IReNA), and in part by the ExtreMe Matter Institute EMMI at the GSI Helmholtzzentrum für Schwerionenforschung in Darmstadt, and the “ChETEC” COST Action (CA16117) supported by COST (European Cooperation in Science and Technology). 
\clearpage
\bibliography{bibs/references, bibs/fxt}

\end{document}